\documentclass{aa}  
\usepackage{graphicx}
\usepackage{txfonts}
\usepackage{natbib}
\usepackage{booktabs} 
\usepackage{hyperref}
\usepackage{amsmath} 
\usepackage[switch]{lineno}
\usepackage{xcolor}
\usepackage{lscape}
\usepackage{placeins}
\usepackage{cuted}
\usepackage{stfloats}
\usepackage{subcaption}
\usepackage{microtype} 
\usepackage[justification=raggedright, singlelinecheck=false]{caption}

\hypersetup{
    colorlinks=true,
    linkcolor=blue,
    citecolor=blue,
    filecolor=magenta,
    urlcolor=blue
}

\begin{document}

   \title{Environmental dependence of Type Ia supernova standardisation on the local luminosity-weighted age}
\titlerunning{Environmental dependence of Type Ia supernova standardisation on the local LWA}

\author{
    Yuhui Zhang\inst{1,2,3}\fnmsep\thanks{Corresponding authors: zhangyuhui@ynao.ac.cn; xiangcunmeng@ynao.ac.cn}
    \and
    Xiangcun Meng\inst{1,2}\fnmsep\protect\footnotemark[1]
    \and
    Jingxiao Luo\inst{1,2,3}
    \and
    Xiejin Li\inst{1,2,3}
    \and
    Yunkun Han\inst{1,2}
    \and
    Fenghui Zhang\inst{1,2}
}

   \institute{Yunnan Observatories, Chinese Academy of Sciences (CAS), Kunming 650216, People's Republic of China
         \and
         International Centre of Supernovae (ICESUN), Yunnan Key Laboratory of Supernova Research, Kunming 650216, People's Republic of China
         \and
             University of Chinese Academy of Sciences, Beijing 100049, PR China
             }

   \date{Received 6 February 2026 / Accepted 4 September 2026}

 \abstract
{The dependence of the Type Ia supernova (SN~Ia) standardised luminosity on host galaxy properties is a major source of uncertainty in cosmology. As next-generation surveys like LSST and HLTDS reduce statistical uncertainties, cosmological precision will be dominated by astrophysical systematic errors. However, the widely used empirical mass step, acting as an indirect global proxy, obscures the direct physical link to the progenitor's environment, thereby potentially limiting the precision of SN~Ia luminosity standardisation.}
{We test whether the local luminosity-weighted age (LWA), used here as a statistical proxy for the age of the stellar population associated with SN~Ia progenitors, is more closely related to the physical driver of standardised-luminosity variations than host galaxy mass.}
{Using SDSS-MaNGA Pipe3D, we extracted the local LWA within a 1~kpc aperture for 56 SNe~Ia and performed a joint likelihood analysis to disentangle the impacts of local age and mass on Hubble residuals.}
{We find that SNe~Ia in younger environments are significantly fainter than those in older ones, exhibiting a step amplitude of $0.163$~mag ($5.2\sigma$) after standardisation. 
While global and local mass steps are initially observed ($0.071$~mag, $2.0\sigma$ and $0.087$~mag, $2.4\sigma$, respectively), both become statistically insignificant when fitting for age and mass simultaneously: the global mass step decreases to $0.028$~mag ($0.9\sigma$), while the corresponding age step remains robust at $0.156$~mag ($4.9\sigma$). 
Similarly, the local mass step becomes statistically insignificant ($0.012$~mag, $0.3\sigma$), whereas the age step persists at $0.157$~mag ($4.4\sigma$). 
Furthermore, the weighted Hubble residual decreases from $0.1550$ to $0.1376$~mag after the local LWA step is introduced.}
{Our results provide evidence that the mass step is substantially reduced when the local LWA is included. Furthermore, the inferred dark energy equation of state parameter ($w$) could be affected by the treatment of the local LWA, highlighting the potential importance of environmental effects for next-generation SN~Ia cosmology.}


   \maketitle
   \nolinenumbers

\section{Introduction}

Type Ia supernovae (SNe~Ia), owing to the high consistency of their peak absolute magnitudes, are widely used as distance indicators. They are thus referred to as standard candles and occupy a central role in modern cosmology \citep{Phillips1993, Hamuy1995, Riess1996, Tripp1998}. Observations of SNe~Ia revealed the accelerating expansion of the Universe \citep{Riess1998, Schmidt1998, Perlmutter1999a} and remain core probes for constraining the nature of dark energy and its evolution \citep{Garnavich1998, Perlmutter1999b, Brout2019, Riess2022}. With the imminent commencement of fourth-generation large-scale survey projects, including the Legacy Survey of Space and Time (LSST) \cite[]{Ivezic2019} and the High-Latitude Time-Domain Survey (HLTDS) \cite[]{Spergel2015}, the number of known SNe~Ia is expected to increase substantially. To match the statistical precision of future experiments, resolve the Hubble tension between current local Hubble constant measurements and cosmic microwave background (CMB) inferred values \citep{Planck2020, DiValentino2021, Freedman2021, Riess2022,Cai2026}, and achieve high precision in cosmological distance measurements, a strict control of systematic errors arising from astrophysical factors during the luminosity standardisation is imperative.

Despite the success of luminosity standardisation \citep{Phillips1993, Tripp1998, Riess1998, Perlmutter1999a}, SNe~Ia still exhibit significant luminosity dispersion driven by host galaxy properties. Crucially, these environmental dependences pose a risk beyond merely increasing the scatter in the Hubble diagram. Since the properties of host galaxies -- particularly the stellar population age -- evolve with cosmic time (known as progenitor drift), any uncorrected dependence of SN~Ia luminosity on its local environment will manifest as a redshift-dependent systematic bias \citep[e.g.][]{Branch2001, Rigault2020, Son2025}. Such biases could mimic cosmological signals, potentially leading to significant shifts in the inference of the dark energy equation of state parameter, $w$ \citep{Campbell2016, Lee2022}. Therefore, identifying the fundamental physical driver of these environmental effects is essential for unbiased cosmology. 

Recent studies have found that the standardised luminosities of SNe~Ia are correlated with the global host galaxy mass, an effect known as the `mass step': SNe~Ia in lower-mass host galaxies are fainter than those in higher-mass host galaxies after luminosity standardisation \citep[e.g.][]{Kelly2010, Sullivan2010, Lampeitl2010, Gupta2011, Childress2013a, Kim2019, Smith2020, Rigault2020, Briday2022}. Although the mass step has been widely applied \citep{Betoule2014, Scolnic2018,BroutScolnic2021}, current research indicates that galaxy colour also exhibits a step \citep[e.g.][]{Rigault2013, Roman2018, Kelsey2021}. Correlations between standardised luminosity and host galaxy stellar-population age \citep{Sullivan2010, Gupta2011, Roman2018, Rigault2020}, metallicity \citep{DAndrea2011, Childress2013a, Hayden2013, Campbell2016}, and dust properties \citep{BroutScolnic2021,Popovic2021, Popovic2023, Meldorf2023} have been reported. In addition, the inferred amplitudes of environmental steps can depend on the adopted light-curve standardisation model. In particular, the conventional single-$\alpha$ correction does not capture the non-linear stretch--luminosity relation \citep{Ginolin2025b}. Joint standardisation using the host galaxy mass and the local stellar age has also been shown to improve distance precision, indicating some combined effect of the host properties on SN Ia luminosity \citep{Rose2021}. However, the interpretation of age dependence remains contentious; for instance, while \citet{Son2025} used host galaxy age as a proxy to claim a strong dependence on progenitor age, \citet{Wiseman2026} argue that host galaxy age is not a reliable tracer for the progenitor age, and that conflating global host age with the age of the specific SN~Ia progenitor could lead to erroneous conclusions. Currently, the physical origin of the mass step remains debated \citep{ThorpMandel2022, Wiseman2023,Popovic2025}, and the understanding of the dependence of standardised luminosity on host galaxy properties is incomplete. Therefore, further research into the dependence of SN~Ia luminosity on a host galaxy's environment is necessary \citep{BroutScolnic2021, Briday2022}. 

From a physical perspective, these environmental dependences could be used to constrain the SN~Ia progenitor systems. The debate over progenitor channels -- specifically between single-degenerate accretion and double-degenerate merger scenarios -- remains unresolved \citep[e.g.][]{Hillebrandt2013, Maoz2014, RuiterSeitenzahl2025}. Theoretical models predict that these channels lead to distinct explosion mechanisms, ranging from near-Chandrasekhar mass ($M_{\rm Ch}$) delayed detonations \citep{Khokhlov1991, HoeflichKhokhlov1996} to sub-$M_{\rm Ch}$ double detonations \citep{Fink2010, Sim2010} and super-$M_{\rm Ch}$ mergers \citep{Pakmor2012}. Crucially, these diverse progenitor channels are believed to operate on distinct timescales, meaning their contributions may evolve with the age of the stellar population. This temporal connection is constrained by the SN~Ia delay-time distribution (DTD): early `prompt + delayed' parameterisations \citep[e.g.][]{Mannucci2005, Scannapieco2005, Sullivan2006, Aubourg2008} and later results favouring a smoother $\propto t^{-1}$ DTD \citep[e.g.][]{Maoz2014} imply that observed samples mix progenitors with a wide range of ages. This naturally produces young and old modes in the progenitor age distribution when convolved with galaxy star formation histories \citep{Childress2014}. This physical link strongly suggests that the observed environmental dependences are driven by the diversity of progenitor ages.

However, relying on global host galaxy properties as proxies for the SN environment is often insufficient. Massive galaxies typically exhibit complex star formation histories, where global properties can dilute the true physical signal at the specific explosion site. For instance, a SN exploding in a localised star-forming region at the edge of a passive massive galaxy experiences an environment vastly different from the galaxy's global average \citep{Rigault2013, Roman2018, Meng2019}. Consequently, recent research has shifted towards analysing local environmental parameters within a few kiloparsecs of the SN \citep[]{Stanishev2012, Galbany2014}. Pioneering studies demonstrated that corrections based on the local specific star formation rate (sSFR) significantly outperform global mass corrections \citep[]{Rigault2013, Rigault2020}, strongly suggesting that the mass step is driven by the local stellar population age. However, establishing the robustness of this age-driven hypothesis requires probing the stellar population age across multiple dimensions using independent physical indicators.

Although the LsSFR is an effective tracer of very recent star formation and can broadly separate environments dominated by young or old populations \citep{Kennicutt1998}, it does not provide a quantitative constraint on the age distribution of the SN~Ia progenitor population -- especially in passive galaxies where star formation has largely ceased. To overcome this limitation, we need an indicator derived from stellar population synthesis models that can reconstruct the star formation history and hence constrain the stellar population age even in the absence of ongoing star formation \citep[e.g.][]{Conroy2013}.
The Sloan Digital Sky Survey (SDSS) Mapping Nearby Galaxies at Apache Point Observatory (MaNGA) provides the necessary integral field spectroscopy (IFS) data \citep{Bundy2015, Drory2015, Yan2016,Wake2017, Blanton2017}. Leveraging the MaNGA Pipe3D pipeline \citep{Sanchez2016,Sanchez2018}, we adopted the local luminosity-weighted age (LWA) derived from full-spectrum fitting, which uses both the continuum shape and absorption features to constrain stellar population ages. This allowed us to effectively distinguish between young and old stellar populations over a wide range of stellar-population ages. In this work, we used the local LWA as a statistical proxy for the age of the stellar population associated with SN~Ia progenitors.

We also used the local LWA derived from simple stellar population (SSP) fitting to investigate the environmental dependence of SN~Ia standardisation. We used a wide-field, un-targeted SN sample provided by the \textit{Zwicky} Transient Facility (ZTF; \citealt{Bellm2019,Graham2019}), combined with the MaNGA Pipe3D stellar population products, to obtain environmental parameters within a 1~kpc radius of the SN explosion site. Following the general procedure for studying the magnitude step \citep[e.g.][]{Kelly2010, Sullivan2010, Rigault2020, Burgaz2025}, we divided the SN~Ia sample into young and old subsamples based on their local LWA. By comparing the Hubble residual step analyses of these two groups, we aim to:  (1) verify whether local LWAs based on IFS measurements can effectively separate SN~Ia populations of different luminosities; and (2) compare the effect of local LWA on SN Ia standardisation with the local and global mass step to determine the dominant factor driving the mass step. In Sect. \ref{sec:method} we describe the cross-matching of the Transient Name Server (TNS) catalogue with the MaNGA survey and detail the methodology for deriving local environmental metrics, specifically the local LWA, as well as the global fit framework used to minimise Hubble residuals. Section~\ref{sec:results} presents our main results regarding the correlations between Hubble residuals and environmental properties, focusing on the comparison between the mass step and the age step. In Sect. \ref{sec:robustness} we perform a series of robustness checks to validate the stability of our findings. Finally, we discuss the implications and limitations of our results in Sect.~\ref{sec:5}, and summarise our main conclusions in Sect.~\ref{sec:conclusion}.

\section{Methodology} \label{sec:method}

In this section we outline the data reduction and analysis pipeline developed to investigate the correlation between SN~Ia standardised luminosity and local host galaxy properties. Our workflow proceeds in six main stages: cross-matching the TNS SN~Ia sample with the MaNGA data reduction pipeline map to identify hosts (Sect.~\ref{sec:crossmatch}); correcting for Milky Way extinction and redshift effects (Sect.~\ref{sec:corrections}); acquiring and filtering forced photometry from ZTF to construct high-quality light curves (Sect.~\ref{sec:quality}); fitting the light curves with the SALT2 model to derive standardisation parameters (Sect.~\ref{sec:fitting}); extracting local environmental properties including the stellar mass and local LWA (Sect.~\ref{sec:localparams}); and analysing the impact of local environmental properties on Hubble residuals (Sect.~\ref{sec:hubble}). The rigorous selection criteria applied at each stage ensure a robust sample for cosmological analysis. After subsequent light curve quality filtering (Sect.~\ref{sec:quality}), fitting selection (Sect.~\ref{sec:fitting}), local parameter extraction (Sect.~\ref{sec:localparams}), and Hubble residual fitting (Sect.~\ref{sec:hubble}), the final sample contained 56 SNe~Ia.

\subsection{Cross-matching the TNS with MaNGA} \label{sec:crossmatch}

We first queried the public database of the TNS and obtained a sample containing 11,524 transient sources spectroscopically confirmed as normal SNe~Ia \citep[]{Branch1993}. To obtain detailed host galaxy properties for these SNe, we cross-matched this sample with the Value Added Catalogue (VAC) of the SDSS MaNGA project.

\begin{figure}
       \resizebox{0.9\hsize}{!}{\includegraphics{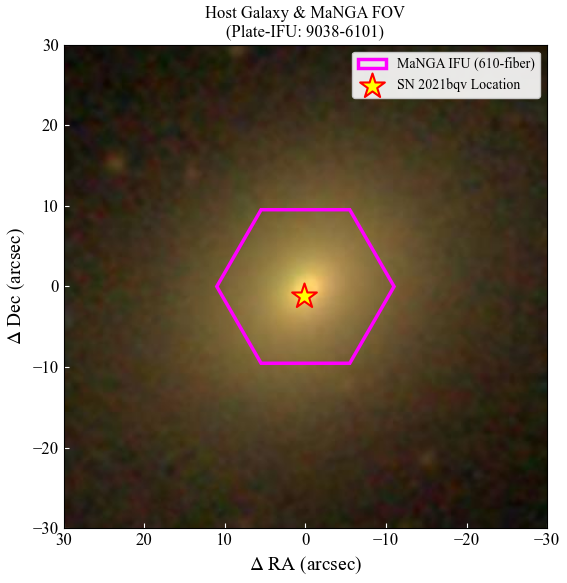}}
      \caption{Illustration of an IFU observational FoV using SN 2021bqv as a representative example. The purple hexagonal frame is the IFU observational FoV coverage of SDSS MaNGA for the host galaxy of this SN, and the yellow star is the location of the SN. (The optical image is from SDSS.)}
   \label{fig1}
      \vspace{-0.33cm}
\end{figure}

We identified host galaxies using a joint strategy based on spatial coordinates and redshift matching. Initially, we defined a conservative search radius of $40\arcsec$ around the coordinates of SNe~Ia provided by the TNS. We selected this threshold to sufficiently encompass the spatial extent of the targeted galaxies. Figure~\ref{fig1} illustrates the spatial coverage of the SDSS MaNGA integral field unit (IFU) using the host galaxy of SN 2021bqv as a representative example. For our full sample, the IFU bundles range from $37$ to $127$ fibres, corresponding to a field of view (FoV) with diameters between $17\arcsec$ and $32\arcsec$ (radii of $\sim 8.5\arcsec$ to $16\arcsec$). Since our $40\arcsec$ search radius significantly exceeds the maximum MaNGA FoV radius, it ensures a complete inclusion of any SN~Ia that falls within the effective IFU coverage, while the subsequent redshift matching effectively filters out chance alignments.

Subsequently, to eliminate chance alignments along the line of sight of the host galaxy, we introduced the directional light radius method \citep{Sullivan2006,Gupta2016}. For each candidate host, we computed the normalised separation between the SN and the galaxy centre in units of the directional light radius, $\mathrm{dDLR}$, based on the galaxy effective size, axis ratio, and position angle. When multiple candidates are present, we kept the one with the smallest $\mathrm{dDLR}$ and required $\mathrm{dDLR} \leq 5.4$ for host assignment. Finally, we applied strict spectral redshift consistency constraints. We required the redshift difference between the SN and the host galaxy to be $|\Delta z| < 0.003$, which corresponds to a velocity tolerance of about $900$ km s$^{-1}$, sufficient to cover the peculiar velocity range of the host galaxy. After these selection steps, this selection yields a final sample containing 166 matched SN~Ia--host pairs. Notably, the redshift differences of our final sample are generally less than 0.001 ($\sim 300$ km s$^{-1}$), further verifying the reliability of host galaxy identification. 

\begin{figure}
    \centering
    \resizebox{\hsize}{!}{\includegraphics{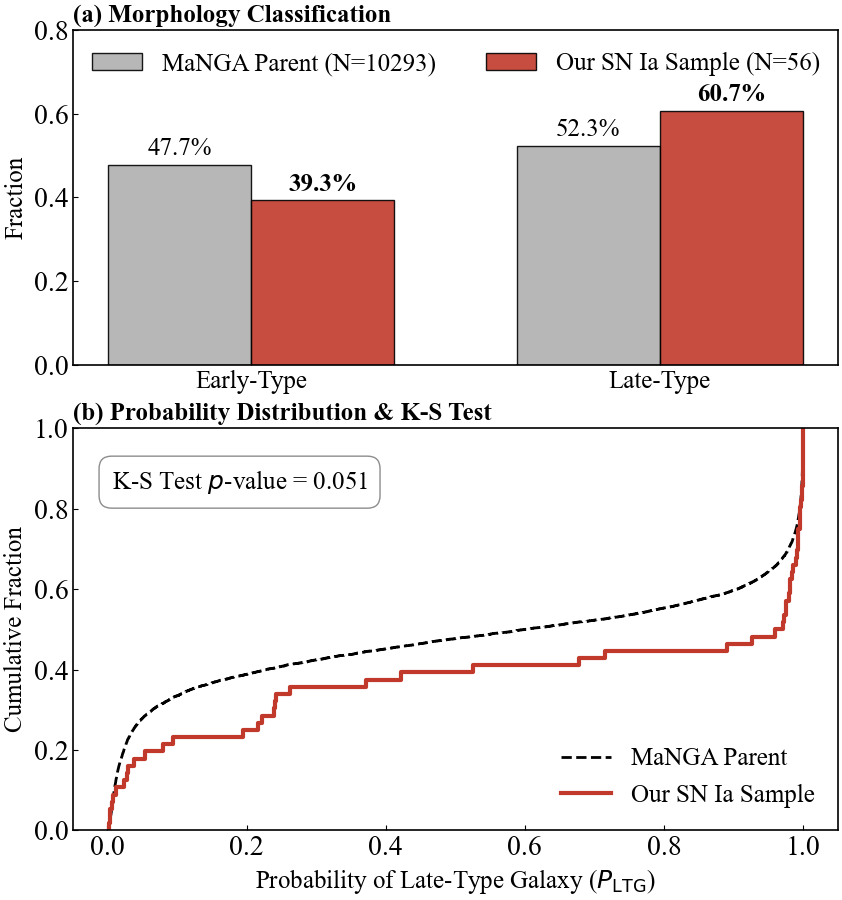}}
    \caption{
    Validation of the host galaxy morphology distribution. 
    \textit{Panel (a)}: Categorical comparison showing the fraction of early-type versus late-type galaxies in the MaNGA parent sample (grey) and our SN~Ia host sample (red). Classifications are based on a probability threshold of $P_{\text{LTG}}=0.5$. 
    \textit{Panel (b)}: Cumulative distribution function (CDF) of the continuous morphology probability, $P_{\text{LTG}}$ (0=pure early-type, 1=pure late-type). The dashed black line traces the parent sample, and the solid red line traces our SN~Ia sample. 
    The red curve lies below the black curve, indicating that our sample accumulates more slowly at low probabilities and is skewed towards higher $P_{\text{LTG}}$ values (i.e. a preference for late-type hosts).}
    \label{fig:morph_check}
    
\end{figure}

Given the established connections between SN~Ia properties and host galaxy morphology \citep{Hamuy1996, Hamuy2000, Pruzhinskaya2020}, it is critical to verify that our sample selection did not introduce morphology-dependent biases. To this end, we compared the host galaxy morphologies of our 56 SNe~Ia with the full MaNGA parent catalogue. Figure \ref{fig:morph_check} presents the comparison using the MaNGA Deep Learning Morphology VAC \citep{DominguezSanchez2018}. As shown in panel (a), our sample exhibits a slightly higher fraction of late-type galaxies (60.7\%) compared with the parent sample (52.3\%). To rigorously quantify this difference, we performed a Kolmogorov-Smirnov (KS) test on the continuous probability distributions, $P_{\mathrm{LTG}}$ (i.e. the probability of being late-type; galaxies with $P_{\mathrm{LTG}} < 0.5$ are classified as early-type and those with $P_{\mathrm{LTG}} \geq 0.5$ as late-type), as shown in panel (b). The test yields a $p$-value of 0.051. Since $p > 0.05$, we cannot reject the null hypothesis that the two samples are drawn from the same underlying distribution. The visible trend in panel (b) -- where the SN~Ia sample (red line) lags below the parent sample (black line) -- indicates a mild distributional shift towards late-type galaxies. This trend is physically expected due to the enhanced SN~Ia rate in star-forming environments \citep{Sullivan2006}, rather than being an artefact of our selection criteria. We therefore find no evidence that our selection procedure introduces a significant morphology-dependent bias.

\subsection{Redshift correction and Milky Way extinction} \label{sec:corrections}

Redshift correction. Given that the sample is in the low-redshift interval ($0.02 < z < 0.08$), the impact of the host galaxy's peculiar velocity on the redshift measurement cannot be ignored. Considering that the uncertainty of the velocity field model dominates the error budget in some regions of the sky, we adopted the CMB-rest-frame redshift $z_{\text{{cmb}}}$ in the subsequent analysis. To statistically eliminate the impact of uncorrected host galaxy peculiar velocities, we added an additional uncertainty term $\sigma_{\text{pec}} = 300$ km s$^{-1}$ (uncertainty floor) to the final error, as adopted in previous works like \cite{Riess2016} and \cite{Brout2019}.

Milky Way extinction. We adopted the all-sky dust reddening map produced by \cite{Schlegel1998} with the recalibration by \cite{Schlafly2011}, named Schlegel-Finkbeiner-Davis (SFD) dust map. This dust map provides all-sky coverage of Milky Way dust column density ($E(B-V)$) based on far-infrared (100~$\mu\mathrm{m}$ and 240~$\mu\mathrm{m}$) COBE/DIRBE and IRAS/ISSA data synthesis. For each SN in the sample, we interpolated the Milky Way colour excess $E(B-V)_{\mathrm{MW}}$ along its line of sight from this map based on its celestial coordinates (RA, Dec). To convert colour excess into extinction values for specific bands, we adopted the parameterised extinction law proposed by \cite{Cardelli1989}. This law describes the shape of the extinction curve from ultraviolet to near-infrared and is a standard choice in extragalactic astronomy.

The purpose of this correction is to remove the foreground extinction effect from Milky Way dust. Without this correction, SNe would appear fainter than they actually are due to obscuration by Milky Way dust, affecting the results of luminosity calibration and cosmological parameter measurements.

\subsection{Raw light-curve data quality filtering and baseline correction} \label{sec:quality}

We used the ZTF Forced Photometry Service (ZFPS) to obtain the raw photometric data for our SN sample \citep{Masci2019,Masci2023}. We emphasise that these data are forced-photometry products generated by the ZFPS on archived ZTF difference images. In this framework, the photometric measurements are obtained through forced point spread function (PSF) fit photometry at the fixed SN coordinates on the difference images, and the primary host galaxy background is suppressed through subtraction of a reference image in the ZTF difference-imaging pipeline. This approach allows us to recover flux measurements below the nominal single-epoch detection threshold and provides a continuous flux history, which is critical for constraining the explosion epoch and assessing the baseline of the differential-flux light curve.

However, the returned fluxes are differential fluxes relative to the reference image rather than absolute host-subtracted source fluxes. As noted in the ZFPS documentation, systematic errors may remain owing to imperfect subtraction, contamination from neighbouring sources, non-photometric observing conditions, or contamination of the reference image itself by transient flux. For this reason, we did not assume that the reference-image subtraction completely removes all host-related or background-related systematic errors; instead, the ZFPS products require subsequent validation, quality filtering, and baseline correction before cosmological use.

In addition, the ZTF survey officially commenced science operations in March 2018. Although our parent sample sourced from the TNS contains records predating this timeline, we excluded targets discovered before 2018 because no ZTF photometric data are available for these events. This cut ensures that all selected targets fall within the ZTF observational window. With our data cutoff set to October 2025, this constraint reduced our initial cross-matched SN~Ia sample to 154 objects.

\subsubsection{Light-curve quality filtering}

Following the ZFPS user guide \citep{Masci2023}, we performed strict quality filtering on the acquired raw light curves. This step eliminates epochs with low signal-to-noise ratio (S/N) affected by observing conditions (such as cloud cover and moonlight pollution) or data processing anomalies. We excluded measurement points that met any of the following conditions: 
\begin{enumerate}
\item Photometric calibration anomaly: The \texttt{infobitssci} flag indicates serious problems with image processing or instrument calibration (usually filtering \texttt{infobitssci} $> 0$ or specific high-order masks). 

\item High background noise: The background noise level \texttt{scisigpix} of the scientific image exceeds the threshold (recommended value $> 25$ DN, where DN denotes data number), which usually implies severe cloud occlusion or moonlight interference. 

\item Poor seeing: The seeing \texttt{sciinpseeing} during the observation is poor (recommended value $> 4$ arcseconds), leading to severe PSF broadening. 

\end{enumerate}

\noindent Although this strict filtering strategy reduces the number of sampling points in the light curve, potentially losing some early or late data obtained in poor weather, it significantly reduces outliers and systematic errors in photometric data. This is crucial for accurately reconstructing SN luminosity evolution using the SALT2 model and precisely determining fitting parameters (Sect.~\ref{sec:fitting}).

\subsubsection{Light-curve baseline correction}

Although the ZFPS pipeline has been preliminarily processed, after completing quality filtering, we still find non-zero baseline offsets in some light curves, typically originating from potential transient source contamination in reference images or incompletely removed systematic errors \cite[]{Masci2019}. Therefore, we implemented a strict baseline-correction procedure, following standard procedures for ZTF photometry \citep[e.g.][]{Yao2019}.

First, to distinguish transient-dominated observations from those
representing the background baseline, we selected measurements with $|S/N|>3$ and sorted them by observation time. Successive selected measurements are placed in the same group if they are separated by no more than 30 days. Using the TNS-reported discovery epoch as a reference, we identified the group whose median observation epoch was closest to the discovery epoch as the SN-dominated time window.

Second, we defined a clean baseline sampling window, selecting epochs at least 50 days before the beginning and at least 50 days after the end of the SN-dominated time window. Before calculating the baseline flux, given that the uncertainty of difference photometry is often underestimated, we followed the empirical advice adopted in \cite{Betoule2014} and \cite{Rigault2025} and introduced an error floor. Specifically, we applied error floor corrections of approximately 2.5\%, 3.5\%, and 6.0\% in the g, r, and i bands, respectively.

Finally, to eliminate outliers we combined median absolute deviation with an iterative sigma-clipping algorithm \cite[]{ivezic2014statistics} to clean the baseline data. Based on the cleaned sample, we used the inverse-variance weighting method \cite[]{bevington2003data} to calculate the average background flux and subtract it from the full time-series data to obtain baseline-corrected light curves.

\subsection{Light-curve fitting} \label{sec:fitting}

We used the SALT2 \cite[]{Guy2007} light-curve fitting model to fit the SN~Ia light curves using the SALT2-T21 version \citep{Taylor2021}. For a detailed description of the model framework, we refer the reader to the original publications; here, we avoid reiterating these details and focus on the parameters relevant to our analysis. This model standardises SN~Ia luminosity using the peak apparent magnitude ($m_{\text{B}}$), light-curve stretch factor ($x_1$), and colour parameter ($c$), where $x_1$ is the light-curve stretch parameter that correlates with the intrinsic luminosity, which describes the brighter-slower relation \cite[]{Rust1974, Pskovskii1977, Phillips1993}, and $c$ is the colour parameter reflecting SN~Ia intrinsic colour variations and host galaxy dust extinction.

Since the peak time ($t_0$) of the ZFPS data is not known a priori, we implemented a two-stage iterative fitting strategy to ensure the robustness of parameter estimation:
The first stage applies Gaussian smoothing and S/N threshold filtering to the raw light curve to roughly estimate the peak time $t_0$. Subsequently, centring on this peak $t_0$, a time window of 20 days before and 50 days after the explosion is selected for preliminary fitting to obtain a high-precision peak time $t_0^{\text{1st}}$. 
 In the second stage, we used the first-stage estimate of the epoch of maximum light, $t_0^{\mathrm{1st}}$, to define a narrower fitting window extending from 10 days before to 40 days after maximum light. We excluded observations at phases earlier than $-10$ days because SALT2-T21 is poorly constrained at these early phases and exhibits systematic residuals, including non-physical negative model fluxes in some cases \citep{Rigault2025}.

\begin{figure}

    \centering
    \resizebox{0.9\hsize}{!}{\includegraphics{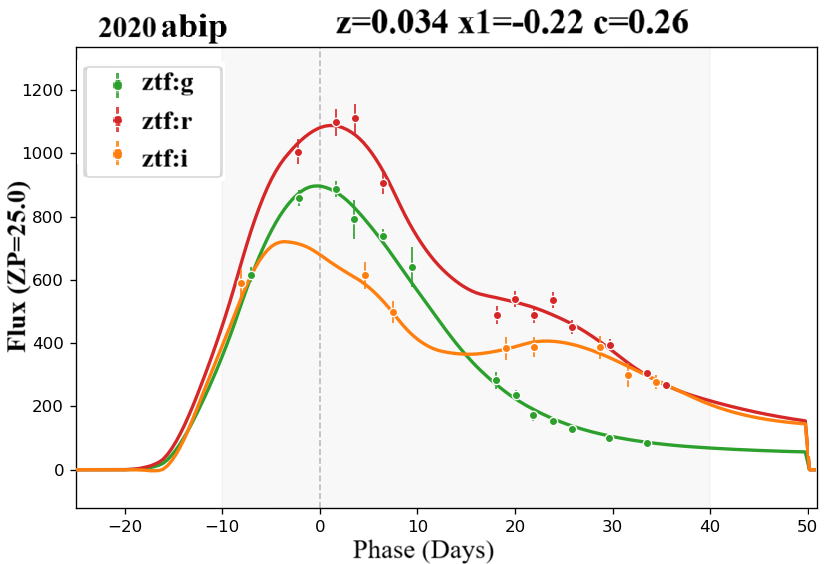}}

    \caption{Example of SALT2 light-curve fitting results for SN 2020abip. Shown are the observed data (points with error bars) and the best-fit curves (solid lines) for the ZTF-$g$ (blue), ZTF-$r$ (green), and ZTF-$i$ (yellow) bands. The fitted parameters ($x_1$ and $c$) are listed at the top.}
   \label{fig2}
   \vspace{-0.33cm}
\end{figure}

Figure \ref{fig2} shows a representative example of the fitting results for SN 2020abip. The model (solid lines) shows excellent agreement with the observed photometric data (points) across the ZTF-$g$, $r$, and $i$ bands.

To ensure the cosmological applicability of the sample, we imposed strict constraints on the light-curve parameters, specifically $|x_1| < 3$ and $|c| < 0.3$ \cite[]{Betoule2014}. These selection criteria exclude peculiar SNe~Ia, ensuring that our final sample consists of normal SNe~Ia. Finally, limited by insufficient data points in the light curves for some targets, 110 out of the 154 initial objects converged successfully. After filtering by goodness of fit (requiring a fit probability $> 0.01$), we finally obtained a high-quality sample of SNe~Ia for the subsequent analysis.

\subsection{Extraction and processing of local environmental parameters} \label{sec:localparams}

To investigate the physical drivers of SN~Ia luminosity variations, we used the spatially resolved capabilities of MaNGA IFU data to derive local environmental properties. Unlike global measurements, local parameter extraction requires careful consideration of the physical resolution and S/N, and the specific nature of the physical quantities involved, such as mass and stellar age. In this section we first detail the definition of the local aperture and the different strategies for the stellar mass and local LWA (Sect.~\ref{sec:2.5.1}), with particular attention to determining the effective physical resolution. Subsequently, we outline the dynamic threshold scanning (Sect.~\ref{sec:classify}) strategy used to objectively classify the SN~Ia sample into distinct populations based on their local stellar population age.

\subsubsection{Definition of the target region and parameter extraction process} \label{sec:2.5.1}

The local environmental properties used in this study were taken from the MaNGA Pipe3D VAC, which contains pixel-level parameters, rather than being re-fitted ourselves from the raw MaNGA spectra. For the DR17 MaNGA Pipe3D products used in this work, the released stellar-population decomposition is based on a MaStar-based SSP library. The corresponding MaStar spectra cover 3622--10354~\AA, and the DR17 MaNGA Pipe3D VAC uses a template set comprising 39 ages and 7 metallicities.

Regarding the age--dust degeneracy, dust extinction is explicitly modelled in Pipe3D through the \cite{Cardelli1989} extinction law, and the initial reduced-template fit is specifically designed to reduce the degeneracy between dust and stellar-population parameters. We therefore did not assume that the age--dust degeneracy was completely removed; rather, it was explicitly modelled and mitigated within the full-spectrum fitting framework.

\begin{figure*}
\sidecaption
   \includegraphics[width=12cm]{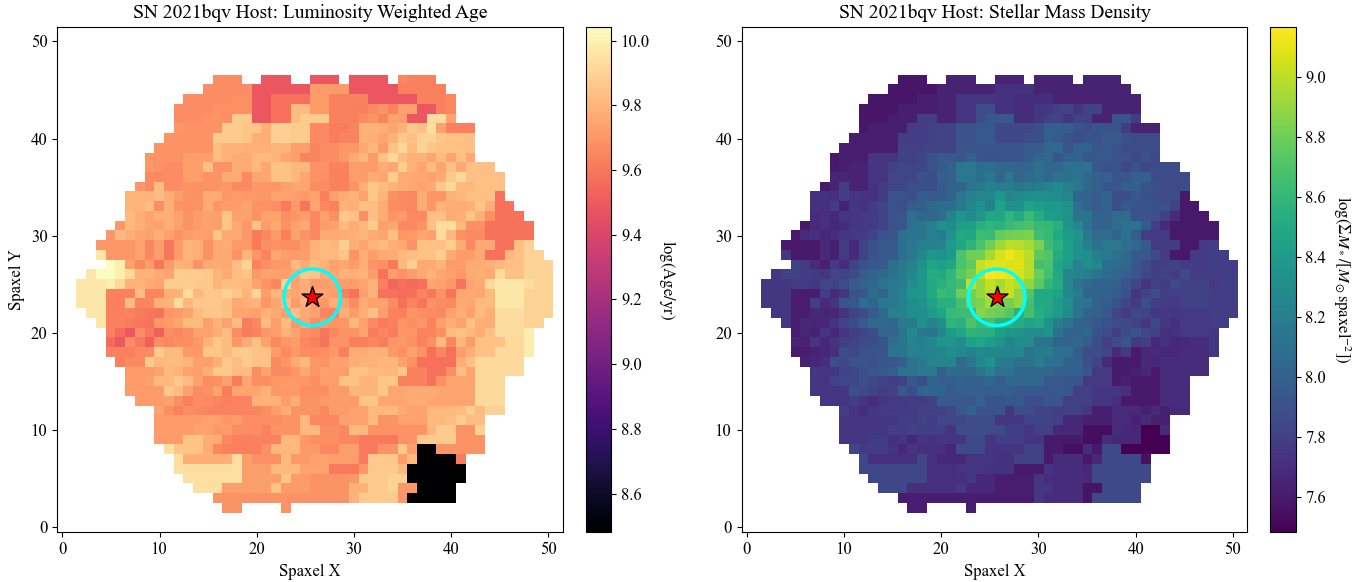}
     \caption{Map of local properties for SN 2021bqv. The maps show the local LWA and stellar mass surface density derived with Pipe3D. The red star marks the SN position, and the blue circle marks the 1 kpc aperture centred on the SN.}
     \label{fig3}
\end{figure*}

To facilitate a consistent comparison among all SNe, we defined the local environment as a circular aperture with a physical radius of 1~kpc centred on the explosion site of each SN \citep[e.g.][]{Rigault2013,Rigault2020}. This aperture provides a common physical scale for all measurements. However, the physical region represented by this aperture is ultimately limited by the spatial resolution of the MaNGA observations.
The effective spatial resolution is determined by two factors:\begin{itemize} 

\item Native spatial resolution limit. The typical full width at half maximum (FWHM) of the PSF of MaNGA data is about $2.5\arcsec$ \citep{Law2016}. Within the redshift range of our sample ($0.02 < z < 0.08$), this corresponds to a physical scale of approximately 0.5 to 4.2~kpc.

\item Spatial binning. Pipe3D first performs Voronoi spatial binning \cite[]{Cappellari2003} on the spectra to achieve the required S/N, so the values of these environmental parameters in a specific bin are equal. This operation further degrades spatial detail information.
\end{itemize}
As shown in Fig. \ref{fig3}, the values we measured from the 1~kpc aperture are essentially mixed information within a larger scale region (usually $> 1$ kpc) composed of multiple pixels (spaxels) centred on the SN position and determined by the effective resolution of the data.

Given the variation of physical resolution with redshift (resolution bias), we adopted different strategies based on the different nature of physical parameters  to ensure physical consistency and robustness of measurements:\begin{itemize}

\item Local stellar mass. We extracted the dust-corrected pixel-level stellar mass surface density from the 19th channel of the SSP extension (HDU1) in the MaNGA Pipe3D data cubes. To eliminate resolution differences caused by redshift, we performed PSF homogenisation \cite[]{Mast2014, Galbany2014}. First, we linearised the logarithmic mass surface density map, then convolved with a Gaussian kernel to degrade the native resolution of low-redshift galaxies to the target physical scale of 2.0 kpc radius. Subsequently, we calculated the mean stellar mass surface density within the aperture, weighted by the fractional coverage of each spaxel. This observed density was then deprojected to a face-on orientation using the galaxy axis ratio (b/a) from the NASA-Sloan Atlas \cite[]{Blanton2011}. Finally, the total local stellar mass was derived by multiplying this face-on surface density by the physical area of the 1~kpc aperture. This approach, combined with our PSF homogenisation, ensures consistent physical resolution and minimises systematic biases caused by varying spatial resolutions.

\item Local LWA. We extracted the local LWA maps from the 5th channel of the SSP extension (HDU1) in the MaNGA Pipe3D data cubes. These parameter maps were derived from Voronoi-binned spectra to ensure sufficient S/N. Unlike the stellar mass surface density, which we linearised and homogenised to a fixed physical PSF, we used the age maps at their native resolution. We adopted this approach because stellar age is an intensive property provided in logarithmic units ($\log(\mathrm{yr})$). Spatial convolution of logarithmic quantities is mathematically non-robust and could introduce biases. To derive the local LWA within our standardised 1~kpc physical aperture, we computed the weighted mean of the logarithmic age for all spaxels enclosed by the aperture. We used the reconstructed V-band flux (SSP layer 0) as the weighting factor. This luminosity weighting ensures that the extracted age faithfully reflects the stellar populations that dominate the optical light output within the local 1~kpc environment, effectively mitigating the resolution variations without artificial smoothing. This procedure adheres to the standard definition by \cite{Sanchez2016}, who emphasise that calculating the weighted mean on a logarithmic scale is necessary to mitigate biases arising from the logarithmic time-sampling of stellar population templates. Consequently, the local LWA is quantified as
\begin{equation}\langle \log \text{Age} \rangle_L = \frac{\displaystyle \sum f_i \cdot \log(\text{Age}_i)}{\displaystyle \sum f_i},
\end{equation}
where $f_i$ represents the flux weight (V-band flux) and $\log(\text{Age}_i)$ denotes the logarithmic age of the $i$-th spaxel within the aperture.

\end{itemize}Notably, for low-redshift galaxies whose native resolution is better than 2~kpc, this processing blurs their information; while for high-redshift galaxies whose native resolution is worse than 2~kpc, their measured values come from regions larger than 2~kpc. Therefore, all measurements based on 1~kpc physical apertures in this study should be understood as smoothed values reflecting the environment within 1~kpc centred on the SN Ia at an effective resolution scale of about 2~kpc.

Finally, through geometric filtering of MaNGA IFU field coverage, we excluded 11 targets whose explosion positions fell outside the observation FoV. This reduces the effective SN~Ia sample size to 68.

\subsubsection{Classifying the SN~Ia sample by the local LWA} \label{sec:classify}

Previous studies have discussed the observed dependence of standardised SN~Ia luminosity on their host galaxy environments, including differences in host galaxy mass, progenitor age, dust properties, and metallicity. Among the various environmental properties, a significant correlation between the Hubble residuals and the LsSFR has been reported by \cite{Rigault2020}; this correlation motivates our focus on local stellar population age, which may be one of the fundamental physical factors driving the dispersion of Hubble residuals and affecting the precision of cosmological distance measurements. Consequently, we used the local LWA as a statistical tracer of SN~Ia progenitor age and divide the sample into younger and older environments.

Due to the limitation of the number of existing spectral samples, artificially specifying a fixed age threshold (such as the median or a specific physical value) could introduce subjective selection bias, thereby masking the true physical signal. To objectively determine the optimal physical boundary distinguishing young and old stellar populations, this study adopts a dynamic threshold scanning strategy.

Specifically, we scanned all possible split points within the 10\% to 90\% quantile range of the local age distribution. This trimming of the sample boundaries ensures that both the young and old subsamples contain a sufficient number of objects to provide reliable statistical power and prevents the threshold search from being dominated by extreme outliers at the distribution tails. For each potential threshold, we divided the sample and calculated the Hubble residual step and its statistical significance. We determined the optimal threshold by maximising this significance; this approach effectively minimises the residual sum of squares of the model, which is statistically equivalent to a likelihood ratio test for threshold models and yields the maximum-likelihood estimate (MLE) for the luminosity step \citep{Hansen2000}. To further ensure the robustness of our results and mitigate interference from small-sample fluctuations, we imposed a strict boundary constraint, requiring each subsample to contain at least 20 SNe.

\begin{table*}[ht!]
\caption{Data used for the analysis in this study (extract).}
\label{tab1}
\centering

\begin{tabular}{l c c c c c c c}
\hline\hline
SN Name & $z_{\text{{cmb}}}$ & $\Delta \mu$ (mag) & $x_1$ & $c$ & Global Mass & Local Mass & local LWA \\
& & & & & $\log(M_*/M_\odot)$ & $\log(M_*/M_\odot)$ & $\log(\mathrm{yr})$ \\
\hline
2022yrt & 0.05042 & $-0.003 \pm 0.144$ & $0.189 \pm 0.236$ & $0.004 \pm 0.034$ & 10.725 & $9.159 \pm 0.007$ & $9.381 \pm 0.016$ \\
2019lpd & 0.06153 & $0.006 \pm 0.143$ & $-0.383 \pm 0.163$ & $0.165 \pm 0.034$ & 10.391 & $9.835 \pm 0.007$ & $8.826 \pm 0.007$ \\
2023wvo & 0.06585 & $0.005 \pm 0.151$ & $-0.913 \pm 0.237$ & $0.081 \pm 0.039$ & 11.061 & $9.807 \pm 0.008$ & $9.598 \pm 0.033$ \\
2024pi & 0.04686 & $-0.009 \pm 0.144$ & $-0.633 \pm 0.191$ & $0.085 \pm 0.032$ & 10.564 & $8.972 \pm 0.007$ & $9.254 \pm 0.015$ \\
2023gdj & 0.04574 & $-0.019 \pm 0.174$ & $-0.155 \pm 0.346$ & $0.005 \pm 0.044$ & 10.025 & $9.453 \pm 0.004$ & $9.380 \pm 0.012$ \\
2021zoj & 0.08260 & $-0.033 \pm 0.144$ & $-0.589 \pm 0.220$ & $-0.064 \pm 0.032$ & 10.830 & $8.088 \pm 0.017$ & $9.013 \pm 0.018$ \\
2018fbh & 0.04111 & $0.034 \pm 0.142$ & $0.324 \pm 0.229$ & $0.112 \pm 0.029$ & 10.568 & $9.318 \pm 0.004$ & $9.030 \pm 0.016$ \\
... & ... & ... & ... & ... & ... & ... & ... \\
\hline
\end{tabular}
\tablefoot{The full table is available at the CDS.}
\end{table*}

\subsection{Hubble residual fitting} \label{sec:hubble}

Following previous studies \citep[e.g.][]{Kelly2010, Sullivan2010, Rigault2020}, to standardise the SN~Ia luminosities and calculate Hubble residuals, we used the model from \citep{Tripp1998}, which relates the standardised distance modulus $\mu$ to the SALT2 light-curve parameters:
\begin{equation}
    \mu_\text{obs}^\text{corr} = m_B + \alpha x_1 - \beta c - M,
    \label{eq2}
\end{equation}
where $x_1$ is the light-curve stretch parameter and $c$ is the colour parameter. The coefficients $\alpha$ and $\beta$ characterise the global luminosity standardisation corrections for stretch and colour, respectively, and $M$ is the absolute B-band magnitude of a fiducial SN~Ia (the absolute magnitude is centred around a standard value of $M \approx -19.1 \mathrm{mag}$ in this study). The peak apparent magnitude $m_B$ is derived from the SALT2 light-curve amplitude $x_0$ via the relation $m_B = -2.5 \log_{10}(x_0) + M_0$. Here, we adopted a fixed magnitude zero-point constant of $M_0 = 10.635$ \cite[]{Betoule2014}, which converts the dimensionless SALT2 flux amplitude $x_0$ to the standard B-band magnitude in the AB system. 

Following previous studies \citep[e.g.][]{Sullivan2010,Betoule2014, Rigault2020}, the Hubble residual, $\Delta\mu$, is defined as the difference between the observed standardised distance modulus and the theoretical distance modulus predicted by the cosmological model:
\begin{equation}
    \Delta\mu = \mu_\text{obs}^\text{corr} - \mu_{\rm model}(z_{\text{cmb}}, \Omega_M),
    \label{eq:hubble_residual}
\end{equation}
where the theoretical distance modulus $\mu_{\rm model}$ is calculated assuming a flat $\Lambda$ cold dark matter ($\Lambda CDM$) cosmology with matter density $\Omega_\mathrm{m} = 0.3$ , Hubble constant $H_0 = 70\,\mathrm{km\,s^{-1}\,Mpc^{-1}}$, and $z_{\text{cmb}}$ is the redshift of the SN in the CMB rest frame.

We determined the best-fit standardisation parameters ($\alpha, \beta$) using MLE. Under the assumption of Gaussian errors, maximising the likelihood function is equivalent to minimising the following $\chi^2$ statistic:
\begin{equation}
\chi^2 = \sum_{i=1}^{N} \frac{[\mu_{\text{obs},i}(\alpha, \beta) - \mu_{\text{model}}(z_i)]^2}{\sigma_{\text{$\mu$,i}}^2 + \sigma_{\text{int}}^2},
\label{eq:chi2_base}
\end{equation}
where $\mu_{\text{obs},i}$ is the observed distance modulus for the $i$-th SN~Ia defined in Eq.~\ref{eq2}, $\mu_{\text{model}}(z_i)$ is the theoretical distance modulus for the $i$-th SN~Ia, and $\sigma_{\mu,i}$ is the measurement uncertainty propagated from the SALT2 parameters. The intrinsic scatter $\sigma_{\text{int}}$ is determined iteratively such that the reduced $\chi^2$ is unity. 

During this fitting process, we applied a 3$\sigma$ clipping procedure to iteratively remove outliers. All of the derived quantities used for this analysis are given in Table \ref{tab1}. Starting from the initial cross-match between the TNS catalogue and the MaNGA survey, we enforced stringent cuts on data quality and SALT2 fit parameters to exclude peculiar SNe~Ia and ensure precise distance measurements. As indicated in the table, the sample is refined through these quality control steps; after applying the final 3$\sigma$ clipping, our sample size for analysis is reduced to 60. However, among these 60 objects, we found that three have exceptionally large total uncertainties, more than three times larger than those of the other objects. After re-examining their standardisation parameters, we found that the uncertainties in their fitted c parameters are extremely large, about 4--10 times those of the rest of the sample. This likely indicates failed or unreliable fits caused by poor quality in the raw light curves, which in turn produced anomalous uncertainty estimates. We therefore excluded these three objects, reducing the sample size to 57. Finally, to avoid contamination from objects with abnormal colour, we excluded one additional object whose fitted colour parameter reached the fitting boundary ($c = 0.3$), leaving a final sample of 56 objects. We verified that the results obtained before and after excluding these four objects are consistent, and our conclusions remain unchanged.

\begin{figure}
    \centering
    \resizebox{0.9\hsize}{!}{\includegraphics{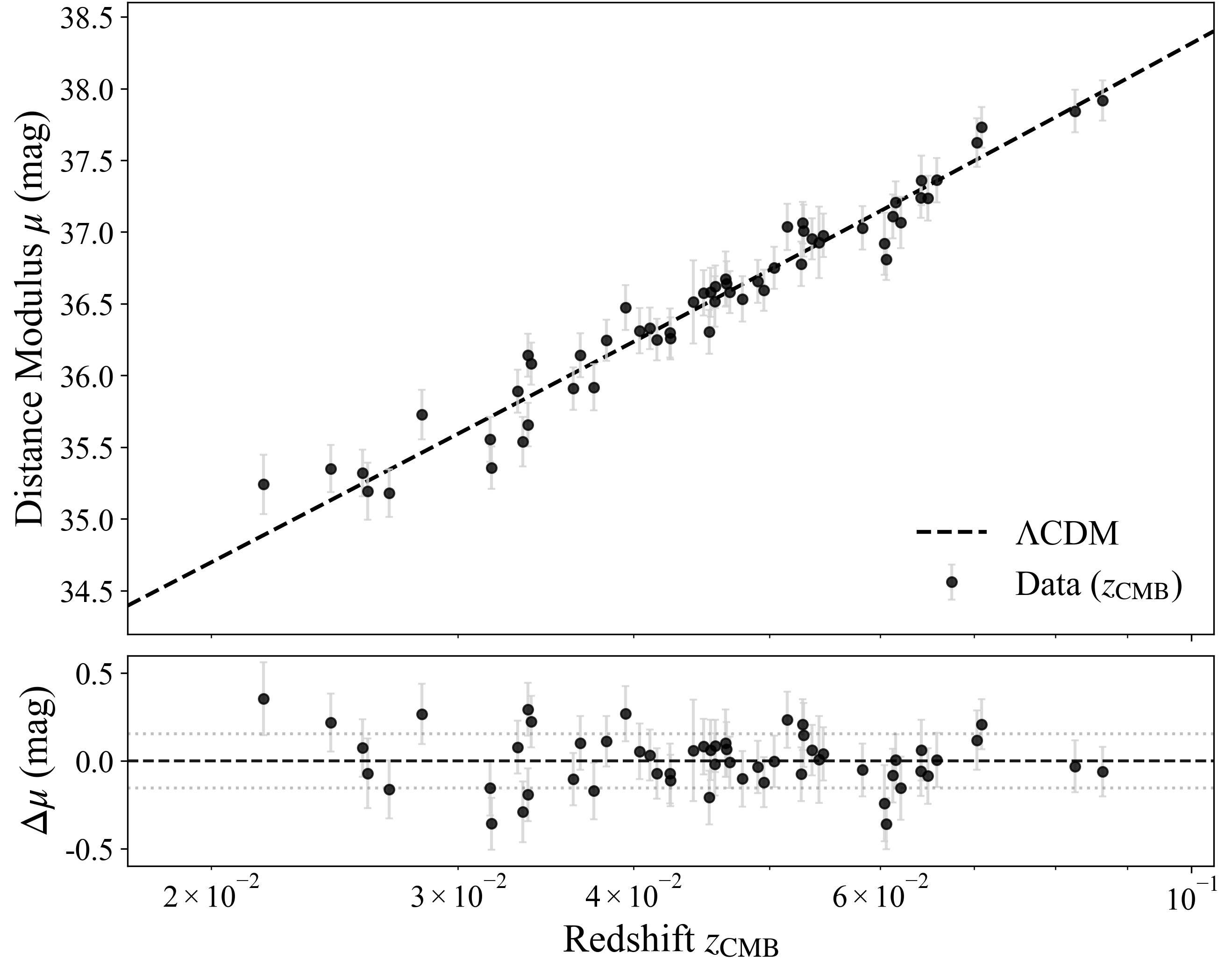}}
    \caption{Baseline Hubble diagram after luminosity standardisation. The theoretical curve (black line) is calculated using a flat $\Lambda CDM$ model with $H_0=70$ km s$^{-1}$ Mpc$^{-1}$ and $\Omega_\mathrm{m}=0.3$.}
    \label{fig4}
\end{figure}

Figure \ref{fig4} presents the results of this baseline likelihood fit. The derived weighted root mean square (wRMS), computed using inverse-variance weights, $w_i = (\sigma_{\mu,i}^2 + \sigma_{\rm int}^2)^{-1}$ consistent with Eq.~(\ref{eq:chi2_base}), is $0.1550 \pm 0.0149$ mag, with an intrinsic scatter of $\sigma_{\rm int}=0.1218 \pm 0.0118$ mag. In our MLE framework, the intrinsic scatter, $\sigma_{\rm int}$, is treated as a free parameter that captures the dispersion not described by the standardisation model. It is adjusted iteratively during the likelihood maximisation until the reduced chi-square ($\chi^2/{\rm ndof}$) approaches unity. Based on these baseline residuals, we observe a systematic trend in which SNe~Ia in young environments appear fainter than those in old environments (appearing as positive residuals). Motivated by this observation, we introduced the local LWA as an additional correction term in the standardisation model for a subsequent fit (Sect.~\ref{sec:3.2.2}).

Notably, while we performed complete covariance propagation during the fitting process, we did not apply the irreducible uncertainty floors of the SALT2 model ($\sigma_{x_1}\sim0.011$ and $\sigma_c\sim0.018$) as suggested by \citet{Rigault2020}. Instead, following the standard practice \cite[]{Conley2011,Betoule2014}, we allowed the global intrinsic dispersion term, $\sigma_{\rm int}$, to account for any remaining unexplained variance. Consequently, including the SALT2 model error floors would mainly shift part of the unexplained dispersion from $\sigma_{\rm int}$ to the SALT2 parameter uncertainties. The fitted value of $\sigma_{\rm int}$ would therefore decrease, while the total uncertainty and the significance of the environmental step would remain nearly unchanged.

\section{Results} \label{sec:results}

We examined the distribution of SN~Ia Hubble residuals in host galaxies with different properties. We analysed the relationship between the local LWA distribution and SN~Ia standardised parameters (Sect.~\ref{sec:SALT}). Then we used the local LWA as an additional parameter to participate in Hubble residual standardisation (Sect.~\ref{sec:standardisation}). Finally, we explored the connection between the local LWA step and the mass step in the Hubble residuals (Sect.~\ref{sec:joint} and ~\ref{sec:localmass}).

\subsection{Light-curve parameters} \label{sec:SALT}

\cite{Tripp1998} showed that SN~Ia standardisation for cosmology relies on the empirical correlations between peak luminosity and light-curve stretch factor ($x_1$) and colour ($c$). Therefore, understanding how $x_1$ and $c$ depend on the local environment is crucial for identifying potential sources of bias. Specifically, correlating these parameters with local LWA allows us to test whether the local stellar-population age is more strongly associated with the light-curve stretch factor ($x_1$) and the dust and intrinsic colour traced by $c$. In this section we examine the dependence of SALT2-T21 parameters on the local LWA, as illustrated in Fig. \ref{fig5}.

\subsubsection{Light-curve stretch factor}

The upper panel of Fig. \ref{fig5} shows the relationship between the SALT2-T21 stretch factor ($x_1$) and the local LWA. We find that the correlation between $x_1$ and the local LWA is very significant. The Spearman rank correlation coefficient is $r_\mathrm{s} = -0.61$, with a significance of 5.7$\sigma$. At the same time, we performed KS test and Student's T test (T test) on the young and old groups grouped by the threshold of 9.084 (see Sect.~\ref{sec:classify}). The results show that the $x_1$ value of the young group is larger than that of the old group by $0.803 \pm 0.236$. The obtained KS test p-value is 0.045, and T test p-value is 0.002. This indicates that the $x_1$ distributions of the young group and old group are statistically different.

This result indicates that SNe~Ia in younger local environments tend to have broader, more slowly declining light curves (larger $x_1$), whereas those in older environments show narrower, faster-declining light curves. This trend is consistent with the long-established environmental dependence of SN~Ia light-curve width. Previous studies already showed that younger stellar populations host more slowly declining and more luminous SNe~Ia \citep{Hamuy1996}, while later studies demonstrated that light-curve width or stretch is closely connected to host galaxy stellar population properties such as the sSFR, host type, stellar mass, and progenitor age  \citep{Sullivan2010,Lampeitl2010,Rigault2013, Rigault2020}. 

For completeness, we also examined the relation between $x_1$ and stellar mass using the same SN~Ia sample. We find moderate anti-correlations between $x_1$ and both global host-mass and local stellar mass, with Spearman coefficients of $r_s=-0.37$ ($p=0.0047$) and $r_s=-0.35$ ($p=0.0080$), respectively. However, both trends are weaker than the correlation between $x_1$ and the local LWA ($r_s=-0.61$). This suggests that the local age is more closely connected to the variation in $x_1$ than stellar mass in our sample.

\subsubsection{Light-curve colour}

\begin{figure}
\centering
  \resizebox{\hsize}{!}{\includegraphics{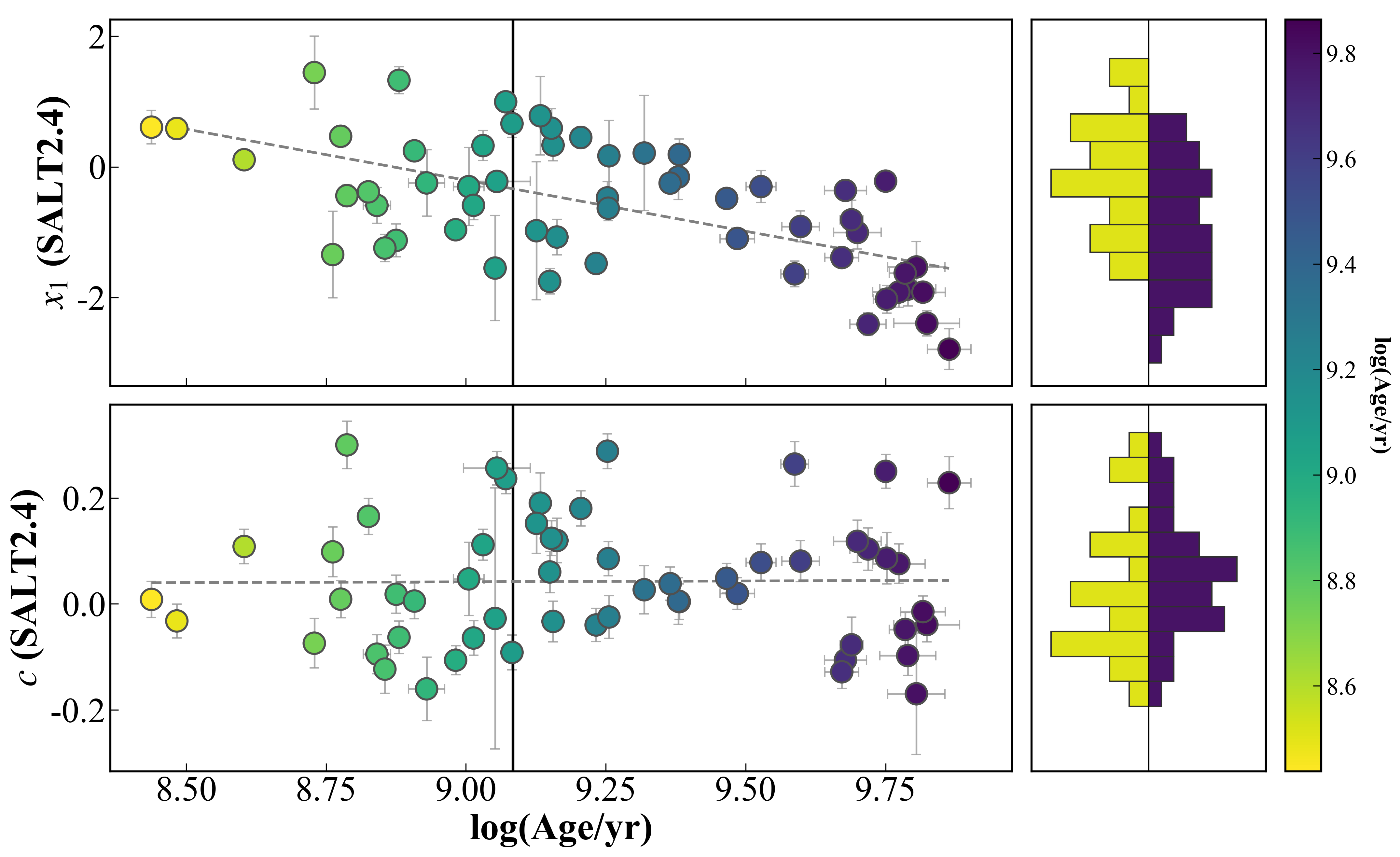}}
    \caption{Relationship between the local LWA and SN~Ia standardisation parameters' stretch factor ($x_1$; \textit{top}) and colour ($c$; \textit{bottom}). The marginal histograms on the right are split to compare the two age groups: the younger group is shown on the left side (yellow) and the older group on the right side (purple).}
   \label{fig5}
\end{figure}

The lower panel of Fig. \ref{fig5} displays the relationship between the SALT2-T21 light curve colour parameter ($c$) and the local LWA.
From this analysis, we find no significant correlation between $c$ and the local LWA within this normal SN~Ia sample, indicated by a Spearman rank correlation coefficient of only $r_\mathrm{s} = 0.01$.
This lack of dependence aligns with established findings in the literature, which consistently report that while the light-curve stretch factor ($x_1$) is strongly driven by progenitor age or mass, the colour parameter ($c$) shows little to no correlation with host galaxy properties \citep[e.g.][]{Sullivan2010,Lampeitl2010,Rigault2020}.
Our results confirm that this independence persists even when explicitly examined through the lens of the local LWA, suggesting that the standardisation colour parameter of normal SNe~Ia is largely insensitive to the age of the progenitor population.

\subsection{Standardisation using the local LWA} \label{sec:standardisation}

For SN~Ia observational results to be used in cosmological research, luminosity standardisation must be performed first. However, systematic dispersion remains in the Hubble residuals of SNe~Ia after standardisation. It is crucial to understand why other astrophysical factors affecting luminosity have not been completely removed. The common standardisation process uses a linear combination of light-curve stretch factor and colour \citep{Tripp1998}. Previous studies have additionally considered the global mass step of host galaxies \citep{Kelly2010, Sullivan2010} and non-linear relationships between other environmental properties and Hubble residuals \citep[e.g.][]{ Childress2013a,Rigault2020}.

Therefore, our environmental dependence study starts with the standardisation of SN~Ia light curves. The luminosity standardisation model is given in Eq.~\ref{eq2} by \cite{Tripp1998}. In the following sections, the local LWA and the global host galaxy mass are introduced as additional standardisation parameters.

\subsubsection{Age step measurement}

We applied the dynamic threshold scanning method described in Sect.~\ref{sec:classify} to partition the sample into young and old environments using a threshold of $\log_{10}(\mathrm{age}/\mathrm{yr}) = 9.084$. To validate the statistical framework of our analysis, we first examined the residual distributions. The Shapiro-Wilk test yields high p-values for both the young ($p=0.83$) and old ($p=0.85$) subsamples, confirming that both populations are consistent with Gaussian distributions.

\begin{figure}
\centering
  \resizebox{\hsize}{!}{\includegraphics{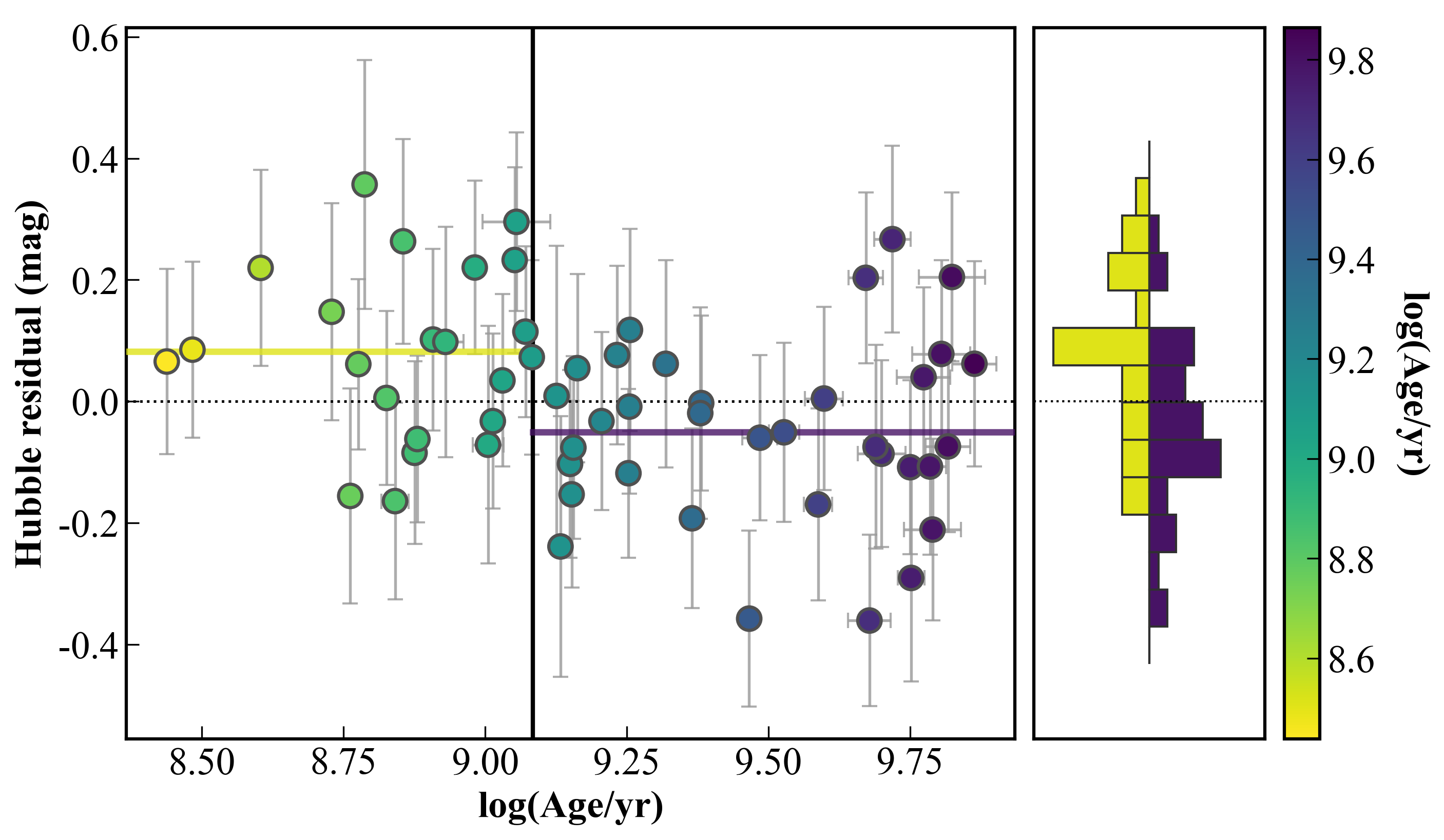}}
    \caption{Hubble residuals as a function of the local LWA. The vertical black line indicates the optimal split threshold ($\sim 9.084$). The horizontal lines represent the mean Hubble residuals for the young (yellow) and old (purple) subgroups. The right panel shows the marginal histograms of the residuals.}
    \label{fig6}
\end{figure}

With the Gaussianity established, we proceeded to quantify the magnitude of the step, $\Delta_\mathrm{A}$. We find a significant magnitude step of $\Delta_\mathrm{A}=0.125 \pm 0.027$ mag ($4.6\sigma$), which is calculated as the difference between the variance-weighted mean Hubble residuals of the two subsamples based on the baseline fit (Eq.~\ref{eq:chi2_base}). To further verify the distinct nature of these distributions, we performed a KS test ($p=0.013$) and a T test ($p=0.002$). Collectively, these results strongly reject the null hypothesis that the young and old subsamples are drawn from the same parent population, confirming that the observed environmental dependence is driven by two distinct underlying distributions rather than random fluctuations. The marginal histograms in Fig. \ref{fig6} visually demonstrate this systematic shift between the two distinct Gaussian profiles.

Our findings provide robust support for the environmental dependence of SN~Ia luminosity, a phenomenon first established by global host galaxy studies \citep[e.g.][]{Kelly2010, Sullivan2010, Gupta2011}. These seminal works, along with studies focusing on global specific star formation rates \citep[e.g.][]{Neill2009,Childress2013a}, demonstrated that SNe~Ia in passive (older) environments appear brighter after standardisation than those in star-forming (younger) ones. In the context of local environments, our results align with the pioneering work of \citet{Rigault2013} and subsequent analysis by \citet{Rigault2020}, who identified a strong correlation using the LsSFR. While \citet{Rigault2020} used the LsSFR as a proxy for progenitor age, we employed the local LWA derived directly from MaNGA Pipe3D stellar population synthesis. Despite the difference in tracers, our detection of a local LWA step provides further evidence of a connection between SN~Ia standardised luminosity and the age of the local stellar population, which suggests that progenitor age is one underlying physical factor.

\subsubsection{Age step as an additional standardisation parameter} \label{sec:3.2.2}

To quantitatively eliminate the environmental dependence of Hubble residuals on age \citep[e.g.][]{Rigault2013, Rigault2020}, we adopted a global fit following the standard methodology of \citet{Sullivan2010} and \citet{Betoule2014} while introducing the age step as an additional standardisation parameter into the distance modulus formula as suggested by recent studies \citep[e.g.][]{Rigault2013, Rigault2020}. Under this framework, $\alpha, \beta$ and the step amplitudes are all free parameters ($\alpha$ and $\beta$ are constrained to be the same in both young and old environments), solving for all parameters simultaneously by minimising $\chi^2$. The standardised formula after introducing the step is defined as follows:

\begin{equation}
\mu_{\mathrm{obs}}^{\mathrm{step}}  = \mu_{\mathrm{obs}}^{\mathrm{corr}}  + \sum_i \Delta_{\text{i}} [\Theta(X_{\text{i}} - X_{\text{th,i}}) - 0.5],
\label{eq4}
\end{equation}
where $\Delta_i$ represents the amplitude of the environmental step correction associated with the $i$-th host galaxy property; $X_{\text{i}}$ is the host galaxy property, and $X_{\text{th}}$ is the optimal splitting threshold determined in Sect.~\ref{sec:classify}; $\Theta$ is the Heaviside step function, taking the value 1 when the host galaxy property value $X_{\text{i}}$ is greater than the threshold $X_{\text{th,i}}$, and 0 otherwise.

The parameters $\alpha$, $\beta$ and $\Delta_i$ are determined simultaneously via minimisation:

\begin{equation}
\chi^{2} =
\sum_{i}
\frac{
\left[
\mu_{\mathrm{obs, i}}^{\mathrm{step}} 
-\mu_{\mathrm{th}}(z_{\text{i}})
\right]^{2}
}
{\sigma_{\text{$\mu$, i}}^{2}+\sigma_{\mathrm{int}}^{2}},
\label{eq:chi2_step}
\end{equation}
where $\mu_{\mathrm{obs}}^{\mathrm{step}}  - \mu_{\mathrm{th}}$ corresponds to the Hubble residual. Therefore, $\Delta_i$ directly measures the difference in the standardised luminosity between the two environmental subsamples.

Notably, we adopt a symmetric correction method, that is, using the $\Theta(X_{\text{i}} - X_{\text{th,i}}) - 0.5$ term in Eq. (\ref{eq4}). This means that for the low-value group ($\Theta=0$), the correction is $-\Delta_{\text{i}}/2$, and for the high-value group ($\Theta=1$), the correction is $+\Delta_{\text{i}}/2$. This processing method is mathematically equivalent to the traditional 0/1 step model.

After introducing the age step parameter $\Delta_\mathrm{A}$ into the standardisation model and re-fitting the full sample using MLE, we obtained an age step amplitude of $0.163 \pm 0.031$~mag ($5.2\,\sigma$). This value is larger than the step measured directly from the initially standardised Hubble residuals, $0.125 \pm 0.027$~mag, which corresponds to a simple mean offset between the young and old subsamples. This change reflects the correlation between the SN~Ia light-curve stretch factor ($x_1$) and the local LWA. Incorporating the age parameter significantly reduces the wRMS from $0.1550 \pm 0.0149$ to $0.1376 \pm 0.0133$. This demonstrates that introducing the age parameter reduces the dispersion of Hubble residuals and improves SN~Ia standardisation precision.

\subsection{Hubble residual correction from the global mass and local LWA} \label{sec:joint}

\begin{figure}
    \centering
    \resizebox{0.9\hsize}{!}{\includegraphics{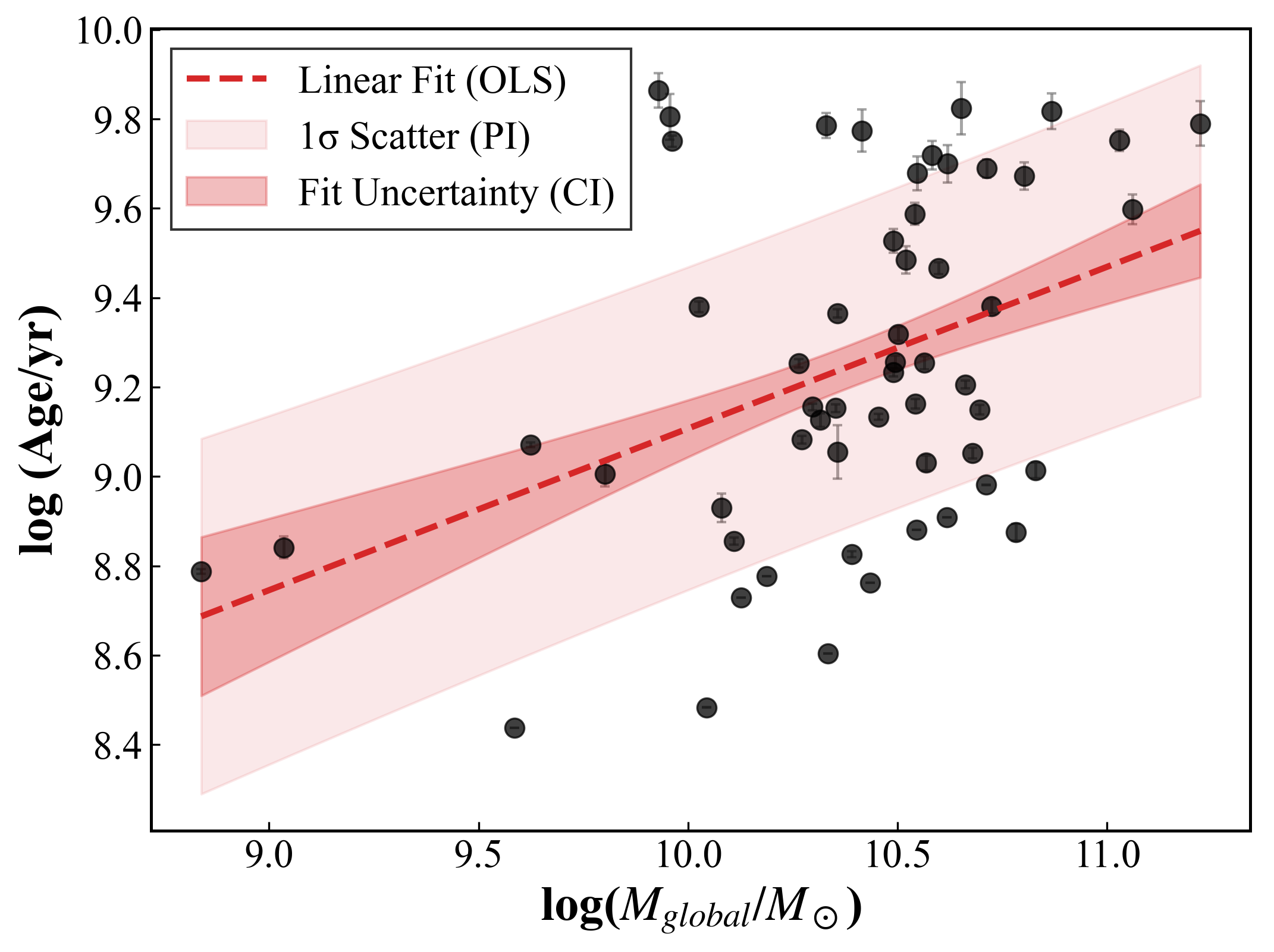}}
    \caption{Correlation between the local LWA and the global mass of the host galaxies. The dashed red line represents the best-fit linear relation derived from ordinary least squares (OLS) regression. The dark and light pink shaded regions indicate the fit uncertainty (95\% confidence interval) and the $1\sigma$ scatter (prediction interval), respectively. The sample shows a positive correlation with a Pearson coefficient of $r=0.41$ and a significance of $3.3\sigma$.}
\label{fig7}
\end{figure}

As shown in Fig. \ref{fig7}, our sample shows a correlation between the local LWA and global host galaxy mass ($r = 0.41, 3.3\sigma$). Previous studies often used global mass as a proxy for environmental correction (the mass step, $\Delta_\mathrm{M}$). However, galaxy evolution theory points out that massive galaxies usually have more complex star formation histories and older average stellar populations, which means there is an intrinsic physical degeneracy between mass and age \cite[e.g.][]{Cowie1996, Gallazzi2005, Thomas2005}. 

To break the degeneracy between global host galaxy mass and local stellar-population age and to identify the primary driver of the luminosity step, we compared three standardisation models: (i) a baseline fit with no step term, (ii) a mass step-only fit including a global host-mass step $\Delta_{\mathrm{M}_{\text{global}}}$, and (iii) a joint fit including both the global host-mass step $\Delta_{\mathrm{M}_{\text{global}}}$ and the local LWA step $\Delta_\mathrm{A}$.

Using a global host galaxy mass threshold of $\log(M_*/M_\odot)=10.278$, the mass step-only fit yields $\Delta_{\mathrm{M}_{\text{global}}} = 0.071 \pm 0.035$~mag ($2.0\,\sigma$) and reduces the wRMS from $0.1550 \pm 0.0149$~mag (baseline) to $0.1520 \pm 0.0146$~mag. When we fit $\Delta_{\mathrm{M}_{\text{global}}}$ and $\Delta_\mathrm{A}$ simultaneously as independent parameters, the inferred mass step is strongly suppressed, $\Delta_{\mathrm{M}_{\text{global}}} = 0.028 \pm 0.033$~mag ($0.9\,\sigma$), while the age step remains highly significant, $\Delta_\mathrm{A} = 0.156 \pm 0.032$~mag ($4.9\,\sigma$). The wRMS further decreases to $0.1372 \pm 0.0133$~mag. 

As shown in Table \ref{tab2}, after considering the influence of the local LWA, the global mass step decreased to an insignificant level. Approximately $60\%$ of the host-mass step amplitude is due to an environmental dependence on the local LWA. This result supports the interpretation that the local LWA is more closely related to the physical driver of the dispersion of luminosity than the mass. Because the local LWA statistically traces the age of the stellar population associated with SN~Ia progenitors, this finding lends support to the possibility that progenitor age contributes to the physical origin of the mass step \cite[e.g.][]{Childress2013a,Rigault2013,Rigault2020,Campbell2016}.

Although a global host galaxy mass threshold of $\log(M_*/M_\odot) = 10.0$ is the canonical choice in cosmological analyses \citep[e.g.][]{Sullivan2010, Suzuki2012, Betoule2014}, we did not adopt this fixed value in our study. Our sample is predominantly composed of massive galaxies, with only a negligible number of objects falling below $10.0$. Enforcing a fixed threshold at $10.0$ would result in a highly unbalanced sample split, rendering the mass step calculation statistically unstable. Instead, we applied a dynamic threshold scanning method to identify the optimal splitting point that maximises the statistical significance of the mass step. A more detailed discussion on this methodology is presented in Sect.~\ref{sec:4.3}.

\begin{table*}
\caption{Environmental step corrections in the SN~Ia luminosity standardisation.}
\label{tab2}
\centering

\begin{tabular}{l c c c c}
\hline\hline
Parameters & wRMS & $\sigma_{\mathrm{int}}$ & $\Delta_\mathrm{M}$ & $\Delta_\mathrm{A}$ \\
\hline
SALT2-T21 & $0.1550 \pm 0.0149$ & $0.1218 \pm 0.0118$ & - & - \\
SALT2-T21+$\Delta_\mathrm{A}$ & $0.1376 \pm 0.0133$ & $0.1002 \pm 0.0098$ & - & $0.163 \pm 0.031$ \\
SALT2-T21+$\Delta_\mathrm{M}$ & $0.1520 \pm 0.0146$ & $0.1198 \pm 0.0117$ & $0.071 \pm 0.035$ & - \\
SALT2-T21+$\Delta_\mathrm{A}$+$\Delta_\mathrm{M}$ & $0.1372 \pm 0.0133$ & $0.1014 \pm 0.0100$ & $0.028 \pm 0.033$ & $0.156 \pm 0.032$ \\
\hline
\end{tabular}
\tablefoot{The table presents the results obtained by adding the $\Delta_{\mathrm{M}}$ and $\Delta_{\mathrm{A}}$ separately and simultaneously to the SN~Ia luminosity standardisation.}
\end{table*}

\subsection{Hubble residual correction from the local mass and local LWA} \label{sec:localmass}

Given that the local LWA we used is the stellar population age within 1~kpc of the SN explosion site, in addition to using global mass for analysis, we used local stellar mass for mass step analysis. After light curve standardisation, we find that when the sample is divided into lower-mass and higher-mass groups using the local stellar mass $\log(M_*/M_\odot)=8.983$ as the threshold in the baseline Hubble residuals, the lower-mass group is fainter than the higher-mass group by $\Delta_{\mathrm{M}_{\text{local}}} = 0.087 \pm 0.035 mag$, 2.4$\sigma$. After we added $\Delta_{\mathrm{M}_{\text{local}}}$ and $\Delta_\mathrm{A}$ simultaneously as additional parameters to the light curve standardisation, $\Delta_{\mathrm{M}_{\text{local}}}$ decreased to $0.012 \pm 0.037$ mag, 0.3$\sigma$, while $\Delta_\mathrm{A}$ remained at $\Delta_\mathrm{A} = 0.157 \pm 0.036$ mag, 4.4$\sigma$. This indicates that the local mass step is almost entirely caused by the local LWA.


\section{Robustness checks} \label{sec:robustness}

We find a significant correlation between the local LWA and SN~Ia Hubble residuals. However, as this study explores an environmental effect rather than a fundamental parameter derived from a physical law, it is imperative to determine whether the appearance of this signal could be induced by systematic errors or methodological artefacts. In this section we test the stability of our results to ensure the observed local LWA step is not a consequence of dilution from PSF homogenisation, nor a byproduct of degeneracy with host galaxy dust. Furthermore, we discuss the impact of mass threshold selection and verified that the signal persists across different SN~Ia sub-populations (classified by stretch factor and colour) and remains statistically significant under subsample resampling.

\subsection{Dilution from PSF homogenisation}

The extraction of all local environmental parameters in this study is built on the framework described in Sect.~\ref{sec:localparams}: defining a target region with a physical radius of 1~kpc. However, limited by the spatial resolution of MaNGA data, what is actually measured represents the mixed information of this target region and its surrounding larger range (scale $\geq$ 2~kpc). This unavoidable dilution effect could affect our results.

First, the most direct impact is the dilution of extreme local environmental signals. If a SN progenitor was born in a young star cluster with a scale much smaller than 1~kpc, or located on a boundary with a very steep age gradient, the local LWA we measure will be closer to the average value of the larger scale stellar population around it. This means that we may underestimate the extreme values of environmental parameters in the sample, and the correlation coefficient between the observed environmental parameters and Hubble residuals could be stronger than what we measure. In other words, our results could be viewed as a conservative yet still significant environmental correlation detected on the kiloparsec-scale smoothing.

Second, this measurement scale affects our interpretation of young and old environmental classification. The optimal age threshold we determined using the maximum significance method (see Sect.~\ref{sec:classify}) is essentially based on smoothed data. Therefore, this threshold effectively divides the sample into environments that are relatively young on the kiloparsec-scale smoothing and relatively old on the kiloparsec-scale smoothing. This definition is compatible with current research based on nearby-galaxy survey (such as MaNGA), allowing our conclusions to be compared with similar studies \cite[e.g.][]{Rigault2013,Rigault2020,Galbany2014,Roman2018}.

Our work shows that even using ground-based survey data with resolution smoothed to kpc levels, this environmental signal is clearly distinguishable. This not only verifies the feasibility of the studies of environmental effects but also highlights their universality and robustness.

\subsection{Impact of the host galaxy's dust} \label{sec:4.2}

Since the colour coefficient, $\beta$, for SNe~Ia originates physically from host galaxy dust and intrinsic SN colour, recent studies have indicated that dust properties could vary with the environment \citep{BroutScolnic2021, Popovic2021, GonzalezGaitan2021, Meldorf2023, Wojtak2023, Ginolin2025a}. Variations in dust properties with the environment could induce colour biases in SNe, which in turn lead to biases in the standardised luminosity. Furthermore, given that dust properties are correlated with many host galaxy properties \citep[e.g.][]{Tremonti2004}, the calculation of luminosity-weighted properties, such as stellar mass and age, relies on dust extinction corrections. Consequently, the influence of dust must be treated with caution when assessing and interpreting the environmental effects on SN~Ia luminosity.

Although the impact of dust cannot be ignored in standardisation, \citet{Childress2013a} pointed out that the variation of dust extinction with global host galaxy mass is too smooth to be the primary driver of the mass step. \citet{Ginolin2025a} noted that they found no evolution of steps as a function of light-curve colour. This implies that the environmental dependence of standardised luminosity is not simply explained by the evolution of host galaxy dust properties. While host galaxy dust properties are known to evolve with galaxy mass and age, potentially biasing SN~Ia standardisation, our results suggest that dust is not the primary driver of the observed age step in this sample. As demonstrated in Sect.~\ref{sec:SALT} (Fig.~\ref{fig5}), we find no significant correlation between the local LWA and the SALT2 colour parameter $c$ (Spearman $r_\mathrm{s} = 0.01$). Since the parameter $c$ captures the combined effects of intrinsic SN colour and dust reddening, its lack of correlation with the local LWA implies that the dust properties of the progenitor environments in our sample are statistically decoupled from their stellar population age. Consequently, the significant correlation observed between Hubble residuals and the local LWA cannot be attributed to a hidden dust-age degeneracy. Based on this orthogonality, we focused the subsequent analysis on the direct impact of stellar age, treating it as an independent physical driver distinct from colour-based dust corrections.

We note that this interpretation is further supported by the independent fits to the young and old subsamples presented in Sect.~\ref{sec:4.5}. Although the best-fit nuisance parameters differ moderately between the two groups ($\beta_{\mathrm{young}} = 3.363$, $\beta_{\mathrm{old}} = 2.985$), the age step remains significant at $0.1607 \pm 0.0345$\, mag when the two subsamples are fitted separately, arguing against a purely dust-driven origin.

\subsection{Impact of the mass threshold selection}\label{sec:4.3}

\begin{figure}
    \centering
    \resizebox{0.9\hsize}{!}{\includegraphics{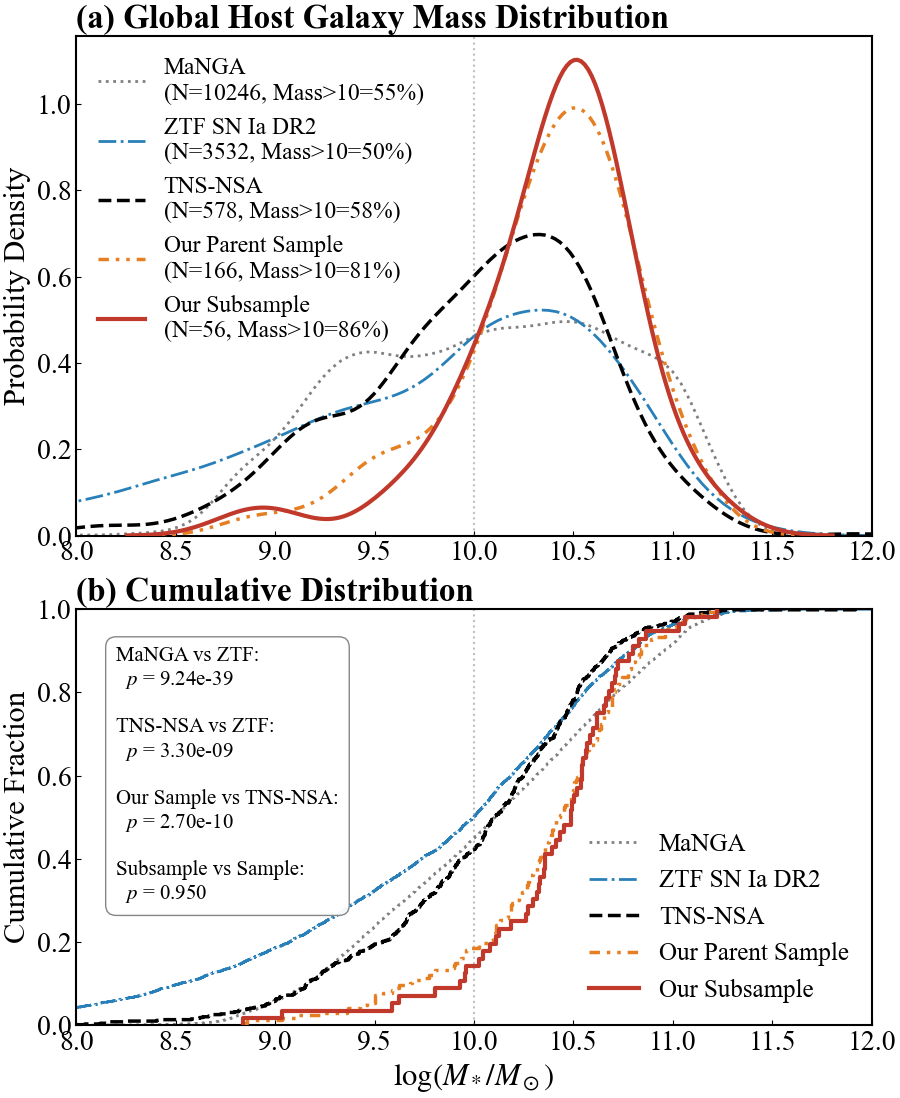}}
\caption{Comparison of global host galaxy mass distributions. 
\textit{Panel (a)}: Probability density functions normalised to unit area. \textit{Panel (b)}: Corresponding CDFs. 
The samples are as follows: 
the MaNGA DR17 parent galaxy catalogue (dotted grey line); 
the unbiased ZTF DR2 SN~Ia sample (dash-dotted blue line); 
the TNS SN~Ia candidates cross-matched with the NSA catalogue (dashed black line); 
our full parent sample ($N=166$, orange dash-dot-dot line); 
and our final cosmological subsample ($N=56$; solid red line). 
The vertical dotted grid line marks the global mass threshold at $\log(M_*/M_\odot) = 10.0$. 
The inset tables list the $p$-values from KS tests between different samples.}
\label{fig:mass_dist}
\end{figure}

The global host galaxy mass threshold of $\log(M_*/M_\odot) = 10.0$ is conventionally adopted in large-scale SN~Ia cosmological samples \citep[e.g.][]{Sullivan2010, Betoule2014}. To assess the applicability of this convention to our sample, we performed a comparative analysis of global host galaxy mass distributions in Fig.~\ref{fig:mass_dist}.

First, we examined the unbiased ZTF DR2 sample (panel a, dash-dotted blue line), which serves as an ideal reference. It exhibits a balanced distribution roughly symmetric around $\log(M_*/M_\odot) = 10.0$, with exactly 50\% of events exceeding this mass. To further evaluate the inherent selection tendency of the TNS catalogue, we constructed a baseline sample by cross-matching the full TNS SN~Ia catalogue with the NASA-Sloan Atlas (NSA) catalogue (dashed~black line; $N=578$). This TNS--NSA general sample is moderately skewed towards higher global host galaxy mass, with 58\% of objects lying above $\log(M_*/M_\odot)=10.0$, while retaining a relatively broad coverage across the mass range.

In contrast, our sample, constructed by cross-matching TNS with MaNGA, exhibits a pronounced concentration at the higher-mass end. As shown in panel (b), a KS test reveals a significant difference between our TNS-MaNGA parent sample ($N=166$; `Our Parent Sample' in Fig.~\ref{fig:mass_dist}) and the broader TNS-NSA baseline ($p \approx 2.7 \times 10^{-10}$). This comparison demonstrates that the requirement for MaNGA IFU coverage introduces a strong selection preference for massive galaxies beyond the inherent tendency of the TNS catalogue. Crucially, however, our subsequent data reduction and quality cuts (Sect.~\ref{sec:method}) did not introduce any additional bias. The mass distribution of our final cosmological subset ($N=56$; `Our Subsample' in Fig.~\ref{fig:mass_dist}) is statistically indistinguishable from the initial TNS-MaNGA parent sample, with a KS test $p$-value of 0.950.

While the typical threshold of 10.0 effectively splits the ZTF DR2 sample, applying this fixed threshold to our limited dataset ($N=56$) is statistically unstable. The skewness results in an extreme imbalance in our sample, isolating only eight events in the lower mass bin compared to 48 events in the higher mass bin. From a statistical perspective, a subsample size of $N=8$ is insufficient to robustly constrain the mean Hubble residual, as it leaves the measurement highly susceptible to individual outliers and small-number fluctuations.

Indeed, applying this threshold yields a statistically insignificant mass step ($0.040 \pm 0.046$ mag, $<1\sigma$), which vanishes completely in the joint fit ($0.000 \pm 0.042$ mag). Therefore, the choice to employ a dynamic threshold scanning method (Sect.~\ref{sec:classify}) is driven by statistical necessity; it allows us to identify a split point where both subsamples retain sufficient statistical weight to provide a stable and meaningful comparison, rather than strictly adhering to a convention that is ill-suited for our sample distribution. Crucially, we emphasise that our core conclusion remains robust regardless of this threshold selection: the mass step analysis consistently yields the same qualitative results whether using the fixed literature value ($10.0$) or our optimised dynamic threshold.

\subsection{Robustness after classification by the stretch factor or colour} \label{sec:4.4}

\citet{Sullivan2010} proposed another way to test for potential non-uniformity in the standardisation, which examined the variation of the brightness offset between SNe~Ia in low- and high-mass hosts when splitting the sample at $c = 0$ or $x_1 = 0$ \cite[]{Rigault2020}. To perform these tests, we classified samples by colour and stretch factor to assess the significance of the age step as in \cite[]{Sullivan2010}. We obtain $\Delta_\mathrm{A} = 0.168 \pm 0.055 mag$ for 21 SNe with $c < 0$, and $\Delta_\mathrm{A} = 0.156 \pm 0.034 mag$ in 35 SNe with $c > 0$. In 39 SNe with $x_1 < 0$, we obtain $\Delta_\mathrm{A} = 0.177 \pm 0.040 mag$, and in 17 SNe with $x_1 > 0$, we obtain $\Delta_\mathrm{A} = 0.144 \pm 0.039 mag$. The age step in samples with $c < 0$ is slightly larger than in samples with $c > 0$, and the age step in samples with $x_1 < 0$ is slightly larger than in samples with $x_1 > 0$.

The results demonstrate that the local LWA step clearly exists on both sides of these dividing lines for both the stretch factor and colour parameter. Moreover, the size of the local-LWA step is consistent between these subsets.
This indicates that the environmental dependence cannot be removed by the standard linear SALT2-T21 corrections based only on $x_1$ and $c$.
We test this interpretation more directly in Sect.~\ref{sec:4.5} by allowing the young and old subsamples to have independent standardisation coefficients.

\subsection{Robustness check of  the subsample fitting} \label{sec:4.5}

Although the global joint fit preliminarily identifies the local LWA as the primary driver of SN~Ia luminosity variations -- with a significance exceeding that of the mass step -- this conclusion rests on the assumption that light curve standardisation parameters ($\alpha$ and $\beta$) are universal across both young and old environments. However, dust properties (affecting $\beta$) or intrinsic luminosity relations (affecting $\alpha$) in different stellar population environments differ; for example, host-dependent variations in $\beta$ have been reported when dividing by global host properties \citep[e.g.][]{Sullivan2010}. Consequently, enforcing a universal set of $(\alpha,\beta)$ for the entire sample could introduce a systematic bias.

To mitigate potential parameter degeneracies, we adopted an independent MLE subsample-fitting strategy, similar to the approach in \cite{Rigault2020}. We divided the sample into young and old subsamples based on the age threshold determined in Sect.~\ref{sec:classify}. For each subsample $k \in \{\text{young, old}\}$, we defined the Hubble residual of the $i$-th SN~Ia as
\begin{equation}
    \Delta\mu_{i,k} = \mu_{i, \text{obs}}(\alpha_k, \beta_k) - \mu_{i, \text{model}} - \mu_{0,k},
\end{equation}
where $\alpha_k$ and $\beta_k$ are the standardisation coefficients specific to each subsample.

Critically, we introduced a group-specific global offset parameter, $\mu_{0,k}$. While the linear terms $\alpha x_1$ and $\beta c$ compensate for the correlation between luminosity and light-curve shape or colour, they do not guarantee that the resulting residuals are strictly centred at zero for a given environment. The parameter $\mu_{0,k}$ absorbs any residual mean bias in the standardised distance modulus, allowing us to isolate the intrinsic difference in luminosity between the two environments under their respective optimal standardisation.

The parameters $\{\alpha_k, \beta_k, \mu_{0,k}\}$ and the intrinsic scatter $\sigma_{\text{int},k}$ are determined simultaneously by minimising the negative log-likelihood, corresponding to
\begin{equation}
    \chi^2_k = \sum_{i} \frac{\Delta\mu_{i,k}^2}{\sigma_{i, \text{obs}}^2 + \sigma_{\text{int},k}^2},
\end{equation}
where $\sigma_{i, \text{obs}}$ represents the propagated measurement uncertainties. In this framework, the age step is not an explicit fitting parameter but is instead derived from the difference between the two independent offsets:
\begin{equation}
    \Delta_{\rm {A}} = \mu_{0, \text{young}} - \mu_{0, \text{old}},
\end{equation}
with the uncertainty calculated as $\sigma_{\Delta} = \sqrt{\sigma_{\mu_{0, \text{young}}}^2 + \sigma_{\mu_{0, \text{old}}}^2}$.

We still find $\Delta_{\rm {A}} = 0.1607 \pm 0.0345~\rm {mag}$, corresponding to a significance of $4.7\sigma$ after subsample-fitting. Therefore, the age-dependent offset persists even when the two subsamples are allowed to have different standardisation coefficients ($\alpha_{\mathrm{young}} = 0.156 \pm 0.025, \beta_{\mathrm{young}} = 3.363 \pm 0.171$, and $\alpha_{\mathrm{old}} = 0.190 \pm 0.020, \beta_{\mathrm{old}} = 2.985 \pm 0.157$), supporting a physical origin of the age step and precluding an artefact induced by potential differences in $(\alpha,\beta)$.

Taken together, Sects.~\ref{sec:4.4} and \ref{sec:4.5} show that the local-LWA step is robust both across $x_1$- and $c$-defined SN~Ia sub-populations, and also when the young and old subsamples are allowed to have independent standardisation coefficients.
Therefore, the observed luminosity offset is unlikely to be an artefact of linear SALT2-T21 coefficients, and instead supports a genuine environmental dependence linked to the local LWA.

\subsection{Robustness of the inferred mass and age steps} \label{sec:4.6}

\begin{table*} 
\centering
\caption{Mass and age steps under different fitting criteria and redshift prescriptions.} 

\label{tab3}
\begin{tabular}{lccccccc}
\hline
Sample & Redshift & $N$ & Threshold $(M_{\rm thr}, A_{\rm thr})$ & Mass-only step & Age-only step & Joint mass step & Joint age step\\
\hline
\multicolumn{8}{c}{Default single-$\alpha$ analysis} \\
\hline
Narrow-$c$ & $z_{\rm cmb}$ & 56
& $(10.278,\,9.084)$
& $0.071 \pm 0.035$
& $0.163 \pm 0.031$
& $0.028 \pm 0.033$
& $0.156 \pm 0.032$
\\

Narrow-$c$ & $z_{\rm cor}$ & 54
& $(10.276,\,9.083)$
& $0.110 \pm 0.034$
& $0.148 \pm 0.030$
& $0.059 \pm 0.035$
& $0.128 \pm 0.033$
\\

Broad-$c$  & $z_{\rm cmb}$ & 63
& $(10.188,\,9.085)$
& $0.112 \pm 0.039$
& $0.166 \pm 0.034$
& $0.050 \pm 0.040$
& $0.147 \pm 0.037$
\\

Broad-$c$  & $z_{\rm cor}$ & 61
& $(10.195,\,9.086)$
& $0.114 \pm 0.034$
& $0.165 \pm 0.031$
& $0.057 \pm 0.034$
& $0.144 \pm 0.033$
\\

\hline
\multicolumn{8}{c}{Literature broken-$\alpha$ analysis (Ginolin et al.)} \\
\hline
Narrow-$c$ & $z_{\rm cmb}$ & 56
& $(10.278,\,9.084)$
& $0.118 \pm 0.052$
& $0.193 \pm 0.042$
& $0.049 \pm 0.049$
& $0.178 \pm 0.045$
\\

Narrow-$c$ & $z_{\rm cor}$ & 54
& $(10.276,\,9.083)$
& $0.139 \pm 0.052$
& $0.190 \pm 0.041$
& $0.049 \pm 0.053$
& $0.171 \pm 0.046$
\\

Broad-$c$  & $z_{\rm cmb}$ & 63
& $(10.188,\,9.085)$
& $0.133 \pm 0.054$
& $0.184 \pm 0.043$
& $0.042 \pm 0.057$
& $0.167 \pm 0.049$
\\

Broad-$c$  & $z_{\rm cor}$ & 61
& $(10.195,\,9.086)$
& $0.138 \pm 0.051$
& $0.198 \pm 0.040$
& $0.052 \pm 0.050$
& $0.179 \pm 0.044$
\\
\hline

\end{tabular}

\tablefoot{Narrow-$c$ corresponds to our fitting colour range, $c \in (-0.3,\,0.3)$, whereas Broad-$c$ follows the fitting colour range adopted by \cite{Ginolin2025b}, $c \in (-0.2,\,0.8)$. We use $z_{\rm cmb}$ for the redshift in the CMB frame and $z_{\rm cor}$ for the redshift corrected for host galaxy peculiar velocity. The upper block lists the default single-$\alpha$ analysis, while the lower block shows the corresponding results obtained with the literature broken-$\alpha$ prescription following \cite{Ginolin2025b}}

\end{table*}

As an additional robustness test, we examined how the inferred mass and age steps depend on both the light-curve-selection criteria and the redshift prescription used to compute the Hubble residuals. We considered two sample definitions: (i) a conservative one with a narrower colour range and a stricter fit-probability requirement ($-0.3 < c < 0.3$ and $\mathrm{fitprob} > 10^{-3}$) and (ii)  a broader one with a wider colour range and a looser fit-probability cut ($-0.2 < c < 0.8$ and $\mathrm{fitprob} > 10^{-7}$). The latter follows the baseline selection adopted in previous volume-limited ZTF SN~Ia DR2 analyses of environmental steps, in particular \cite{Ginolin2025b}. For each sample definition, we repeated the analysis using both the CMB-frame redshift, $z_{\rm cmb}$, with a conservative $300\,\mathrm{km\,s^{-1}}$ peculiar-velocity uncertainty, and the peculiar-velocity-corrected redshift, $z_{\rm cor}$, derived from the 2M++ density field \cite{Carrick2015}.

As shown in Table~\ref{tab3}, for the conservative sample, the analysis includes 56 SNe when using $z_{\rm cmb}$ and 54 SNe when using $z_{\rm cor}$. In this case, the mass step increases from $0.071 \pm 0.035$~mag for $z_{\rm cmb}$ to $0.110 \pm 0.034$~mag for $z_{\rm cor}$, while the age step remains comparatively stable, changing from $0.163 \pm 0.031$~mag to $0.148 \pm 0.030$~mag. In the joint fit that includes both mass and age, the mass step is reduced, to $0.028 \pm 0.033$~mag for $z_{\rm cmb}$ and $0.059 \pm 0.035$~mag for $z_{\rm cor}$, whereas the age step remains significant at $0.156 \pm 0.032$~mag and $0.128 \pm 0.033$~mag, respectively.

For the broader sample, the analysis includes 63 SNe for $z_{\rm cmb}$ and 61 SNe for $z_{\rm cor}$. In this setup, the mass step is $0.112 \pm 0.039$~mag for $z_{\rm cmb}$ and $0.114 \pm 0.034$~mag for $z_{\rm cor}$. At the same time, the age step remains significant and stable, with amplitudes of $0.166 \pm 0.034$~mag and $0.165 \pm 0.031$~mag, respectively. In the corresponding joint fits, the mass step is again reduced, to $0.050 \pm 0.040$~mag for $z_{\rm cmb}$ and $0.057 \pm 0.034$~mag for $z_{\rm cor}$, while the age step remains significant at $0.147 \pm 0.037$~mag and $0.144 \pm 0.033$~mag, respectively.

These tests show that the standalone mass step amplitude is not identical under different colour-range selections and redshift prescriptions. The inferred amplitudes vary across the four setups, indicating that both the adopted $c$ range and the treatment of peculiar velocities can affect the numerical value of the mass step. However, our conclusion remains unchanged: in all four cases, once the age information is included, the conventional mass step is reduced, whereas the local LWA-based age step remains significant. Our interpretation therefore does not rely on any particular combination of sample-selection criteria or redshift prescription.

In addition, the Hubble residuals are modelled with a single linear stretch coefficient $\alpha$ in the preceding analysis. To make the robustness tests fully explicit, we adopted the broken-$\alpha$ stretch standardisation introduced by \cite{Ginolin2025b}, who found that the usual single-$\alpha$ model is insufficient because the stretch--luminosity relation could be non-linear. In their formulation, the stretch coefficient is instead modelled as a piecewise function of $x_1$:
\begin{equation}
\alpha(x_1)=
\begin{cases}
\alpha_{\rm low}, & x_1 < x_1^0,\\
\alpha_{\rm high}, & x_1 \ge x_1^0~.
\end{cases}
\end{equation}
The best-fit parameters reported in that work are $x_1^0=-0.48\pm0.08$, $\alpha_{\rm low}=0.271\pm0.011$, and $\alpha_{\rm high}=0.083\pm0.009$. In our robustness test, we adopted these literature best-fit values as fixed inputs and applied the corresponding broken-$\alpha$ law as a fixed re-standardisation of the Hubble residuals before fitting any environmental step. Accordingly, the broken-$\alpha$ prescription is treated here as an $x_1$ correction, while the fitted free parameters in each step model are limited to the zero-point offset and the environmental step amplitude. This choice ensures that the environmental mass and age do not absorb the stretch standardisation itself.

As also summarised in Table~\ref{tab3}, adopting this literature broken-$\alpha$ prescription systematically increases both the mass step and LWA-step amplitudes relative to the default single-$\alpha$ case. For the narrow-$c$ sample, the mass-only step becomes $0.118 \pm 0.052$~mag for $z_{\rm cmb}$ and $0.139 \pm 0.052$~mag for $z_{\rm cor}$, while the age-only step is $0.193 \pm 0.042$~mag and $0.190 \pm 0.041$~mag, respectively. In the corresponding joint fits, the mass step is reduced to $0.049 \pm 0.049$~mag and $0.049 \pm 0.053$~mag, whereas the age step remains significant at $0.178 \pm 0.045$~mag and $0.171 \pm 0.046$~mag. A similar pattern is found for the broader-$c$ sample: the mass-only step reaches $0.133 \pm 0.054$~mag for $z_{\rm cmb}$ and $0.138 \pm 0.051$~mag for $z_{\rm cor}$, but drops to $0.042 \pm 0.057$~mag and $0.052 \pm 0.050$~mag in the joint fit, while the joint age step remains at $0.167 \pm 0.049$~mag and $0.179 \pm 0.044$~mag, respectively. Therefore, adopting the literature broken-$\alpha$ increases the amplitude of the standalone mass step, but does not alter our main conclusion that, once the global mass and local LWA are modelled jointly, the mass step is substantially weakened whereas the local LWA step remains significant.

To assess whether our local LWA step is driven by the host-mass distribution, we performed two complementary re-weighting tests; the detailed weighting procedure is presented in Appendix~\ref{appendix}. First, we re-weighted our SN~Ia sample so that its host-mass distribution matches that of the ZTF SN~Ia DR2 reference sample. Second, we repeated the same set of fits using the MaNGA volume weight \texttt{ESWEIGHT} from the MaNGA \texttt{drpall} catalogue as an approximate correction for the MaNGA selection function.

As shown in Table~\ref{tab4}, for our sample, the ZTF-re-weighted fits give $\Delta_A = 0.145 \pm 0.049$ and $\Delta_{A,\mathrm{joint}} = 0.142 \pm 0.055$ for $z_{\rm cmb}$, and $\Delta_A = 0.118 \pm 0.052$ and $\Delta_{A,\mathrm{joint}} = 0.108 \pm 0.062$ for $z_{\rm cor}$, while the corresponding joint mass step are reduced to $0.007 \pm 0.050$ and $0.017 \pm 0.055$, respectively. Under the MaNGA volume-weight, the age step remains comparably strong, with $\Delta_A = 0.186 \pm 0.049$ and $\Delta_{A,\mathrm{joint}} = 0.162 \pm 0.057$ for $z_{\rm cmb}$, and $\Delta_A = 0.156 \pm 0.052$ and $\Delta_{A,\mathrm{joint}} = 0.118 \pm 0.066$ for $z_{\rm cor}$. Therefore, although the numerical value of the mass step depends sensitively on the adopted weighting scheme, the age step persists across these tests and is not removed by explicitly rebalancing the host-mass distribution.

\begin{table*}
\centering
\caption{Robustness tests of the mass and age steps under different re-weighting schemes.}
\label{tab4}
\begin{tabular}{lccccccc}
\hline
Weight scheme & Redshift & $N$ & Threshold $(M_{\rm thr}, A_{\rm thr})$ & Mass-only step & Age-only step & Joint mass step & Joint age step\\
\hline
unweighted & $z_{\rm cmb}$ & 56
& $(10.278,\,9.084)$
& $0.071 \pm 0.035$
& $0.163 \pm 0.031$
& $0.028 \pm 0.033$
& $0.156 \pm 0.032$
\\

unweighted & $z_{\rm cor}$ & 54
& $(10.276,\,9.083)$
& $0.110 \pm 0.034$
& $0.148 \pm 0.030$
& $0.059 \pm 0.035$
& $0.128 \pm 0.033$
\\

\hline
ZTF weighted & $z_{\rm cmb}$ & 56
& $(10.278,\,9.084)$
& $0.063 \pm 0.048$
& $0.145 \pm 0.049$
& $0.007 \pm 0.050$
& $0.142 \pm 0.055$
\\

ZTF weighted & $z_{\rm cor}$ & 54
& $(10.339,\,9.129)$
& $0.068 \pm 0.048$
& $0.118 \pm 0.052$
& $0.017 \pm 0.055$
& $0.108 \pm 0.062$
\\

\hline
MaNGA weighted & $z_{\rm cmb}$ & 54
& $(10.191,\,9.089)$
& $0.117 \pm 0.048$
& $0.186 \pm 0.049$
& $0.043 \pm 0.052$
& $0.162 \pm 0.057$
\\

MaNGA weighted & $z_{\rm cor}$ & 52
& $(10.279,\,9.130)$
& $0.123 \pm 0.049$
& $0.156 \pm 0.052$
& $0.059 \pm 0.060$
& $0.118 \pm 0.066$
\\
\hline
\end{tabular}
\tablefoot{The first block lists the original unweighted results. The second block gives the ZTF host-mass re-weighted results using the weighted-scan thresholds. The third block lists the MaNGA volume-weighted results using the weighted-scan thresholds.}
\end{table*}

\begin{figure}
    \centering
    \resizebox{0.8\hsize}{!}{\includegraphics{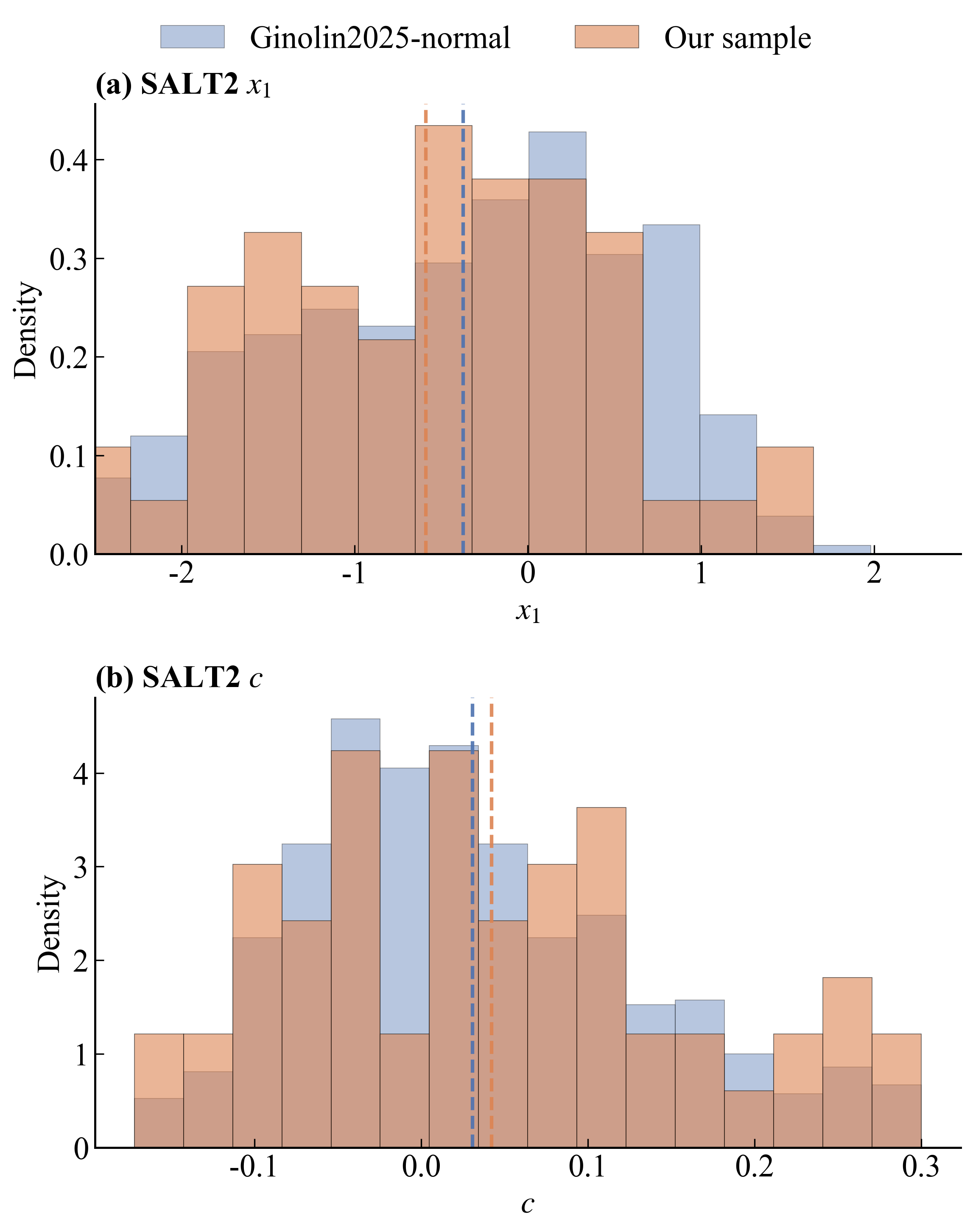}}
    \caption{
Comparison of the SALT2 $x_1$ and $c$ distributions between our sample (Narrow-$c$) and the volume-limited ZTF reference sample (Narrow-$c$ with Normal). 
The upper and lower panels show the normalised histograms of the $x_1$ and $c$ distributions, respectively. 
The vertical~dashed lines mark the sample means.
}
    \label{fig:selection_check}
\end{figure}

Finally, we examined whether differences in the SALT2 parameter distributions between our sample and the ZTF reference samples could introduce additional selection biases. We compared the SALT2 $x_1$ and $c$ distributions of our sample with those of the volume-limited ZTF reference samples constructed following the selection adopted in previous ZTF SN~Ia DR2 environmental step analyses. Two SN~Ia reference samples are considered: the full reference sample, which includes both normal SNe~Ia and SN~Ia-91T, and the normal reference sample, which contains only normal SNe~Ia. The comparison with the normal-SN~Ia reference sample is shown in Fig.~\ref{fig:selection_check}, while the results for different fitting colour ranges and SN~Ia samples are summarised in Table~\ref{tab:selection_check}.

Compared with the full ZTF reference samples, our sample shows a moderate shift towards lower $x_1$, from -0.328 to -0.352 and KS $p$-values from 0.002 to 0.011, depending on the fitting colour range. However, this comparison involves different SN~Ia populations, since the full reference samples include both normal SNe~Ia and SN~Ia-91T, whereas our sample consists only of normal SNe~Ia.

When compared with the normal-SN~Ia reference samples, the offset in stretch is reduced. For our Narrow-$c$ sample, the difference becomes statistically insignificant for the Narrow-$c$ with normal reference sample ($\Delta\langle x_1\rangle=-0.215$, $p=0.086$) and also for the broader-c with normal reference sample comparison ($\Delta\langle x_1\rangle=-0.233$, $p=0.080$). For our broad-c sample, a modest difference remains ($\Delta\langle x_1\rangle=-0.241$, $p=0.026$). These results indicate that the inclusion of SN~Ia-91T objects in the full reference samples contributes to the $x_1$ offset, while the difference becomes insignificant when the comparison is restricted to normal SNe~Ia, with the $x_1$ distributions becoming statistically indistinguishable for our Narrow-$c$ sample. Since the SALT2 stretch parameter $x_1$ has been found to correlate with host galaxy properties, including stellar mass \cite[e.g.][]{Sullivan2010,Rigault2013, Rigault2020}, the remaining difference may partly reflect differences in the environment between our sample and the ZTF reference samples.

The comparison of the colour distributions gives a subtler result. Although our sample applies a stricter colour range ($c<0.3$), the comparison with the ZTF broad-colour reference sample ($c<0.8$) does not show a statistically significant difference (KS $p=0.597$). This is because most objects still occupy the moderate-colour region, and the additional high-$c$ tail does not substantially alter the overall colour distribution for the broad-$c$ reference sample. In all comparisons, we find no statistically significant difference in the colour distributions, with KS $p$-values from 0.389 to 0.746 and small mean offsets (0.012 to 0.032).

\begin{table*} 
    \centering 
    \caption{SALT2 $x_1$ and $c$ distributions in our sample and the ZTF DR2 reference samples.} 
    \label{tab:selection_check} 
    \begin{tabular}{llrrrrrr} 
        \hline 
        Our sample & Reference & 
        $N_{\rm sam}$ & $N_{\rm ref}$ & $\Delta\langle x_1\rangle$ & KS $p(x_1)$ & $\Delta\langle c\rangle$ & KS $p(c)$\\ 
        \hline 
        Narrow-$c$ & Narrow-$c$ with Full &56&868 &-0.328&0.011 &+0.014&0.389 \\ Narrow-$c$ & Narrow-$c$ with Normal &56&711 &-0.215&0.086 &+0.012&0.414 \\ 
        \hline 
        Narrow-$c$ & Broad-c with Full &56&967 &-0.344&0.009 &-0.032&0.597 \\ Narrow-$c$ & Broad-c with Normal &56&782 &-0.233&0.080 &-0.028&0.746 \\ 
        \hline 
        Broad-$c$ & Broad-c with Full &63&967 &-0.352&0.002 &+0.015&0.580 \\ Broad-$c$ & Broad-c with Normal &63&782 &-0.241&0.026 &+0.019&0.508 \\ 
        \hline 
    \end{tabular} 
    \tablefoot{The reference samples are constructed with two colour selections: Narrow-$c$ ($c<0.3$) and Broad-$c$ ($c<0.8$), and with two SN~Ia sample definitions: the full sample including normal SNe~Ia and SN~Ia-91T, and the normal sample containing only normal SNe~Ia. $\Delta$ is defined as the mean value of our sample minus that of the corresponding reference sample.}
\end{table*}

In summary, selection effects are unavoidable. However, all of the robustness checks performed here consistently show that they do not change our main conclusion: the conventional mass step is reduced once age information is included, whereas the local LWA step remains significant.


\section{Discussion} \label{sec:5}

\subsection{Environmental dependence of SN~Ia standardisation}

Differences in progenitor ages could be one of the underlying physical origins of the host-mass step \cite[e.g.][]{Rigault2020}. This motivated us to explore whether more physically motivated environmental properties can be obtained with the advent of high-quality IFS. Full-spectrum fitting of IFS data provides detailed properties of the local stellar populations. Compared with previous studies that mainly relied on global galaxy properties (e.g. stellar mass, age, and metallicity), local stellar population parameters may provide deeper insights into the physical origin of environmental dependences.

In this work, we used the MaNGA Pipe3D products to derive the local LWA of the stellar population within a 1~kpc aperture around the SN~Ia explosion sites. We used this quantity as a statistical tracer of the age of the progenitor stellar population and investigated its correlation with standardised luminosities of SNe~Ia. We find that the standardised luminosities of SNe~Ia in younger environments are systematically fainter than those in older environments by $0.163$ mag, with a significance of $5.2\sigma$. When the global stellar mass and local LWA are simultaneously included in the standardisation, the wRMS decreases from $0.155$ mag to $0.137$ mag. Meanwhile, the mass step becomes statistically insignificant, while the age step remains significant. This result provides further evidence that the local LWA is more closely related to the physical driver than the mass and lends support to the idea that progenitor age is that driver.

It should be noted that the local LWA derived from SPS fitting should not be interpreted as a direct measurement of the true progenitor age, but rather as a statistical tracer of the age of the progenitor stellar population around the SN position. The limitations associated with this interpretation are discussed in Sect.~\ref{sec:limitations}.

\subsection{Impact of the local LWA on the dark energy equation of state parameter}
\label{sec:5.2}

Previous studies have shown that SNe~Ia in younger environments are systematically fainter than those in older environments after standardisation \citep{Kang2020,Rigault2020,Lee2022}. This residual correlation indicates that the luminosities of SNe~Ia require additional corrections beyond the standard light-curve standardisation; their standardised brightness retains a dependence on their host galaxies' properties or progenitor physics. 

Recent work has quantified this effect. While \citet{Nicolas2021} observed redshift-dependent evolution in light-curve parameters, \citet{Chung2025} provided evidence that the standardised luminosity itself correlates strongly with host galaxy age, suggesting that SN~Ia from younger environments appear systematically fainter. Our detection of a significant local LWA step is consistent with such an environmental age dependence (Sect.~\ref{sec:3.2.2}).

Such environmental dependences may become a potential cosmological systematic bias through progenitor drift when extending SN~Ia samples to higher redshift \citep{Branch2001}. If the fraction of SNe Ia occurring in younger environments increases with redshift, a redshift-dependent bias in the Hubble diagram could be introduced if an age-dependent term is ignored. This could bias the inferred dark energy equation of state parameter, $w$, with the sign and amplitude depending on the sample redshift distribution and the adopted treatment of the age-dependent luminosity correction \citep{Campbell2016,Kang2020,Lee2022,Son2025,Hoang2026}.

Our present sample spans only a limited redshift range
($0.02<z<0.08$) and therefore does not directly constrain the
redshift evolution of the different local environments. Moreover, the limited redshift prevents this sample from providing strong constraints on the dark energy equation of state parameter, $w$. We therefore used the following
analysis to test the sensitivity of the inferred $w$ to the local LWA investigated in Sect.~\ref{sec:3.2.2}, rather than to provide
a precision cosmological measurement. The likelihood analysis is
described in Appendix~\ref{appendixB}.

\begin{figure}
    \centering
    \resizebox{0.9\hsize}{!}{\includegraphics{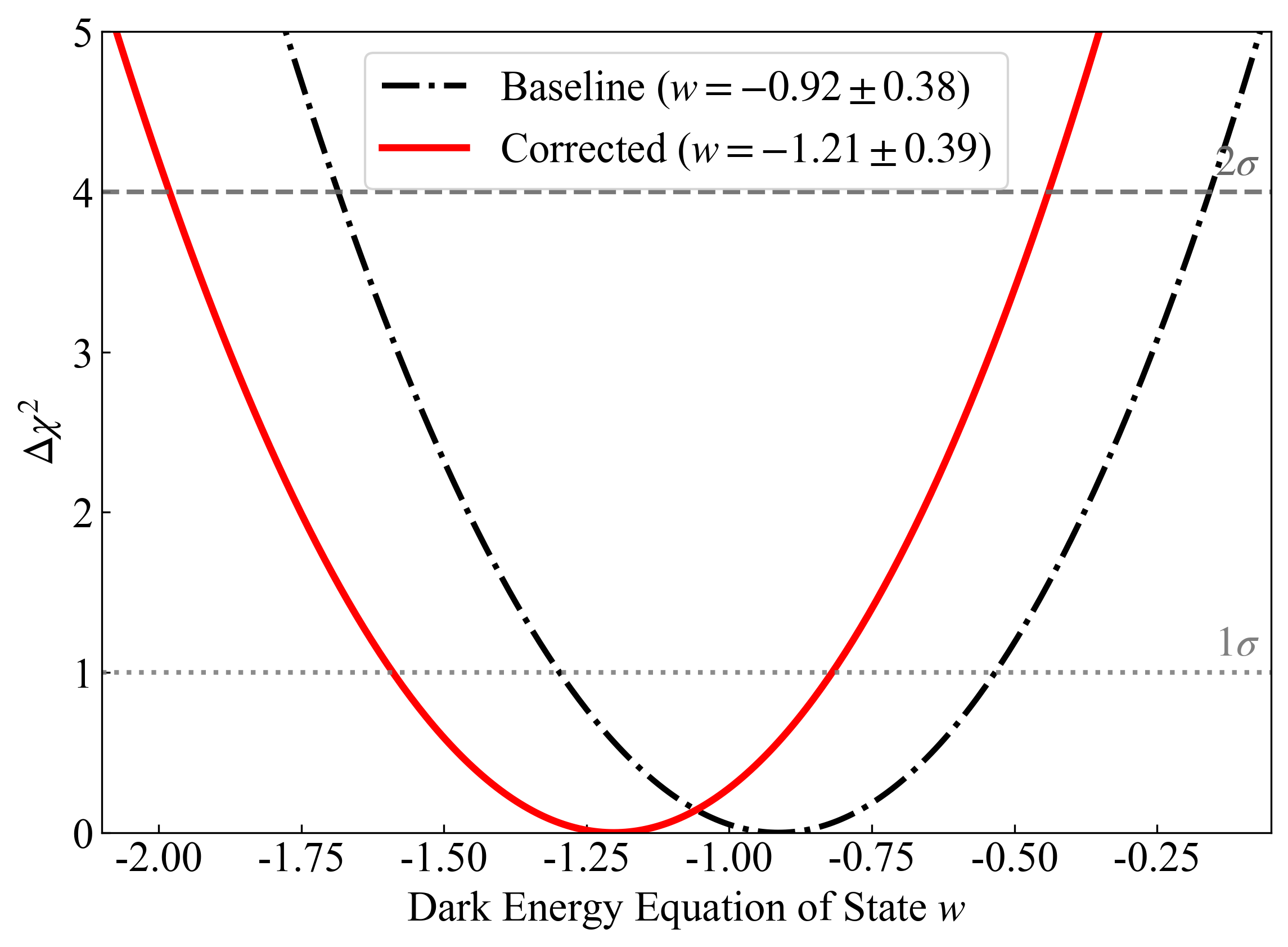}}
\caption{
Likelihood profiles of the dark energy equation of state parameter ($w$) for the $z_{\rm cmb}$ analysis under the fixed-baseline treatment.
The curves show the $\Delta\chi^2$ profiles for the baseline model
(dashed black) and for the case in which the local-LWA step is applied
on top of the baseline calibration (solid~red). The likelihood minimum
shifts from $w\approx-0.92$ to $w\approx-1.21$, corresponding to
$\Delta w\approx-0.29$. This illustrates the sensitivity of the inferred
$w$ to the local-LWA correction when the baseline calibration is fixed. Horizontal dotted lines indicate the $1\sigma$ and $2\sigma$ confidence levels.
}
    \label{fig8}
\end{figure}

We considered two cases of the local LWA step in the SN~Ia
standardisation. In the first case, we applied the age step to a standardisation that did not originally include it (see `Age step added'
in Table~\ref{tab:w_summary}). The baseline analysis gives
$w=-0.92\pm0.38$ for $z_{\rm cmb}$ and
$w=-0.88\pm0.39$ for $z_{\rm cor}$. After introducing the local LWA
step, the corresponding values become
$w=-1.21\pm0.39$ and $w=-1.08\pm0.38$, respectively,
corresponding to shifts of $\Delta w=-0.29$ and $-0.20$.
For the $z_{\rm cmb}$ case, the corresponding likelihood profiles are
shown in Fig.~\ref{fig8}.

\begin{table*}
\centering

\caption{Constraints on $w$ for different treatments of the local LWA step.}
\label{tab:w_summary}
\begin{tabular}{@{}l|ccc|ccc|ccc@{}}
\hline
Redshift
& \multicolumn{3}{c|}{Baseline}
& \multicolumn{3}{c|}{Age step added}
& \multicolumn{3}{c}{Age step jointly fitted}\\
\cline{2-10}
& $w$ & $\Delta_A$ & $\Delta w$
& $w$ & $\Delta_A$ & $\Delta w$
& $w$ & $\Delta_A$ & $\Delta w$\\
\hline
$z_{\rm cmb}$
& $-0.92\pm0.38$ & -- & --
& $-1.21\pm0.39$ & $0.127\pm0.041$ & $-0.29$
& $-0.97\pm0.35$ & $0.163\pm0.031$ & $-0.05$\\

$z_{\rm cor}$
& $-0.88\pm0.39$ & -- & --
& $-1.08\pm0.38$ & $0.115\pm0.041$ & $-0.20$
& $-0.94\pm0.35$ & $0.148\pm0.030$ & $-0.06$\\
\hline
\end{tabular}
\tablefoot{The `Age step added' case applies
the age correction on top of a fixed baseline calibration, while the `Age step
jointly fitted' case simultaneously fits the age step with the standardisation
parameters before releasing $w$.}
\end{table*}

This case produces a large shift of $\Delta w \simeq -0.3$. Because the standardisation parameters are fixed in this test, however, this value should not be interpreted as a quantitative cosmological bias. We therefore considered a second case, denoted `Age step jointly fitted' in Table~\ref{tab:w_summary}, in which the local LWA step is simultaneously fitted with the standardisation parameters $\alpha$, $\beta$, $M$, and $\sigma_{\rm int}$, assuming $w=-1$. We subsequently released $w$ and repeated the likelihood analysis using the resulting age-corrected Hubble residuals. The fitted local LWA step is consistent with the value obtained in the environmental analysis in Sect.~\ref{sec:3.2.2}. The resulting cosmological constraints are
$w=-0.97\pm0.35$ for $z_{\rm cmb}$ and
$w=-0.94\pm0.35$ for $z_{\rm cor}$, corresponding to shifts of
$\Delta w=-0.05$ and $-0.06$ relative to the baseline values.

The shifts obtained from the two cases are therefore not contradictory. Rather, they correspond to different ways of introducing the local LWA step into the SN~Ia standardisation. We therefore regard the $\Delta w \approx -0.3$ as a clue: it indicates the potential sensitivity of the inferred $w$ to the environmental corrections, and highlights the potential importance of accounting for the local LWA in future SN~Ia standardisation analyses.

\begin{figure}
    \centering
    \resizebox{0.95\hsize}{!}{\includegraphics{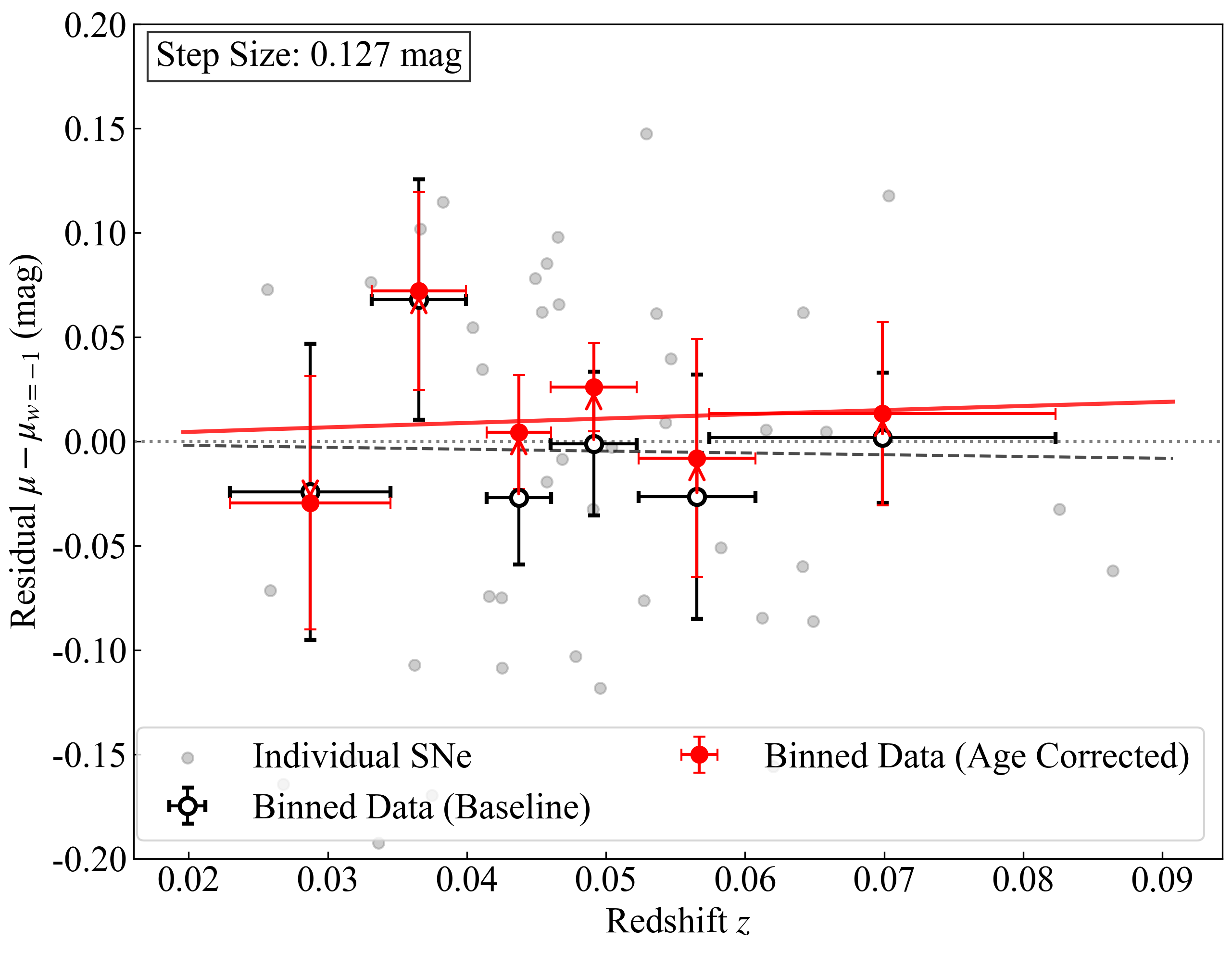}}
\caption{
Schematic illustration of a possible pathway through which environmental dependence could affect cosmological inference.
This figure illustrates the age step correction applied to individual SNe. If the relative fractions of SNe~Ia occurring in different local stellar environments vary with redshift, an un-modelled luminosity difference between these populations could introduce a redshift-dependent offset in the Hubble diagram and alter the inferred value of $w$. The age step amplitude used in this example is taken from the `Age step added' case in Table~\ref{tab:w_summary}, $\Delta_A=0.127$ mag. The redshift evolution of the relative population fractions shown here is illustrative and is not constrained by our present low-redshift sample.
}
    \label{fig9}
\end{figure}

Figure~\ref{fig9} schematically illustrates the possible
pathway through which such an environmental dependence could affect cosmological inference. If the fractions of SNe~Ia in younger and older environments vary with redshift, an un-modelled difference in their standardised luminosities could introduce a redshift-dependent offset in the Hubble diagram and consequently alter the inferred value of $w$. We emphasise that our low-redshift sample does not constrain such population evolution; Fig.~\ref{fig9} is intended only to illustrate the possible mechanism.

To examine whether this sensitivity is specific to the local LWA, we also
repeated the analysis using global stellar mass as the environmental
tracer. As summarised in Table~\ref{tab:compare}, including the
local LWA step alone changes the best-fitting value of $w$ by
approximately $\Delta w=-0.05$ for $z_{\rm cmb}$ and
$\Delta w=-0.06$ for $z_{\rm cor}$, whereas including the mass step alone gives shifts of approximately $\Delta w=-0.04$ and $-0.09$, respectively. Thus, within the statistical precision of the present sample, the mass step correction can produce a change in the inferred $w$ that is comparable in magnitude to that produced by the local LWA step correction.

\begin{table}[htbp]
\centering

\caption{Impact of environmental step corrections on $w$.}
\label{tab:compare}

\begin{tabular}{lccc}
\hline
\multicolumn{4}{c}{$z_{\rm cmb}$}\\
\hline
Case & $\Delta_A$ & $\Delta_M$ & $w$\\
\hline

Baseline & -- & -- & $-0.92\pm0.38$\\

$\Delta_A$ & $0.163\pm0.031$ & -- & $-0.97\pm0.35$\\

$\Delta_M$ & -- & $0.071\pm0.035$ & $-0.96\pm0.39$\\

$\Delta_A+\Delta_M$
& $0.156\pm0.032$ & $0.028\pm0.033$
& $-0.98\pm0.36$\\

\hline
\multicolumn{4}{c}{$z_{\rm cor}$}\\
\hline

Baseline & -- & -- & $-0.88\pm0.39$\\

$\Delta_A$
& $0.148\pm0.030$ & --
& $-0.94\pm0.35$\\

$\Delta_M$
& -- & $0.110\pm0.034$
& $-0.97\pm0.37$\\

$\Delta_A+\Delta_M$
& $0.128\pm0.033$ & $0.059\pm0.035$
& $-0.98\pm0.35$\\

\hline
\end{tabular}
\tablefoot{The fitted environmental step amplitudes $\Delta_\mathrm{A}$ and $\Delta_\mathrm{M}$ and the resulting values of $w$ are shown for two redshift corrections.}
\end{table}

The magnitude of the shift in $w$ alone therefore does not distinguish which environmental tracer is more closely related to the underlying dependence. The evidence favouring the local LWA over global stellar mass instead comes from the environmental-step analysis in Sect.~\ref{sec:results}. When the local LWA and global stellar mass are fitted simultaneously, the mass step is substantially reduced, whereas the local-LWA step remains significant. For example, for $z_{\rm cmb}$ the global mass step decreases from $0.071\pm0.035$ mag to $0.028\pm0.033$ mag when the local LWA is included, while the local LWA step remains $0.156\pm0.032$ mag. The cosmological test and the environmental step test therefore address different aspects of the problem: the former quantifies the sensitivity of the inferred $w$ to the adopted environmental treatment, whereas the latter indicates that the local LWA contains environmental information that is not fully represented by global stellar mass.

Taken together, these tests show that the inferred value of $w$ in our
low-redshift sample could be affected by the treatment of environmental dependences. When the local LWA step is introduced on top of a fixed baseline calibration, the shift can reach $\Delta w\simeq-0.3$, whereas fitting the environmental term simultaneously with the SN~Ia standardisation parameters yields substantially smaller shifts of approximately $-0.05$--$-0.06$. In the simultaneous-fit case, the shift in $w$ induced by the mass step correction is comparable in magnitude to that induced by the local LWA term. Given the limited sample size and redshift range, these values should not be interpreted as precise predictions of the cosmological bias in current large SN~Ia surveys. Instead, they demonstrate the potential impact of environmental dependences on the inferred cosmological parameters.

\subsection{Limitations} \label{sec:limitations}

This work still has several limitations. Firstly, the local LWA is not a direct measurement of the true SN~Ia progenitor age. Currently, it remains extremely challenging to directly observe the progenitor systems of SNe Ia, and therefore the physical properties of SN~Ia progenitors remain uncertain \cite[e.g.][]{Li2011, McCully2014}. Although recent studies combining IFS with SPS modelling can provide detailed stellar population properties of SN environments, these measurements characterise the average properties of mixed stellar populations around the explosion sites rather than the properties of the single progenitor population of SN~Ia. Therefore, the local LWA used in this work should be regarded as a tracer of the age of the stellar population associated with SN~Ia progenitors rather than the true progenitor age. Nevertheless, the local LWA provides a statistically meaningful characterisation of the stellar population environment in which SNe Ia explode. Future studies combining larger IFU surveys and more sophisticated stellar population models will further constrain the physical properties of SN~Ia progenitors.

Second, the sample size in this work is limited. The SN sample is obtained by cross-matching with MaNGA galaxies, and the number of available SNe is restricted by the MaNGA survey size (approximately 10,000 galaxies) and its spatial coverage. For some SNe located in the outer regions of their host galaxies, the SN positions fall outside the MaNGA FoV, preventing us from obtaining local environmental properties. The finite spatial coverage of MaNGA therefore introduces a potential radial selection effect, because SNe at larger distances from the host galaxy centre may be more likely to be excluded from the environmental analysis. Previous studies have reported radial differences in Hubble residuals and environmental-step amplitudes \citep[e.g.][]{Toy2025}. At the same time, spatially resolved stellar-population studies have shown that stellar age can also vary systematically with galactic radius, with younger stellar populations commonly found at larger distances from the galaxy centre, particularly in disc and late-type galaxies \citep[e.g.][]{KangX2016,KangX2017,Goddard2017}. Therefore, part of the reported radial dependence may reflect variations in the local stellar population rather than an independent effect of radial position itself. If such a radial dependence is partly associated with local stellar-population age, the age-related component would already be probed in our analysis, because we directly measured the local LWA at each SN position rather than using radial position as a proxy for the local environment. \cite{Toy2025} found that the mass step is weaker in their outer SN subsample; however, the physical origin of this radial dependence remains unclear. Although radial variations in stellar age provide one possible explanation for part of this dependence, other environmental properties, such as dust or metallicity, may also vary with radial position. Although the age-related component of the radial dependence is directly probed by the local LWA in our analysis, the loss of outer SNe may still affect the sampled distribution of local environments.
In addition, the quality of ZTF light curves is not uniformly high, and some SNe are excluded during light-curve fitting, further reducing the sample.

Thirdly, the redshift range of our sample limits the cosmological analysis. Constraining cosmological parameters is a critical aspect of SN cosmology; however, different cosmological models predict very similar luminosity distances at low-redshift and only become significantly distinguishable at higher redshift. Since nearby galaxy surveys such as MaNGA cannot provide resolved environmental information for large SN samples at high-redshift, our sample is limited to the low-redshift range ($0.02<z<0.08$). Therefore, we cannot directly determine the behaviour of environmental dependences at higher redshift. Given this limitation, we performed sensitivity tests to explore the possible influence of environmental effects on cosmological parameter inference (see Sect.~\ref{sec:5.2}).

 Future large-scale IFU surveys combined with next-generation SN surveys will provide resolved environmental information for much larger SN samples. Such datasets may enable more robust investigations of the physical origin of environmental dependences and their impact on precision cosmology.

\section{Conclusion} \label{sec:conclusion}

We cross-matched the SDSS MaNGA with TNS samples and find a significant statistical correlation between the environmental properties within a 1~kpc physical aperture of the SN Ia explosion site and the Hubble residuals. By constructing a joint likelihood analysis that contains the full covariance matrix, we find that the local LWA is more closely associated with the environmental dependence of SN~Ia standardised luminosity than either the global or local stellar mass.\begin{itemize} 

\item Light-curve parameters and the local LWA: We find a significant negative correlation between the SN~Ia light-curve stretch factor ($x_1$) and the local LWA, confirming that younger environments tend to host SNe~Ia with broader, more slowly declining light curves, as discovered previously \cite[e.g.][]{Hamuy1995,Howell2001,Lampeitl2010,Sullivan2010,Rigault2013,Rigault2020}. In contrast, no significant correlation is found between the colour parameter ($c$) and the local LWA, suggesting that SN~Ia colour is not strongly driven by the age of the progenitor population in our sample.

\item Local LWA step: After standardising the light curves, SNe~Ia in young environments are significantly fainter than those in old environments. By incorporating an age step, $\Delta_\mathrm{A}$, into the standardisation model, we find a step amplitude of $0.163$~mag with a significance of $5.2\sigma$. This inclusion significantly reduces the dispersion of Hubble residuals (the wRMS decreases from 0.155 to 0.138~mag, i.e. by about 10$\%$), demonstrating that the local LWA is a more effective standardisation parameter than mass.

\item Host galaxy mass step: While a global mass step of $0.071$ mag ($2.0\sigma$) is observed, it is largely explained by the age difference. When the age step and the global mass step are fitted simultaneously, the host-mass step amplitude decreases by approximately 60\% (from 0.071 to 0.028~mag) and becomes statistically insignificant ($0.9\sigma$), whereas the age step remains robust at $ 4.9\sigma$. This result supports the idea that the widely used global mass step traces the differences in stellar populations potentially associated with progenitor age \cite[e.g.][]{Galbany2014, Roman2018, Rigault2020}.

\item Local mass step: The local stellar mass step is substantially reduced when the local LWA is included. When accounting for the local LWA, the local mass step almost vanishes (from $0.087$ to $0.012$ mag, $0.3\sigma$), while the age step signal remains strong. This indicates that local mass does not provide an additional constraint on SN~Ia luminosity when we include the local LWA.

\item Potential impact on the dark energy equation of state parameter:
We assessed the sensitivity of the inferred dark energy equation of state parameter ($w$) to the local-LWA dependence. Applying the age correction produces a shift of $\Delta w\approx-0.05$ to $-0.06$. These results suggest that the inferred $w$ can be affected by different treatments of environmental corrections.

\end{itemize}In summary, our results indicate that the local LWA is more closely related to the physical driver of SN~Ia standardised-luminosity variations than the global host galaxy mass in our sample. The substantial reduction of the mass step when the local LWA is included suggests that the conventional mass step  partly traces underlying stellar-population differences. These results motivate further investigations of resolved local environmental properties using larger SN~Ia samples, particularly for assessing their potential impact on precision cosmology.

\section*{Data availability}

The full version of Table~\ref{tab1} is available in electronic form at the CDS via anonymous ftp to cdsarc.u-strasbg.fr (130.79.128.5) or via http://cdsweb.u-strasbg.fr/cgi-bin/qcat?J/A+A/. 

\begin{acknowledgements}
      We thank the referee for constructive comments that helped improve this work. We also thank Cheng Li for valuable discussion. This work is supported by the National Natural Science Foundation of China (NSFC Nos. 12288102, 12333008 and12125303), the National Key R\&D Program of China (grant Nos. 2021YFA1600401, 2021YFA1600403), the Strategic Priority Research Program of the Chinese Academy of Sciences (grant Nos. XDB1160303, XDB1160000, XDB1160200, XDB1160201, XDB1160300), Yunnan Fundamental Research Projects (NOs. 202401BC070007), the Yunnan Revitalization Talent Support Program-Science \& Technology Champion Project (No. 202305AB350003), the New Cornerstone Science Foundation through the XPLORER PRIZE, the CAS Project for Young Scientists in Basic Research (YSBR-148) and International Centre of Supernovae (ICESUN), Yunnan Key Laboratory of Supernova Research (No. 202505AV340004), the Yunnan Revitalization Talent Support Program ``YunLing Scholar'' project, and the China Manned Space Program with grant No. CMS-CSST-2025-A13.
\end{acknowledgements}

\bibliographystyle{aa} 
\bibliography{paper1_references}

@article{Bellm2019,
    author = {{Bellm}, E.~C. and {Kulkarni}, S.~R. and {Graham}, M.~J. and others},
    title = "{The Zwicky Transient Facility: System Overview, Performance, and First Results}",
    journal = {\pasp},
    year = 2019,
    volume = {131},
    pages = {018002},
    doi = {10.1088/1538-3873/aaecbe},
    adsurl = {https://ui.adsabs.harvard.edu/abs/2019PASP..131a8002B}
}

@article{Betoule2014,
    author = {{Betoule}, M. and {Kessler}, R. and {Guy}, J. and others},
    title = "{Improved cosmological constraints from a joint analysis of the SDSS-II and SNLS supernova samples}",
    journal = {\aap},
    year = 2014,
    volume = {568},
    eid = {A22},
    pages = {A22},
    doi = {10.1051/0004-6361/201423413},
    adsurl = {https://ui.adsabs.harvard.edu/abs/2014A%26A...568A..22B}
}

@article{Branch2001,
    author = {{Branch}, D. and {Perlmutter}, S. and {Baron}, E. and others},
    title = "{Type Ia Supernovae as Probes of the Universe}",
    journal = {arXiv e-prints},
    year = 2001,
    eprint = {astro-ph/0109070},
    archivePrefix = {arXiv},
    adsurl = {https://ui.adsabs.harvard.edu/abs/2001astro.ph..9070B}
}

@ARTICLE{Briday2022,
       author = {{Briday}, M. and {Rigault}, M. and {Graziani}, R. and {Copin}, Y. and {Aldering}, G. and {Amenouche}, M. and {Brinnel}, V. and {Kim}, A.~G. and {Kim}, Y.-L. and {Lezmy}, J. and {Nicolas}, N. and {Nordin}, J. and {Perlmutter}, S. and {Rosnet}, P. and {Smith}, M.},
        title = "{Accuracy of environmental tracers and consequences for determining the Type Ia supernova magnitude step}",
      journal = {\aap},
         year = 2022,
        month = jan,
       volume = {657},
          eid = {A22},
        pages = {A22},
          doi = {10.1051/0004-6361/202141160},
archivePrefix = {arXiv},
       eprint = {2109.02456},
 primaryClass = {astro-ph.CO},
       adsurl = {https://ui.adsabs.harvard.edu/abs/2022A&A...657A..22B}
}

@ARTICLE{BroutScolnic2021,
       author = {{Brout}, Dillon and {Scolnic}, Daniel},
        title = "{It's Dust: Solving the Mysteries of the Intrinsic Scatter and Host-galaxy Dependence of Standardized Type Ia Supernova Brightnesses}",
      journal = {\apj},
         year = 2021,
        month = mar,
       volume = {909},
       number = {1},
          eid = {26},
        pages = {26},
          doi = {10.3847/1538-4357/abd69b},
archivePrefix = {arXiv},
       eprint = {2004.10206},
 primaryClass = {astro-ph.CO},
       adsurl = {https://ui.adsabs.harvard.edu/abs/2021ApJ...909...26B}
}

@ARTICLE{Brout2019,
       author = {{Brout}, D. and {Scolnic}, D. and {Kessler}, R. and {D'Andrea}, C.~B. and {Davis}, T.~M. and {Gupta}, R.~R. and {Hinton}, S.~R. and {Kim}, A.~G. and {Lasker}, J. and {Lidman}, C. and {Macaulay}, E. and {M{\"o}ller}, A. and {Nichol}, R.~C. and {Sako}, M. and {Smith}, M. and {Sullivan}, M. and {Zhang}, B. and {Andersen}, P. and {Asorey}, J. and {Avelino}, A. and {Bassett}, B.~A. and {Brown}, P. and {Calcino}, J. and {Carollo}, D. and {Challis}, P. and {Childress}, M. and {Clocchiatti}, A. and {Filippenko}, A.~V. and {Foley}, R.~J. and {Galbany}, L. and {Glazebrook}, K. and {Hoormann}, J.~K. and {Kasai}, E. and {Kirshner}, R.~P. and {Kuehn}, K. and {Kuhlmann}, S. and {Lewis}, G.~F. and {Mandel}, K.~S. and {March}, M. and {Miranda}, V. and {Morganson}, E. and {Muthukrishna}, D. and {Nugent}, P. and {Palmese}, A. and {Pan}, Y.-C. and {Sharp}, R. and {Sommer}, N.~E. and {Swann}, E. and {Thomas}, R.~C. and {Tucker}, B.~E. and {Uddin}, S.~A. and {Wester}, W. and {Abbott}, T.~M.~C. and {Allam}, S. and {Annis}, J. and {Avila}, S. and {Bechtol}, K. and {Bernstein}, G.~M. and {Bertin}, E. and {Brooks}, D. and {Burke}, D.~L. and {Carnero Rosell}, A. and {Carrasco Kind}, M. and {Carretero}, J. and {Castander}, F.~J. and {Cunha}, C.~E. and {da Costa}, L.~N. and {Davis}, C. and {De Vicente}, J. and {DePoy}, D.~L. and {Desai}, S. and {Diehl}, H.~T. and {Doel}, P. and {Drlica-Wagner}, A. and {Eifler}, T.~F. and {Estrada}, J. and {Fernandez}, E. and {Flaugher}, B. and {Fosalba}, P. and {Frieman}, J. and {Garc{\'\i}a-Bellido}, J. and {Gruen}, D. and {Gruendl}, R.~A. and {Gutierrez}, G. and {Hartley}, W.~G. and {Hollowood}, D.~L. and {Honscheid}, K. and {Hoyle}, B. and {James}, D.~J. and {Jarvis}, M. and {Jeltema}, T. and {Krause}, E. and {Lahav}, O. and {Li}, T.~S. and {Lima}, M. and {Maia}, M.~A.~G. and {Marriner}, J. and {Marshall}, J.~L. and {Martini}, P. and {Menanteau}, F. and {Miller}, C.~J. and {Miquel}, R. and {Ogando}, R.~L.~C. and {Plazas}, A.~A. and {Romer}, A.~K. and {Roodman}, A. and {Rykoff}, E.~S. and {Sanchez}, E. and {Santiago}, B. and {Scarpine}, V. and {Schubnell}, M. and {Serrano}, S. and {Sevilla-Noarbe}, I. and {Smith}, R.~C. and {Soares-Santos}, M. and {Sobreira}, F. and {Suchyta}, E. and {Swanson}, M.~E.~C. and {Tarle}, G. and {Thomas}, D. and {Troxel}, M.~A. and {Tucker}, D.~L. and {Vikram}, V. and {Walker}, A.~R. and {Zhang}, Y. and {DES Collaboration}},
        title = "{First Cosmology Results Using SNe Ia from the Dark Energy Survey: Analysis, Systematic Uncertainties, and Validation}",
      journal = {\apj},
         year = 2019,
        month = apr,
       volume = {874},
       number = {2},
          eid = {150},
        pages = {150},
          doi = {10.3847/1538-4357/ab08a0},
archivePrefix = {arXiv},
       eprint = {1811.02377},
 primaryClass = {astro-ph.CO},
       adsurl = {https://ui.adsabs.harvard.edu/abs/2019ApJ...874..150B}
}

@article{Bundy2015,
    author = {{Bundy}, K. and {Bershady}, M.~A. and {Law}, D.~R. and others},
    title = "{Overview of the SDSS-IV MaNGA Survey}",
    journal = {\apj},
    year = 2015,
    volume = {798},
    pages = {7},
    doi = {10.1088/0004-637X/798/1/7}
}

@article{Campbell2016,
    author = {Campbell, H. and Fraser, M. and Gilmore, G.},
    title = {How SN Ia host-galaxy properties affect cosmological parameters},
    journal = {\mnras},
    volume = {457},
    number = {4},
    pages = {3470-3491},
    year = {2016},
    month = {04},
    issn = {0035-8711},
    doi = {10.1093/mnras/stw115},
    url = {https://doi.org/10.1093/mnras/stw115},
    eprint = {https://academic.oup.com/mnras/article-pdf/457/4/3470/18755090/stw115.pdf},
}

@article{Cardelli1989,
    author = {{Cardelli}, J.~A. and {Clayton}, G.~C. and {Mathis}, J.~S.},
    title = "{The relationship between infrared, optical, and ultraviolet extinction}",
    journal = {\apj},
    year = 1989,
    volume = {345},
    pages = {245},
    doi = {10.1086/167900}
}

@article{Carrick2015,
    author = {{Carrick}, J. and {Turnbull}, S.~J. and {Lavaux}, G. and others},
    title = "{Cosmological parameters from the comparison of peculiar velocities}",
    journal = {\mnras},
    year = 2015,
    volume = {450},
    pages = {317}
}

@article{Childress2013a,
    author = {{Childress}, M. and {Aldering}, G. and {Antilogus}, P. and others},
    title = "{Host Galaxy Properties and Hubble Residuals of Type Ia Supernovae}",
    journal = {\apj},
    year = 2013,
    volume = {770},
    pages = {108},
    doi = {10.1088/0004-637X/770/2/108}
}

@article{DAndrea2011,
    author = {{D'Andrea}, C.~B. and {Gupta}, R.~R. and {Sako}, M. and others},
    title = "{Type Ia Supernova Host Galaxy Properties in SDSS-II}",
    journal = {\apj},
    year = 2011,
    volume = {743},
    pages = {172},
    doi = {10.1088/0004-637X/743/2/172}
}

@article{DiValentino2021,
    author = {{Di Valentino}, E. and {Mena}, O. and {Pan}, S. and others},
    title = "{In the realm of the Hubble tension}",
    journal = {Class. Quant. Grav.},
    year = 2021,
    volume = {38},
    pages = {153001},
    doi = {10.1088/1361-6382/ac086d}
}

@article{Freedman2021,
    author = {{Freedman}, W.~L.},
    title = "{Measurements of the Hubble Constant: Tensions in Perspective}",
    journal = {\apj},
    year = 2021,
    volume = {919},
    pages = {16},
    doi = {10.3847/1538-4357/ac0e95}
}

@ARTICLE{Galbany2014,
       author = {{Galbany}, L. and {Stanishev}, V. and {Mour{\~a}o}, A.~M. and {Rodrigues}, M. and {Flores}, H. and {Garc{\'\i}a-Benito}, R. and {Mast}, D. and {Mendoza}, M.~A. and {S{\'a}nchez}, S.~F. and {Badenes}, C. and {Barrera-Ballesteros}, J. and {Bland-Hawthorn}, J. and {Falc{\'o}n-Barroso}, J. and {Garc{\'\i}a-Lorenzo}, B. and {Gomes}, J.~M. and {Gonz{\'a}lez Delgado}, R.~M. and {Kehrig}, C. and {Lyubenova}, M. and {L{\'o}pez-S{\'a}nchez}, A.~R. and {de Lorenzo-C{\'a}ceres}, A. and {Marino}, R.~A. and {Meidt}, S. and {Moll{\'a}}, M. and {Papaderos}, P. and {P{\'e}rez-Torres}, M.~A. and {Rosales-Ortega}, F.~F. and {van de Ven}, G.},
        title = "{Nearby supernova host galaxies from the CALIFA Survey. I. Sample, data analysis, and correlation to star-forming regions}",
      journal = {\aap},
         year = 2014,
        month = dec,
       volume = {572},
          eid = {A38},
        pages = {A38},
          doi = {10.1051/0004-6361/201424717},
archivePrefix = {arXiv},
       eprint = {1409.1623},
 primaryClass = {astro-ph.GA},
       adsurl = {https://ui.adsabs.harvard.edu/abs/2014A&A...572A..38G}
}

@article{Garnavich1998,
    author = {{Garnavich}, P.~M. and {Jha}, S. and {Challis}, P. and others},
    title = "{Supernova Limits on the Cosmic Equation of State}",
    journal = {\apj},
    year = 1998,
    volume = {509},
    pages = {74},
    doi = {10.1086/306495}
}

@ARTICLE{Ginolin2025b,
       author = {{Ginolin}, M. and {Rigault}, M. and {Smith}, M. and {Copin}, Y. and {Ruppin}, F. and {Dimitriadis}, G. and {Goobar}, A. and {Johansson}, J. and {Maguire}, K. and {Nordin}, J. and {Amenouche}, M. and {Aubert}, M. and {Barjou-Delayre}, C. and {Betoule}, M. and {Burgaz}, U. and {Carreres}, B. and {Deckers}, M. and {Dhawan}, S. and {Feinstein}, F. and {Fouchez}, D. and {Galbany}, L. and {Ganot}, C. and {Harvey}, L. and {de Jaeger}, T. and {Kenworthy}, W.~D. and {Kim}, Y.-L. and {Kowalski}, M. and {Kuhn}, D. and {Lacroix}, L. and {M{\"u}ller-Bravo}, T.~E. and {Nugent}, P. and {Popovic}, B. and {Racine}, B. and {Rosnet}, P. and {Rosselli}, D. and {Sollerman}, J. and {Terwel}, J.~H. and {Townsend}, A. and {Brugger}, J. and {Bellm}, E.~C. and {Kasliwal}, M.~M. and {Kulkarni}, S. and {Laher}, R.~R. and {Masci}, F.~J. and {Riddle}, R.~L. and {Sharma}, Y.},
        title = "{ZTF SN Ia DR2: Environmental dependencies of stretch and luminosity for a volume-limited sample of 1000 type Ia supernovae}",
      journal = {\aap},
         year = 2025,
        month = mar,
       volume = {695},
          eid = {A140},
        pages = {A140},
          doi = {10.1051/0004-6361/202450378},
archivePrefix = {arXiv},
       eprint = {2405.20965},
 primaryClass = {astro-ph.CO},
       adsurl = {https://ui.adsabs.harvard.edu/abs/2025A&A...695A.140G}
}

@ARTICLE{Ginolin2025a,
       author = {{Ginolin}, M. and {Rigault}, M. and {Copin}, Y. and {Popovic}, B. and {Dimitriadis}, G. and {Goobar}, A. and {Johansson}, J. and {Maguire}, K. and {Nordin}, J. and {Smith}, M. and {Aubert}, M. and {Barjou-Delayre}, C. and {Burgaz}, U. and {Carreres}, B. and {Dhawan}, S. and {Deckers}, M. and {Feinstein}, F. and {Fouchez}, D. and {Galbany}, L. and {Ganot}, C. and {de Jaeger}, T. and {Kim}, Y.-L. and {Kuhn}, D. and {Lacroix}, L. and {M{\"u}ller-Bravo}, T.~E. and {Nugent}, P. and {Racine}, B. and {Rosnet}, P. and {Rosselli}, D. and {Ruppin}, F. and {Sollerman}, J. and {Terwel}, J.~H. and {Townsend}, A. and {Dekany}, R. and {Graham}, M. and {Kasliwal}, M. and {Groom}, S.~L. and {Purdum}, J. and {Rusholme}, B. and {van der Walt}, S.},
        title = "{ZTF SN Ia DR2: Colour standardisation of type Ia supernovae and its dependence on the environment}",
      journal = {\aap},
         year = 2025,
        month = feb,
       volume = {694},
          eid = {A4},
        pages = {A4},
          doi = {10.1051/0004-6361/202450943},
archivePrefix = {arXiv},
       eprint = {2406.02072},
 primaryClass = {astro-ph.CO},
       adsurl = {https://ui.adsabs.harvard.edu/abs/2025A&A...694A...4G}
}

@ARTICLE{GonzalezGaitan2021,
       author = {{Gonz{\'a}lez-Gait{\'a}n}, S. and {de Jaeger}, T. and {Galbany}, L. and {Mour{\~a}o}, A. and {Paulino-Afonso}, A. and {Filippenko}, A.~V.},
        title = "{The effects of varying colour-luminosity relations on Type Ia supernova science}",
      journal = {\mnras},
         year = 2021,
        month = dec,
       volume = {508},
       number = {3},
        pages = {4656-4666},
          doi = {10.1093/mnras/stab2802},
archivePrefix = {arXiv},
       eprint = {2009.13230},
 primaryClass = {astro-ph.CO},
       adsurl = {https://ui.adsabs.harvard.edu/abs/2021MNRAS.508.4656G}
}

@article{Graham2019,
    author = {{Graham}, M.~J. and {Kulkarni}, S.~R. and {Bellm}, E.~C. and others},
    title = "{The Zwicky Transient Facility: Science Objectives}",
    journal = {\pasp},
    year = 2019,
    volume = {131},
    pages = {078001}
}

@article{Gupta2011,
    author = {{Gupta}, R.~R. and {D'Andrea}, C.~B. and {Sako}, M. and others},
    title = "{Improved Constraints on Type Ia Supernova Host Galaxy Properties}",
    journal = {\apj},
    year = 2011,
    volume = {740},
    pages = {92},
    doi = {10.1088/0004-637X/740/2/92}
}

@article{Gupta2016,
    author = {{Gupta}, R.~R. and {Kuhlmann}, S. and {Kovacs}, E. and others},
    title = "{Host galaxy identification for supernova surveys}",
    journal = {\mnras},
    year = 2016,
    volume = {458},
    pages = {453}
}

@article{Guy2007,
    author = {{Guy}, J. and {Astier}, P. and {Baumont}, S. and others},
    title = "{SALT2: using distant supernovae to improve the use of Type Ia supernovae as distance indicators}",
    journal = {\aap},
    year = 2007,
    volume = {466},
    pages = {11},
    doi = {10.1051/0004-6361:20066930}
}

@article{Hamuy1995,
    author = {{Hamuy}, M. and {Phillips}, M.~M. and {Maza}, J. and others},
    title = "{A Hubble Diagram of Distant Type Ia Supernovae}",
    journal = {\aj},
    year = 1995,
    volume = {109},
    pages = {1}
}

@article{Hansen2000,
    author = {{Hansen}, B.~E.},
    title = "{Sample Splitting and Threshold Estimation}",
    journal = {Econometrica},
    year = 2000,
    volume = {68},
    pages = {575}
}

@article{Hayden2013,
    author = {{Hayden}, B.~T. and {Gupta}, R.~R. and {Garnavich}, P.~M. and others},
    title = "{The Fundamental Metallicity Relation Reduces Type Ia SN Hubble Residuals More Than Host Mass Alone}",
    journal = {\apj},
    year = 2013,
    volume = {764},
    pages = {191}
}

@article{Howell2001,
    author = {{Howell}, D.~A.},
    title = "{The Progenitors of Subluminous Type Ia Supernovae}",
    journal = {\apjl},
    year = 2001,
    volume = {554},
    pages = {L193}
}

@article{Kelly2010,
    author = {{Kelly}, P.~L. and {Hicken}, M. and {Burke}, D.~L. and others},
    title = "{Hubble Residuals of Nearby Type Ia Supernovae are Correlated with Host Galaxy Mass}",
    journal = {\apj},
    year = 2010,
    volume = {715},
    pages = {743},
    doi = {10.1088/0004-637X/715/2/743}
}

@article{Kelsey2021,
       author = {{Kelsey}, L. and {Sullivan}, M. and {Smith}, M. and {Wiseman}, P. and {Brout}, D. and {Davis}, T.~M. and {Frohmaier}, C. and {Galbany}, L. and {Grayling}, M. and {Guti{\'e}rrez}, C.~P. and {Hinton}, S.~R. and {Kessler}, R. and {Lidman}, C. and {M{\"o}ller}, A. and {Sako}, M. and {Scolnic}, D. and {Uddin}, S.~A. and {Vincenzi}, M. and {Abbott}, T.~M.~C. and {Aguena}, M. and {Allam}, S. and {Annis}, J. and {Avila}, S. and {Bacon}, D. and {Bertin}, E. and {Brooks}, D. and {Burke}, D.~L. and {Carnero Rosell}, A. and {Carrasco Kind}, M. and {Carretero}, J. and {Castander}, F.~J. and {Costanzi}, M. and {da Costa}, L.~N. and {Desai}, S. and {Diehl}, H.~T. and {Doel}, P. and {Everett}, S. and {Ferrero}, I. and {Fert{\'e}}, A. and {Flaugher}, B. and {Fosalba}, P. and {Garc{\'\i}a-Bellido}, J. and {Gerdes}, D.~W. and {Gruen}, D. and {Gruendl}, R.~A. and {Gschwend}, J. and {Gutierrez}, G. and {Hollowood}, D.~L. and {Honscheid}, K. and {James}, D.~J. and {Kim}, A.~G. and {Kuehn}, K. and {Kuropatkin}, N. and {Lahav}, O. and {Lima}, M. and {Marshall}, J.~L. and {Martini}, P. and {Menanteau}, F. and {Miquel}, R. and {Morgan}, R. and {Ogando}, R.~L.~C. and {Palmese}, A. and {Paz-Chinch{\'o}n}, F. and {Plazas}, A.~A. and {Romer}, A.~K. and {S{\'a}nchez}, C. and {Sanchez}, E. and {Serrano}, S. and {Sevilla-Noarbe}, I. and {Suchyta}, E. and {Tarle}, G. and {Thomas}, D. and {To}, C. and {Varga}, T.~N. and {Walker}, A.~R. and {Wilkinson}, R.~D. and {DES Collaboration}},
        title = "{The effect of environment on Type Ia supernovae in the Dark Energy Survey three-year cosmological sample}",
      journal = {\mnras},
         year = 2021,
        month = mar,
       volume = {501},
       number = {4},
        pages = {4861-4876},
          doi = {10.1093/mnras/staa3924},
archivePrefix = {arXiv},
       eprint = {2008.12101},
 primaryClass = {astro-ph.CO},
       adsurl = {https://ui.adsabs.harvard.edu/abs/2021MNRAS.501.4861K}
}

@ARTICLE{Kim2019,
       author = {{Kim}, Young-Lo and {Kang}, Yijung and {Lee}, Young-Wook},
        title = "{Environmental Dependence of Type Ia Supernova Luminosities from the YONSEI Supernova Catalog}",
      journal = {JKAS},
         year = 2019,
        month = oct,
       volume = {52},
        pages = {181-205},
          doi = {10.5303/JKAS.2019.52.5.181},
archivePrefix = {arXiv},
       eprint = {1908.10375},
 primaryClass = {astro-ph.CO},
       adsurl = {https://ui.adsabs.harvard.edu/abs/2019JKAS...52..181K}
}

@article{Lampeitl2010,
    author = {{Lampeitl}, H. and {Smith}, M. and {Nichol}, R.~C. and others},
    title = "{The Effect of Host Galaxies on Type Ia Supernovae in the SDSS-II Supernova Survey}",
    journal = {\apj},
    year = 2010,
    volume = {722},
    pages = {566},
    doi = {10.1088/0004-637X/722/1/566}
}

@article{Law2016,
    author = {{Law}, D.~R. and {Cherinka}, B. and {Yan}, R. and others},
    title = "{The Data Reduction Pipeline for the SDSS-IV MaNGA IFU Survey}",
    journal = {\aj},
    year = 2016,
    volume = {152},
    pages = {83}
}

@ARTICLE{Lee2022,
       author = {{Lee}, Young-Wook and {Chung}, Chul and {Demarque}, Pierre and {Park}, Seunghyun and {Son}, Junhyuk and {Kang}, Yijung},
        title = "{Evidence for strong progenitor age dependence of type Ia supernova luminosity standardization process}",
      journal = {\mnras},
         year = 2022,
        month = dec,
       volume = {517},
       number = {2},
        pages = {2697-2708},
          doi = {10.1093/mnras/stac2840},
archivePrefix = {arXiv},
       eprint = {2107.06288},
 primaryClass = {astro-ph.GA},
       adsurl = {https://ui.adsabs.harvard.edu/abs/2022MNRAS.517.2697L}
}

@misc{Masci2023,
    author = {{Masci}, F.~J. and others},
    title = "{ZTF Forced Photometry Service User Guide}",
    year = 2023,
    url = {http://web.ipac.caltech.edu/staff/fmasci/ztf/forcedphot.pdf}
}

@article{Meldorf2023,
    author = {{Meldorf}, C. and {Palmese}, A. and {Brout}, D. and others},
    title = "{The Dark Energy Survey Supernova Program results: type Ia supernova brightness correlates with host galaxy dust}",
    journal = {\mnras},
    year = 2023,
    volume = {518},
    pages = {1985}
}

@article{Neill2009,
    author = {{Neill}, J.~D. and {Sullivan}, M. and {Howell}, D.~A. and others},
    title = "{The Local Hosts of Type Ia Supernovae}",
    journal = {\apj},
    year = 2009,
    volume = {707},
    pages = {1449}
}

@article{Nicolas2021,
    author = {{Nicolas}, N. and {Rigault}, M. and {Copin}, Y. and others},
    title = "{Redshift evolution of the underlying Type Ia supernova stretch distribution}",
    journal = {\aap},
    year = 2021,
    volume = {649},
    pages = {A74}
}

@article{Perlmutter1999a,
    author = {{Perlmutter}, S. and {Aldering}, G. and {Goldhaber}, G. and others},
    title = "{Measurements of Omega and Lambda from 42 High-Redshift Supernovae}",
    journal = {\apj},
    year = 1999,
    volume = {517},
    pages = {565},
    doi = {10.1086/307221}
}

@article{Perlmutter1999b,
  title = {Constraining Dark Energy with Type Ia Supernovae and Large-Scale Structure},
  author = {Perlmutter, Saul and Turner, Michael S. and White, Martin},
  journal = {Phys. Rev. Lett.},
  volume = {83},
  issue = {4},
  pages = {670-673},
  numpages = {0},
  year = {1999},
  month = {Jul},
  publisher = {American Physical Society},
  doi = {10.1103/PhysRevLett.83.670},
  url = {https://link.aps.org/doi/10.1103/PhysRevLett.83.670}
}

@article{Phillips1993,
    author = {{Phillips}, M.~M.},
    title = "{The Absolute Magnitudes of Type IA Supernovae}",
    journal = {\apjl},
    year = 1993,
    volume = {413},
    pages = {L105},
    doi = {10.1086/186970}
}

@article{Planck2020,
    author = {{Planck Collaboration} and {Aghanim}, N. and {Akrami}, Y. and others},
    title = "{Planck 2018 results. VI. Cosmological parameters}",
    journal = {\aap},
    year = 2020,
    volume = {641},
    pages = {A6},
    doi = {10.1051/0004-6361/201833910}
}

@article{Popovic2021,
    author = {{Popovic}, B. and {Brout}, D. and {Kessler}, R. and others},
    title = "{The Impact of Host Galaxy Dust on Supernova Cosmology}",
    journal = {\apj},
    year = 2021,
    volume = {913},
    pages = {49},
    doi = {10.3847/1538-4357/abf14f}
}

@article{Popovic2023,
doi = {10.3847/1538-4357/aca273},
url = {https://doi.org/10.3847/1538-4357/aca273},
year = {2023},
month = {mar},
publisher = {The American Astronomical Society},
volume = {945},
number = {1},
pages = {84},
author = {Popovic, Brodie and Brout, Dillon and Kessler, Richard and Scolnic, Daniel},
title = {The Pantheon+ Analysis: Forward Modeling the Dust and Intrinsic Color Distributions of Type Ia Supernovae, and Quantifying Their Impact on Cosmological Inferences},
journal = {\apj}
}

@article{Riess1996,
    author = {{Riess}, A.~G. and {Press}, W.~H. and {Kirshner}, R.~P.},
    title = "{A Precise Distance Indicator: Type Ia Supernova Multicolor Light Curve Shapes}",
    journal = {\apj},
    year = 1996,
    volume = {473},
    pages = {88},
    doi = {10.1086/178129}
}

@article{Riess1998,
    author = {{Riess}, A.~G. and {Filippenko}, A.~V. and {Challis}, P. and others},
    title = "{Observational Evidence from Supernovae for an Accelerating Universe}",
    journal = {\aj},
    year = 1998,
    volume = {116},
    pages = {1009},
    doi = {10.1086/300499}
}

@article{Riess2016,
    author = {{Riess}, A.~G. and {Macri}, L.~M. and {Hoffmann}, S.~L. and others},
    title = "{A 2.4\% Determination of the Local Value of the Hubble Constant}",
    journal = {\apj},
    year = 2016,
    volume = {826},
    pages = {56}
}

@article{Riess2022,
    author = {{Riess}, A.~G. and {Yuan}, W. and {Macri}, L.~M. and others},
    title = "{A Comprehensive Measurement of the Local Value of the Hubble Constant}",
    journal = {\apjl},
    year = 2022,
    volume = {934},
    pages = {L7},
    doi = {10.3847/2041-8213/ac5c5b}
}

@article{Rigault2013,
    author = {{Rigault}, M. and {Copin}, Y. and {Aldering}, G. and others},
    title = "{Evidence for environmental dependence of the Type Ia supernova Hubble residual}",
    journal = {\aap},
    year = 2013,
    volume = {560},
    pages = {A66},
    doi = {10.1051/0004-6361/201322104}
}

@article{Rigault2020,
    author = {{Rigault}, M. and {Briday}, M. and {Rocher}, A. and others},
    title = "{Strong dependence of Type Ia supernova standardization on the local specific star formation rate}",
    journal = {\aap},
    year = 2020,
    volume = {644},
    pages = {A176},
    doi = {10.1051/0004-6361/201936335}
}

@ARTICLE{Rigault2025,
       author = {{Rigault}, M. and {Smith}, M. and {Regnault}, N. and {Kenworthy}, W.~D. and {Maguire}, K. and {Goobar}, A. and {Dimitriadis}, G. and {Johansson}, J. and {Amenouche}, M. and {Aubert}, M. and {Barjou-Delayre}, C. and {Bellm}, E.~C. and {Burgaz}, U. and {Carreres}, B. and {Copin}, Y. and {Deckers}, M. and {de Jaeger}, T. and {Dhawan}, S. and {Feinstein}, F. and {Fouchez}, D. and {Galbany}, L. and {Ginolin}, M. and {Graham}, M.~J. and {Kim}, Y.-L. and {Kowalski}, M. and {Kuhn}, D. and {Kulkarni}, S.~R. and {M{\"u}ller-Bravo}, T.~E. and {Nordin}, J. and {Popovic}, B. and {Purdum}, J. and {Rosnet}, P. and {Rosselli}, D. and {Racine}, B. and {Ruppin}, F. and {Sollerman}, J. and {Terwel}, J.~H. and {Townsend}, A.},
        title = "{ZTF SN Ia DR2: Study of Type Ia supernova light-curve fits}",
      journal = {\aap},
         year = 2025,
        month = feb,
       volume = {694},
          eid = {A2},
        pages = {A2},
          doi = {10.1051/0004-6361/202450377},
archivePrefix = {arXiv},
       eprint = {2406.02073},
 primaryClass = {astro-ph.CO},
       adsurl = {https://ui.adsabs.harvard.edu/abs/2025A&A...694A...2R}
}

@article{Roman2018,
    author = {{Roman}, M. and {Hardin}, D. and {Betoule}, M. and others},
    title = "{Dependence of Type Ia supernova luminosities on their local environment}",
    journal = {\aap},
    year = 2018,
    volume = {615},
    pages = {A68},
    doi = {10.1051/0004-6361/201731420}
}

@article{Sanchez2016,
    author = {{S{\'a}nchez}, S.~F. and {P{\'e}rez}, E. and {S{\'a}nchez-Bl{\'a}zquez}, P. and others},
    title = "{Pipe3D: A pipeline to analyze Integral Field Spectroscopy Data}",
    journal = {\rmxaa},
    year = 2016,
    volume = {52},
    pages = {21}
}

@article{Schlegel1998,
    author = {{Schlegel}, D.~J. and {Finkbeiner}, D.~P. and {Davis}, M.},
    title = "{Maps of Dust Infrared Emission for Use in Estimation of Reddening}",
    journal = {\apj},
    year = 1998,
    volume = {500},
    pages = {525},
    doi = {10.1086/305772}
}

@article{Scolnic2018,
    author = {{Scolnic}, D.~M. and {Jones}, D.~O. and {Rest}, A. and others},
    title = "{The Complete Light-curve Sample of Spectroscopically Confirmed SNe Ia}",
    journal = {\apj},
    year = 2018,
    volume = {859},
    pages = {101},
    doi = {10.3847/1538-4357/aab9bb}
}

@article{Smith2020,
    author = {{Smith}, M. and {Sullivan}, M. and {Wiseman}, P. and others},
    title = "{First cosmology results using Type Ia supernovae from the Dark Energy Survey: the effect of host galaxy properties on supernova luminosity}",
    journal = {\mnras},
    year = 2020,
    volume = {494},
    pages = {4426},
    doi = {10.1093/mnras/staa870}
}

@ARTICLE{Son2025,
       author = {{Son}, Junhyuk and {Lee}, Young-Wook and {Chung}, Chul and {Park}, Seunghyun and {Cho}, Hyejeon},
        title = "{Strong progenitor age bias in supernova cosmology ─ II. Alignment with DESI BAO and signs of a non-accelerating universe}",
      journal = {\mnras},
         year = 2025,
        month = nov,
       volume = {544},
       number = {1},
        pages = {975-987},
          doi = {10.1093/mnras/staf1685},
archivePrefix = {arXiv},
       eprint = {2510.13121},
 primaryClass = {astro-ph.CO},
       adsurl = {https://ui.adsabs.harvard.edu/abs/2025MNRAS.544..975S}
}

@article{Stanishev2012,
       author = {{Stanishev}, V. and {Rodrigues}, M. and {Mour{\~a}o}, A. and {Flores}, H.},
        title = "{Type Ia supernova host galaxies as seen with IFU spectroscopy}",
      journal = {\aap},
         year = 2012,
        month = sep,
       volume = {545},
          eid = {A58},
        pages = {A58},
          doi = {10.1051/0004-6361/201219188},
archivePrefix = {arXiv},
       eprint = {1205.5183},
 primaryClass = {astro-ph.CO},
       adsurl = {https://ui.adsabs.harvard.edu/abs/2012A&A...545A..58S}
}

@article{Sullivan2006,
    author = {{Sullivan}, M. and {Le Borgne}, D. and {Pritchet}, C.~J. and others},
    title = "{Rates of SN Ia in passive and star-forming hosts}",
    journal = {\apj},
    year = 2006,
    volume = {648},
    pages = {868},
    doi = {10.1086/506137}
}

@article{Sullivan2010,
    author = {{Sullivan}, M. and {Conley}, A. and {Howell}, D.~A. and others},
    title = "{The dependence of Type Ia Supernova luminosities on their host galaxies}",
    journal = {\mnras},
    year = 2010,
    volume = {406},
    pages = {782},
    doi = {10.1111/j.1365-2966.2010.16731.x}
}

@article{Suzuki2012,
    author = {{Suzuki}, N. and {Rubin}, D. and {Lidman}, C. and others},
    title = "{The Hubble Space Telescope Cluster Supernova Survey. V.}",
    journal = {\apj},
    year = 2012,
    volume = {746},
    pages = {85}
}

@ARTICLE{Taylor2021,
  author={Taylor, G and Lidman, C and Tucker, B E and Brout, D and Hinton, S R and Kessler, R},
  journal={\mnras}, 
  title={A revised SALT2 surface for fitting Type Ia supernova light curves}, 
  year={2021},
  volume={504},
  number={3},
  pages={4111-4122},
  doi={10.1093/mnras/stab962}}

@article{ThorpMandel2022,
    author = {{Thorp}, S. and {Mandel}, K.~S.},
    title = "{BayeSN: a hierarchical Bayesian SED model for Type Ia supernovae}",
    journal = {\mnras},
    year = 2022,
    volume = {517},
    pages = {2360},
    doi = {10.1093/mnras/stac2667}
}

@article{Tremonti2004,
    author = {{Tremonti}, C.~A. and {Heckman}, T.~M. and {Kauffmann}, G. and others},
    title = "{The Origin of the Mass-Metallicity Relation}",
    journal = {\apj},
    year = 2004,
    volume = {613},
    pages = {898},
    doi = {10.1086/423264}
}

@article{Tripp1998,
    author = {{Tripp}, R.},
    title = "{A two-parameter luminosity correction for Type IA supernovae}",
    journal = {\aap},
    year = 1998,
    volume = {331},
    pages = {815},
    adsurl = {https://ui.adsabs.harvard.edu/abs/1998A%26A...331..815T}
}

@article{Wiseman2023,
    author = {{Wiseman}, P. and others},
    title = "{Further evidence that galaxy age drives observed Type Ia supernova luminosity differences}",
    journal = {\mnras},
    year = 2023,
    volume = {520},
    pages = {6214}
}

@ARTICLE{Wiseman2026,
       author = {{Wiseman}, Phil and {Popovic}, Brodie and {Sullivan}, Mark and {Riess}, Adam G. and {Scolnic}, Dan and {Chen}, Rebecca C. and {Davis}, Tamara M. and {Galbany}, Llu{\'\i}s and {Hook}, Isobel M. and {Jha}, Saurabh W. and {Kelsey}, Lisa and {Murakami}, Yukei S. and {Rigault}, Micka{\"e}l and {Rose}, Benjamin M. and {Schmidt}, Brian and {Smith}, Mat and {Vincenzi}, Maria},
        title = "{Still accelerating: type Ia supernova cosmology is robust to host galaxy age evolution}",
      journal = {\mnras},
         year = 2026,
        month = jul,
       volume = {549},
       number = {3},
          eid = {stag797},
        pages = {stag797},
          doi = {10.1093/mnras/stag797},
archivePrefix = {arXiv},
       eprint = {2601.13785},
 primaryClass = {astro-ph.CO},
       adsurl = {https://ui.adsabs.harvard.edu/abs/2026MNRAS.549ag797W}
}

@article{Wojtak2023,
    author = {{Wojtak}, R. and others},
    title = "{Magnification and evolution biases in the supernova Hubble diagram}",
    journal = {\mnras},
    year = 2023,
    volume = {523},
    pages = {2220}
}

@article{Blanton2017,
    author = {Blanton, Michael R. and others},
    title = "{Sloan Digital Sky Survey IV: Mapping the Milky Way, Nearby Galaxies, and the Universe}",
    journal = {\aj},
    year = 2017,
    volume = {154},
    number = {1},
    pages = {28},
    doi = {10.3847/1538-3881/aa7567},
    adsurl = {https://ui.adsabs.harvard.edu/abs/2017AJ....154...28B},
    archivePrefix = {arXiv},
    eprint = {1703.00052},
}

@article{Drory2015,
    author = {Drory, N. and others},
    title = "{The MaNGA Integral Field Unit Fiber Feed System for the Sloan 2.5 m Telescope}",
    journal = {\aj},
    year = 2015,
    volume = {149},
    number = {2},
    pages = {77},
    doi = {10.1088/0004-6256/149/2/77},
    adsurl = {https://ui.adsabs.harvard.edu/abs/2015AJ....149...77D},
    archivePrefix = {arXiv},
    eprint = {1410.5106},
}

@ARTICLE{Yan2016,
       author = {{Yan}, Renbin and {Bundy}, Kevin and {Law}, David R. and {Bershady}, Matthew A. and {Andrews}, Brett and {Cherinka}, Brian and {Diamond-Stanic}, Aleksandar M. and {Drory}, Niv and {MacDonald}, Nicholas and {S{\'a}nchez-Gallego}, Jos{\'e} R. and {Thomas}, Daniel and {Wake}, David A. and {Weijmans}, Anne-Marie and {Westfall}, Kyle B. and {Zhang}, Kai and {Arag{\'o}n-Salamanca}, Alfonso and {Belfiore}, Francesco and {Bizyaev}, Dmitry and {Blanc}, Guillermo A. and {Blanton}, Michael R. and {Brownstein}, Joel and {Cappellari}, Michele and {D'Souza}, Richard and {Emsellem}, Eric and {Fu}, Hai and {Gaulme}, Patrick and {Graham}, Mark T. and {Goddard}, Daniel and {Gunn}, James E. and {Harding}, Paul and {Jones}, Amy and {Kinemuchi}, Karen and {Li}, Cheng and {Li}, Hongyu and {Maiolino}, Roberto and {Mao}, Shude and {Maraston}, Claudia and {Masters}, Karen and {Merrifield}, Michael R. and {Oravetz}, Daniel and {Pan}, Kaike and {Parejko}, John K. and {Sanchez}, Sebastian F. and {Schlegel}, David and {Simmons}, Audrey and {Thanjavur}, Karun and {Tinker}, Jeremy and {Tremonti}, Christy and {van den Bosch}, Remco and {Zheng}, Zheng},
        title = "{SDSS-IV MaNGA IFS Galaxy Survey{\textemdash}Survey Design, Execution, and Initial Data Quality}",
      journal = {\aj},
         year = 2016,
        month = dec,
       volume = {152},
       number = {6},
          eid = {197},
        pages = {197},
          doi = {10.3847/0004-6256/152/6/197},
archivePrefix = {arXiv},
       eprint = {1607.08613},
 primaryClass = {astro-ph.GA},
       adsurl = {https://ui.adsabs.harvard.edu/abs/2016AJ....152..197Y}
}

@ARTICLE{Wake2017,
       author = {{Wake}, David A. and {Bundy}, Kevin and {Diamond-Stanic}, Aleksandar M. and {Yan}, Renbin and {Blanton}, Michael R. and {Bershady}, Matthew A. and {S{\'a}nchez-Gallego}, Jos{\'e} R. and {Drory}, Niv and {Jones}, Amy and {Kauffmann}, Guinevere and {Law}, David R. and {Li}, Cheng and {MacDonald}, Nicholas and {Masters}, Karen and {Thomas}, Daniel and {Tinker}, Jeremy and {Weijmans}, Anne-Marie and {Brownstein}, Joel R.},
        title = "{The SDSS-IV MaNGA Sample: Design, Optimization, and Usage Considerations}",
      journal = {\aj},
         year = 2017,
        month = sep,
       volume = {154},
       number = {3},
          eid = {86},
        pages = {86},
          doi = {10.3847/1538-3881/aa7ecc},
archivePrefix = {arXiv},
       eprint = {1707.02989},
 primaryClass = {astro-ph.GA},
       adsurl = {https://ui.adsabs.harvard.edu/abs/2017AJ....154...86W}
}

@article{Sanchez2018,
       author = {{S{\'a}nchez}, S.~F. and {Avila-Reese}, V. and {Hernandez-Toledo}, H. and {Cortes-Su{\'a}rez}, E. and {Rodr{\'\i}guez-Puebla}, A. and {Ibarra-Medel}, H. and {Cano-D{\'\i}az}, M. and {Barrera-Ballesteros}, J.~K. and {Negrete}, C.~A. and {Calette}, A.~R. and {de Lorenzo-C{\'a}ceres}, A. and {Ortega-Minakata}, R.~A. and {Aquino}, E. and {Valenzuela}, O. and {Clemente}, J.~C. and {Storchi-Bergmann}, T. and {Riffel}, R. and {Schimoia}, J. and {Riffel}, R.~A. and {Rembold}, S.~B. and {Brownstein}, J.~R. and {Pan}, K. and {Yates}, R. and {Mallmann}, N. and {Bitsakis}, T.},
        title = "{SDSS IV MaNGA - Properties of AGN Host Galaxies}",
      journal = {\rmxaa},
         year = 2018,
        month = apr,
       volume = {54},
        pages = {217-260},
          doi = {10.48550/arXiv.1709.05438},
archivePrefix = {arXiv},
       eprint = {1709.05438},
 primaryClass = {astro-ph.GA},
       adsurl = {https://ui.adsabs.harvard.edu/abs/2018RMxAA..54..217S}
}

@article{Masci2019,
    author = {Masci, Frank J. and others},
    title = "{The Zwicky Transient Facility: Data Processing, Products, and Archive}",
    journal = {\pasp},
    year = 2019,
    volume = {131},
    number = {995},
    pages = {018003},
    doi = {10.1088/1538-3873/aae8ac},
    adsurl = {https://ui.adsabs.harvard.edu/abs/2019PASP..131a8003M},
    archivePrefix = {arXiv},
    eprint = {1902.01872},
}

@article{Kang2020,
    author = {Kang, Yijung and others},
    title = "{Early-type Host Galaxies of Type Ia Supernovae. II. Evidence for Luminosity Evolution in Supernova Cosmology}",
    journal = {\apj},
    year = 2020,
    volume = {889},
    number = {1},
    pages = {8},
    doi = {10.3847/1538-4357/ab5afc},
    adsurl = {https://ui.adsabs.harvard.edu/abs/2020ApJ...889....8K},
    archivePrefix = {arXiv},
    eprint = {1912.04903},
}

@article{Ivezic2019,
    author = {{Ivezi{\'c}}, {\v{Z}}. and {Kahn}, S.~M. and {Tyson}, J.~A. and others},
    title = "{LSST: from Science Drivers to Reference Design and Anticipated Data Products}",
    journal = {\apj},
    year = 2019,
    volume = {873},
    number = {2},
    pages = {111},
    doi = {10.3847/1538-4357/ab042c},
    adsurl = {https://ui.adsabs.harvard.edu/abs/2019ApJ...873..111I}
}

@article{Spergel2015,
       author = {{Spergel}, D. and {Gehrels}, N. and {Baltay}, C. and {Bennett}, D. and {Breckinridge}, J. and {Donahue}, M. and {Dressler}, A. and {Gaudi}, B.~S. and {Greene}, T. and {Guyon}, O. and {Hirata}, C. and {Kalirai}, J. and {Kasdin}, N.~J. and {Macintosh}, B. and {Moos}, W. and {Perlmutter}, S. and {Postman}, M. and {Rauscher}, B. and {Rhodes}, J. and {Wang}, Y. and {Weinberg}, D. and {Benford}, D. and {Hudson}, M. and {Jeong}, W.-S. and {Mellier}, Y. and {Traub}, W. and {Yamada}, T. and {Capak}, P. and {Colbert}, J. and {Masters}, D. and {Penny}, M. and {Savransky}, D. and {Stern}, D. and {Zimmerman}, N. and {Barry}, R. and {Bartusek}, L. and {Carpenter}, K. and {Cheng}, E. and {Content}, D. and {Dekens}, F. and {Demers}, R. and {Grady}, K. and {Jackson}, C. and {Kuan}, G. and {Kruk}, J. and {Melton}, M. and {Nemati}, B. and {Parvin}, B. and {Poberezhskiy}, I. and {Peddie}, C. and {Ruffa}, J. and {Wallace}, J.~K. and {Whipple}, A. and {Wollack}, E. and {Zhao}, F.},
        title = "{Wide-Field InfrarRed Survey Telescope-Astrophysics Focused Telescope Assets WFIRST-AFTA 2015 Report}",
      journal = {arXiv e-prints},
         year = 2015,
        month = mar,
          eid = {arXiv:1503.03757},
        pages = {arXiv:1503.03757},
          doi = {10.48550/arXiv.1503.03757},
archivePrefix = {arXiv},
       eprint = {1503.03757},
 primaryClass = {astro-ph.IM},
       adsurl = {https://ui.adsabs.harvard.edu/abs/2015arXiv150303757S}
}

@article{Cappellari2003,
    author = {{Cappellari}, M. and {Copin}, Y.},
    title = "{Adaptive spatial binning of integral-field spectroscopic data using Voronoi tessellations}",
    journal = {\mnras},
    year = 2003,
    volume = {342},
    number = {2},
    pages = {345-354},
    doi = {10.1046/j.1365-8711.2003.06541.x},
    adsurl = {https://ui.adsabs.harvard.edu/abs/2003MNRAS.342..345C}
}

@article{Schlafly2011,
    author = {{Schlafly}, E.~F. and {Finkbeiner}, D.~P.},
    title = "{Measuring Reddening with Sloan Digital Sky Survey Stellar Spectra and Recalibrating SFD}",
    journal = {\apj},
    year = 2011,
    volume = {737},
    number = {2},
    pages = {103},
    doi = {10.1088/0004-637X/737/2/103},
    adsurl = {https://ui.adsabs.harvard.edu/abs/2011ApJ...737..103S}
}

@article{Conley2011,
  title={Supernova Constraints and Systematic Uncertainties from the First Three Years of the Supernova Legacy Survey},
  author={Conley, A. and Guy, J. and Sullivan, M. and Regnault, N. and Astier, P. and Balland, C. and Basa, S. and Carlberg, R. G. and Fouchez, D. and Hardin, D. and others},
  journal={\aaps},
  volume={192},
  number={1},
  pages={1},
  year={2011},
  publisher={IOP Publishing}
}

@article{Childress2014,
  title={Aged to perfection? The history and fate of Type Ia supernova environments},
  author={Childress, M. J. and Wolf, C. and Zahid, H. J.},
  journal={\mnras},
  volume={445},
  number={2},
  pages={1898-1911},
  year={2014},
  publisher={Oxford University Press}
}

@article{Yao2019,
  title={ZTF early observations of Type Ia supernovae. I. Properties of the 2018 sample},
  author={Yao, Yuhan and Miller, AA and Kulkarni, SR and Bulla, M and Masci, FJ and Goldstein, DA and Goobar, A and Nugent, P and Dugas, A and Blagorodnova, N and others},
  journal={\apj},
  volume={886},
  number={2},
  pages={152},
  year={2019},
  publisher={IOP Publishing}
}

@article{Blanton2011,
  title={Improved background subtraction for the Sloan Digital Sky Survey images},
  author={Blanton, Michael R and Kazin, Eyal and Muna, Demitri and others},
  journal={\aj},
  volume={142},
  number={1},
  pages={31},
  year={2011},
  publisher={IOP Publishing}
}

@article{Mast2014,
    author = {{Mast}, D. and {Rosales-Ortega}, F.~F. and {S{\'a}nchez}, S.~F. and {Vilchez}, J.~M. and {Walcher}, C.~J. and {Husemann}, B. and {Marino}, R.~A. and {Kehrig}, C. and {Ceperdello}, L. and {Monreal-Ibero}, A. and {Roth}, M.~M. and {Papaderos}, P.},
    title = "{The effects of spatial resolution on integral field spectrograph surveys at different redshifts. The CALIFA perspective}",
    journal = {\aap},
    year = 2014,
    volume = 561,
    eid = {A129},
    pages = {A129},
    doi = {10.1051/0004-6361/201322195}
}

@article{DominguezSanchez2018,
    author = {{Dom{\'{\i}}nguez S{\'a}nchez}, H. and {Huertas-Company}, M. and {Bernardi}, M. and {Tucci}, M. and {Fischer}, J.~L.},
    title = "{Improving galaxy morphologies for SDSS with Deep Learning}",
    journal = {\mnras},
    year = 2018,
    month = jun,
    volume = {476},
    number = {3},
    pages = {3661-3676},
    doi = {10.1093/mnras/sty338},
    archivePrefix = {arXiv},
    eprint = {1801.04966},
    primaryClass = {astro-ph.GA},
    adsurl = {https://ui.adsabs.harvard.edu/abs/2018MNRAS.476.3661D}
}

@book{bevington2003data,
  title={Data reduction and error analysis for the physical sciences},
  author={Bevington, Philip R and Robinson, D Keith},
  edition={3},
  year={2003},
  publisher={McGraw-Hill},
  address={New York}
}

@book{ivezic2014statistics,
  title={Statistics, data mining, and machine learning in astronomy: a practical Python guide for the analysis of survey data},
  author={Ivezi{\'c}, {\v{Z}}eljko and Connolly, Andrew J and VanderPlas, Jacob T and Gray, Alexander},
  year={2014},
  publisher={Princeton University Press}
}

@ARTICLE{Hamuy1996,
       author = {{Hamuy}, M. and {Phillips}, M.~M. and {Suntzeff}, N.~B. and {Schommer}, R.~A. and {Maza}, J. and {Antezan}, A.~R. and {Wischnjewsky}, M. and {Valladares}, G. and {Muena}, C. and {Gonzales}, L.~E. and {Aviles}, R. and {Wells}, L.~A. and {Smith}, R.~C. and {Navarrete}, M. and {Covarrubias}, R. and {Williger}, G.~M. and {Walker}, A.~R. and {Layden}, A.~C. and {Elias}, J.~H. and {Baldwin}, J.~A. and {Hernandez}, M. and {Tirado}, H. and {Ugarte}, P. and {Elston}, R. and {Saavedra}, N. and {Barrientos}, F. and {Costa}, E. and {Lira}, P. and {Ruiz}, M.~T. and {Anguita}, C. and {Gomez}, X. and {Ortiz}, P. and {della Valle}, M. and {Danziger}, J. and {Storm}, J. and {Kim}, Y.-C. and {Bailyn}, C. and {Rubenstein}, E.~P. and {Tucker}, D. and {Cersosimo}, S. and {Mendez}, R.~A. and {Siciliano}, L. and {Sherry}, W. and {Chaboyer}, B. and {Koopmann}, R.~A. and {Geisler}, D. and {Sarajedini}, A. and {Dey}, A. and {Tyson}, N. and {Rich}, R.~M. and {Gal}, R. and {Lamontagne}, R. and {Caldwell}, N. and {Guhathakurta}, P. and {Phillips}, A.~C. and {Szkody}, P. and {Prosser}, C. and {Ho}, L.~C. and {McMahan}, R. and {Baggley}, G. and {Cheng}, K.-P. and {Havlen}, R. and {Wakamatsu}, K. and {Janes}, K. and {Malkan}, M. and {Baganoff}, F. and {Seitzer}, P. and {Shara}, M. and {Sturch}, C. and {Hesser}, J. and {Hartigan}, P. and {Hughes}, J. and {Welch}, D. and {Williams}, T.~B. and {Ferguson}, H. and {Francis}, P.~J. and {French}, L. and {Bolte}, M. and {Roth}, J. and {Odewahn}, S. and {Howell}, S. and {Krzeminski}, W.},
        title = "{BVRI Light Curves for 29 Type IA Supernovae}",
      journal = {\aj},
         year = 1996,
        month = dec,
       volume = {112},
        pages = {2408},
          doi = {10.1086/118192},
archivePrefix = {arXiv},
       eprint = {astro-ph/9609064},
 primaryClass = {astro-ph},
       adsurl = {https://ui.adsabs.harvard.edu/abs/1996AJ....112.2408H}
}

@ARTICLE{Hamuy2000,
       author = {{Hamuy}, Mario and {Trager}, S.~C. and {Pinto}, Philip A. and {Phillips}, M.~M. and {Schommer}, R.~A. and {Ivanov}, Valentin and {Suntzeff}, Nicholas B.},
        title = "{A Search for Environmental Effects on Type IA Supernovae}",
      journal = {\aj},
         year = 2000,
        month = sep,
       volume = {120},
       number = {3},
        pages = {1479-1486},
          doi = {10.1086/301527},
archivePrefix = {arXiv},
       eprint = {astro-ph/0005213},
 primaryClass = {astro-ph},
       adsurl = {https://ui.adsabs.harvard.edu/abs/2000AJ....120.1479H}
}

@ARTICLE{Kennicutt1998,
       author = {{Kennicutt}, Robert C., Jr.},
        title = "{Star Formation in Galaxies Along the Hubble Sequence}",
      journal = {\araa},
         year = 1998,
       volume = {36},
        pages = {189-232},
          doi = {10.1146/annurev.astro.36.1.189},
       adsurl = {https://ui.adsabs.harvard.edu/abs/1998ARA%26A..36..189K}
}

@ARTICLE{Conroy2013,
       author = {{Conroy}, Charlie},
        title = "{Modeling the Panchromatic Spectral Energy Distributions of Galaxies}",
      journal = {\araa},
         year = 2013,
       volume = {51},
        pages = {393-455},
          doi = {10.1146/annurev-astro-082812-141017},
       adsurl = {https://ui.adsabs.harvard.edu/abs/2013ARA%26A..51..393C}
}

@article{Gallazzi2005,
  author  = {Gallazzi, Anna and Charlot, St{\'e}phane and Brinchmann, Jarle and White, Simon D. M. and Tremonti, Christy A.},
  title   = {The ages and metallicities of galaxies in the local universe},
  journal = {\mnras},
  volume  = {362},
  number  = {1},
  pages   = {41-58},
  year    = {2005},
  doi     = {10.1111/j.1365-2966.2005.09321.x},
  eprint  = {astro-ph/0506539},
  archivePrefix = {arXiv}
}

@article{Cowie1996,
  author  = {Cowie, Lennox L. and Songaila, Antoinette and Hu, Esther M. and Cohen, Judith G.},
  title   = {New Insight on Galaxy Formation and Evolution from Keck Spectroscopy of the Hawaii Deep Fields},
  journal = {\aj},
  volume  = {112},
  pages   = {839},
  year    = {1996},
  doi     = {10.1086/118058},
  eprint  = {astro-ph/9606079},
  archivePrefix = {arXiv}
}

@article{Thomas2005,
  author  = {Thomas, Daniel and Maraston, Claudia and Bender, Ralf and {de Oliveira}, Claudia Mendes},
  title   = {The Epochs of Early-Type Galaxy Formation as a Function of Environment},
  journal = {\apj},
  volume  = {621},
  number  = {2},
  pages   = {673-694},
  year    = {2005},
  doi     = {10.1086/426932},
  eprint  = {astro-ph/0410209},
  archivePrefix = {arXiv}
}

@article{Meng2019,
doi = {10.3847/1538-4357/ab4e10},
url = {https://doi.org/10.3847/1538-4357/ab4e10},
year = {2019},
month = {nov},
publisher = {The American Astronomical Society},
volume = {886},
number = {1},
pages = {58},
author = {Meng, Xiang-Cun},
title = {High-velocity Feature as the Indicator of the Stellar Population of Type Ia Supernovae},
journal = {\apj}
}

@article{Branch1993,
  author  = {Branch, David and Fisher, Adam and Nugent, Peter},
  title   = {On the Relative Frequencies of Spectroscopically Normal and Peculiar Type Ia Supernovae},
  journal = {\aj},
  year    = {1993},
  volume  = {106},
  pages   = {2383-2391},
  doi     = {10.1086/116810},
  adsurl  = {https://ui.adsabs.harvard.edu/abs/1993AJ....106.2383B}
}

@article{Schmidt1998,
doi = {10.1086/306308},
url = {https://doi.org/10.1086/306308},
year = {1998},
month = {nov},
publisher = {},
volume = {507},
number = {1},
pages = {46},
author = {Schmidt, Brian P. and Suntzeff, Nicholas B. and Phillips, M. M. and Schommer, Robert A. and Clocchiatti, Alejandro and Kirshner, Robert P. and Garnavich, Peter and Challis, Peter and Leibundgut, B. and Spyromilio, J. and Riess, Adam G. and Filippenko, Alexei V. and Hamuy, Mario and Smith, R. Chris and Hogan, Craig and Stubbs, Christopher and Diercks, Alan and Reiss, David and Gilliland, Ron and Tonry, John and Maza, José and Dressler, A. and Walsh, J. and Ciardullo, R.},
title = {The High-Z Supernova Search: Measuring Cosmic Deceleration and Global Curvature of the Universe Using Type Ia Supernovae*},
journal = {\apj}
}

@ARTICLE{Rose2021,
       author = {{Rose}, B.~M. and {Rubin}, D. and {Strolger}, L. and {Garnavich}, P.~M.},
        title = "{Host Galaxy Mass Combined with Local Stellar Age Improve Type Ia Supernovae Distances}",
      journal = {\apj},
         year = 2021,
        month = mar,
       volume = {909},
       number = {1},
          eid = {28},
        pages = {28},
          doi = {10.3847/1538-4357/abd550},
archivePrefix = {arXiv},
       eprint = {2012.01460},
 primaryClass = {astro-ph.GA},
       adsurl = {https://ui.adsabs.harvard.edu/abs/2021ApJ...909...28R}
}

@ARTICLE{Pskovskii1977,
       author = {{Pskovskii}, Iu. P.},
        title = "{Light curves, color curves, and expansion velocity of type I supernovae as functions of the rate of brightness decline}",
      journal = {\sovast},
         year = 1977,
        month = dec,
       volume = {21},
        pages = {675},
       adsurl = {https://ui.adsabs.harvard.edu/abs/1977SvA....21..675P}
}

@phdthesis{Rust1974,
       author = {{Rust}, B.~W.},
        title = "{The Use of Supernovae Light Curves for Testing the Expansion Hypothesis and Other Cosmological Relations.}",
       school = {Oak Ridge National Laboratory, Tennessee},
         year = 1974,
        month = jan,
       adsurl = {https://ui.adsabs.harvard.edu/abs/1974PhDT.........7R}
}

@article{Hillebrandt2013,
  author       = {Hillebrandt, Wolfgang and Kromer, Markus and R{\"o}pke, Friedrich K. and Ruiter, Ashley J.},
  title        = {Towards an understanding of Type Ia supernovae from a synthesis of theory and observations},
  journal      = {Front. Phys.},
  year         = {2013},
  volume       = {8},
  number       = {2},
  pages        = {116-143},
  doi          = {10.1007/s11467-013-0303-2}
}

@article{Maoz2014,
  author       = {Maoz, Dan and Mannucci, Filippo and Nelemans, Gijs},
  title        = {Observational Clues to the Progenitors of Type Ia Supernovae},
  journal      = {\araa},
  year         = {2014},
  volume       = {52},
  pages        = {107-170},
  doi          = {10.1146/annurev-astro-082812-141031}
}

@article{RuiterSeitenzahl2025,
       author = {{Ruiter}, Ashley Jade and {Seitenzahl}, Ivo Rolf},
        title = "{Type Ia supernova progenitors: a contemporary view of a long-standing puzzle}",
      journal = {\aapr},
         year = 2025,
        month = dec,
       volume = {33},
       number = {1},
          eid = {1},
        pages = {1},
          doi = {10.1007/s00159-024-00158-9},
archivePrefix = {arXiv},
       eprint = {2412.01766},
 primaryClass = {astro-ph.SR},
       adsurl = {https://ui.adsabs.harvard.edu/abs/2025A&ARv..33....1R}
}

@article{Khokhlov1991,
  author       = {Khokhlov, A. M.},
  title        = {Delayed detonation model for Type Ia supernovae},
  journal      = {\aap},
  year         = {1991},
  volume       = {245},
  pages        = {114-128}
}

@article{HoeflichKhokhlov1996,
  author       = {H{\"o}flich, Peter and Khokhlov, Alexander M.},
  title        = {Explosion Models for Type IA Supernovae: A Comparison with Observed Light Curves, Distances, H$_0$, and Q$_0$},
  journal      = {\apj},
  year         = {1996},
  volume       = {457},
  pages        = {500-528},
  doi          = {10.1086/176748}
}

@article{Fink2010,
  author       = {Fink, M. and R{\"o}pke, F. K. and Hillebrandt, W. and Seitenzahl, I. R. and Sim, S. A. and Kromer, M.},
  title        = {Double-detonation sub-Chandrasekhar supernovae: can minimum helium shell masses detonate the core?},
  journal      = {\aap},
  year         = {2010},
  volume       = {514},
  pages        = {A53},
  doi          = {10.1051/0004-6361/200913892}
}

@article{Sim2010,
  author       = {Sim, S. A. and R{\"o}pke, F. K. and Hillebrandt, W. and Kromer, M. and Pakmor, R. and Fink, M. and Ruiter, A. J. and Seitenzahl, I. R.},
  title        = {Detonations in Sub-Chandrasekhar-mass C+O White Dwarfs},
  journal      = {\apjl},
  year         = {2010},
  volume       = {714},
  number       = {1},
  pages        = {L52-L57},
  doi          = {10.1088/2041-8205/714/1/L52}
}

@article{Pakmor2012,
  author       = {Pakmor, R. and Kromer, M. and Taubenberger, S. and Sim, S. A. and R{\"o}pke, F. K. and Hillebrandt, W.},
  title        = {Normal Type Ia supernovae from violent mergers of white dwarf binaries},
  journal      = {\apjl},
  year         = {2012},
  volume       = {747},
  number       = {1},
  pages        = {L10},
  doi          = {10.1088/2041-8205/747/1/L10}
}

@ARTICLE{Mannucci2005,
       author = {{Mannucci}, F. and {Della Valle}, M. and {Panagia}, N. and {Cappellaro}, E. and {Cresci}, G. and {Maiolino}, R. and {Petrosian}, A. and {Turatto}, M.},
        title = "{The supernova rate per unit mass}",
      journal = {\aap},
         year = 2005,
        month = apr,
       volume = {433},
       number = {3},
        pages = {807-814},
          doi = {10.1051/0004-6361:20041411},
archivePrefix = {arXiv},
       eprint = {astro-ph/0411450},
 primaryClass = {astro-ph},
       adsurl = {https://ui.adsabs.harvard.edu/abs/2005A&A...433..807M}
}

@ARTICLE{Scannapieco2005,
       author = {{Scannapieco}, Evan and {Bildsten}, Lars},
        title = "{The Type Ia Supernova Rate}",
      journal = {\apjl},
         year = 2005,
        month = aug,
       volume = {629},
       number = {2},
        pages = {L85-L88},
          doi = {10.1086/452632},
archivePrefix = {arXiv},
       eprint = {astro-ph/0507456},
 primaryClass = {astro-ph},
       adsurl = {https://ui.adsabs.harvard.edu/abs/2005ApJ...629L..85S}
}

@article{Aubourg2008,
	author = {{Aubourg, É.} and {Tojeiro, R.} and {Jimenez, R.} and {Heavens, A.} and {Strauss, M. A.} and {Spergel, D. N.}},
	title = {Evidence of short-lived SN Ia progenitors},
	DOI= "10.1051/0004-6361:200809796",
	url= "https://doi.org/10.1051/0004-6361:200809796",
	journal = {\aap},
	year = 2008,
	volume = 492,
	number = 3,
	pages = "631-636",
}

@article{Pruzhinskaya2020,
    author = {Pruzhinskaya, M. V. and Novikova, A. P. and Paeltz, L. and Gangler, E. and Rosnet, P. and Blinnikov, S. I.},
    title = "{The influence of host galaxy morphology on the properties of Type Ia supernovae from the Pantheon sample}",
    journal = {\mnras},
    year = 2020,
    month = dec,
    volume = {499},
    number = {4},
    pages = {5121-5135},
    doi = {10.1093/mnras/staa3151},
    adsurl = {https://ui.adsabs.harvard.edu/abs/2020MNRAS.499.5121P},
}

@ARTICLE{Burgaz2025,
       author = {{Burgaz}, U. and {Maguire}, K. and {Dimitriadis}, G. and {Smith}, M. and {Sollerman}, J. and {Galbany}, L. and {Rigault}, M. and {Goobar}, A. and {Johansson}, J. and {Kim}, Y.-L. and {Alburai}, A. and {Amenouche}, M. and {Deckers}, M. and {Ginolin}, M. and {Harvey}, L. and {Muller-Bravo}, T.~E. and {Nordin}, J. and {Phan}, K. and {Rosnet}, P. and {Nugent}, P.~E. and {Terwel}, J.~H. and {Graham}, M. and {Hale}, D. and {Kasliwal}, M.~M. and {Laher}, R.~R. and {Neill}, J.~D. and {Purdum}, J. and {Rusholme}, B.},
        title = "{ZTF SN Ia DR2: Properties of the low-mass host galaxies of Type Ia supernovae in a volume-limited sample}",
      journal = {\aap},
         year = 2025,
        month = feb,
       volume = {694},
          eid = {A13},
        pages = {A13},
          doi = {10.1051/0004-6361/202452571},
archivePrefix = {arXiv},
       eprint = {2412.14262},
 primaryClass = {astro-ph.HE},
       adsurl = {https://ui.adsabs.harvard.edu/abs/2025A&A...694A..13B}
}

@ARTICLE{Popovic2025,
       author = {{Popovic}, B. and {Rigault}, M. and {Smith}, M. and {Ginolin}, M. and {Goobar}, A. and {Kenworthy}, W.~D. and {Ganot}, C. and {Ruppin}, F. and {Dimitriadis}, G. and {Johansson}, J. and {Amenouche}, M. and {Aubert}, M. and {Barjou-Delayre}, C. and {Burgaz}, U. and {Carreres}, B. and {Feinstein}, F. and {Fouchez}, D. and {Galbany}, L. and {de Jaeger}, T. and {Kim}, Y.-L. and {Lacroix}, L. and {Nugent}, P.~E. and {Racine}, B. and {Rosselli}, D. and {Rosnet}, P. and {Sollerman}, J. and {Hale}, D. and {Laher}, R. and {M{\"u}ller-Bravo}, T.~E. and {Reed}, R. and {Rusholme}, B. and {Terwel}, J.},
        title = "{ZTF SN Ia DR2: Evidence of changing dust distribution with redshift using type Ia supernovae}",
      journal = {\aap},
         year = 2025,
        month = feb,
       volume = {694},
          eid = {A5},
        pages = {A5},
          doi = {10.1051/0004-6361/202450391},
archivePrefix = {arXiv},
       eprint = {2406.06215},
 primaryClass = {astro-ph.CO},
       adsurl = {https://ui.adsabs.harvard.edu/abs/2025A&A...694A...5P}
}

@ARTICLE{Cai2026,
    author = "Cai, Rong-Gen and Wang, Shao-Jiang",
    title = "{The Hubble tension: A decade review}",
    eprint = "2606.20434",
    archivePrefix = "arXiv",
    primaryClass = "astro-ph.CO",
    doi = "10.1088/1674-4527/ae842f",
    journal = "RAA",
    volume = "26",
    pages = "084011",
    year = "2026"
}

@ARTICLE{Li2011,
       author = {{Li}, Weidong and {Bloom}, Joshua S. and {Podsiadlowski}, Philipp and {Miller}, Adam A. and {Cenko}, S. Bradley and {Jha}, Saurabh W. and {Sullivan}, Mark and {Howell}, D. Andrew and {Nugent}, Peter E. and {Butler}, Nathaniel R. and {Ofek}, Eran O. and {Kasliwal}, Mansi M. and {Richards}, Joseph W. and {Stockton}, Alan and {Shih}, Hsin-Yi and {Bildsten}, Lars and {Shara}, Michael M. and {Bibby}, Joanne and {Filippenko}, Alexei V. and {Ganeshalingam}, Mohan and {Silverman}, Jeffrey M. and {Kulkarni}, S.~R. and {Law}, Nicholas M. and {Poznanski}, Dovi and {Quimby}, Robert M. and {McCully}, Curtis and {Patel}, Brandon and {Maguire}, Kate and {Shen}, Ken J.},
        title = "{Exclusion of a luminous red giant as a companion star to the progenitor of supernova SN 2011fe}",
      journal = {\nat},
         year = 2011,
        month = dec,
       volume = {480},
       number = {7377},
        pages = {348-350},
          doi = {10.1038/nature10646},
archivePrefix = {arXiv},
       eprint = {1109.1593},
 primaryClass = {astro-ph.CO},
       adsurl = {https://ui.adsabs.harvard.edu/abs/2011Natur.480..348L}
}

@ARTICLE{McCully2014,
       author = {{McCully}, Curtis and {Jha}, Saurabh W. and {Foley}, Ryan J. and {Bildsten}, Lars and {Fong}, Wen-Fai and {Kirshner}, Robert P. and {Marion}, G.~H. and {Riess}, Adam G. and {Stritzinger}, Maximilian D.},
        title = "{A luminous, blue progenitor system for the type Iax supernova 2012Z}",
      journal = {\nat},
         year = 2014,
        month = aug,
       volume = {512},
       number = {7512},
        pages = {54-56},
          doi = {10.1038/nature13615},
archivePrefix = {arXiv},
       eprint = {1408.1089},
 primaryClass = {astro-ph.SR},
       adsurl = {https://ui.adsabs.harvard.edu/abs/2014Natur.512...54M}
}

@ARTICLE{Toy2025,
       author = {{Toy}, M. and {Wiseman}, P. and {Sullivan}, M. and {Scolnic}, D. and {Vincenzi}, M. and {Brout}, D. and {Davis}, T.~M. and {Frohmaier}, C. and {Galbany}, L. and {Lidman}, C. and {Lee}, J. and {Kelsey}, L. and {Kessler}, R. and {M{\"o}ller}, A. and {Popovic}, B. and {S{\'a}nchez}, B.~O. and {Shah}, P. and {Smith}, M. and {Aguena}, M. and {Allam}, S. and {Alves}, O. and {Bacon}, D. and {Brooks}, D. and {Burke}, D.~L. and {Rosell}, A. Carnero and {Carretero}, J. and {da Costa}, L.~N. and {Pereira}, M.~E.~S. and {Desai}, S. and {Diehl}, H.~T. and {Doel}, P. and {Drlica-Wagner}, A. and {Everett}, S. and {Ferrero}, I. and {Flaugher}, B. and {Frieman}, J. and {Garc{\'\i}a-Bellido}, J. and {Gatti}, M. and {Gaztanaga}, E. and {Giannini}, G. and {Gruendl}, R.~A. and {Gutierrez}, G. and {Hinton}, S.~R. and {Hollowood}, D.~L. and {Honscheid}, K. and {James}, D.~J. and {Kuehn}, K. and {Lahav}, O. and {Lee}, S. and {Marshall}, J.~L. and {Mena-Fern{\'a}ndez}, J. and {Miquel}, R. and {Palmese}, A. and {Pieres}, A. and {Malag{\'o}n}, A.~A. Plazas and {Romer}, A.~K. and {Samuroff}, S. and {Sanchez}, E. and {Cid}, D. Sanchez and {Schubnell}, M. and {Suchyta}, E. and {Swanson}, M.~E.~C. and {Tarle}, G. and {Tucker}, D.~L. and {Vikram}, V. and {Walker}, A.~R. and {Weaverdyck}, N.},
        title = "{Reduction of the type Ia supernova host galaxy step in the outer regions of galaxies}",
      journal = {\mnras},
         year = 2025,
        month = mar,
       volume = {538},
       number = {1},
        pages = {181-197},
          doi = {10.1093/mnras/staf248},
archivePrefix = {arXiv},
       eprint = {2408.03749},
 primaryClass = {astro-ph.CO},
       adsurl = {https://ui.adsabs.harvard.edu/abs/2025MNRAS.538..181T}
}

@article{KangX2017,
    author = {Kang, Xiaoyu and Zhang, Fenghui and Chang, Ruixiang},
    title = {The role of environment on the star formation history of disc galaxies},
    journal = {\mnras},
    volume = {469},
    number = {2},
    pages = {1636-1646},
    year = {2017},
    month = {08},
    issn = {0035-8711},
    doi = {10.1093/mnras/stx1001},
    url = {https://doi.org/10.1093/mnras/stx1001},
    eprint = {https://academic.oup.com/mnras/article-pdf/469/2/1636/17416754/stx1001.pdf},
}

@ARTICLE{Goddard2017,
       author = {{Goddard}, D. and {Thomas}, D. and {Maraston}, C. and {Westfall}, K. and {Etherington}, J. and {Riffel}, R. and {Mallmann}, N.~D. and {Zheng}, Z. and {Argudo-Fern{\'a}ndez}, M. and {Lian}, J. and {Bershady}, M. and {Bundy}, K. and {Drory}, N. and {Law}, D. and {Yan}, R. and {Wake}, D. and {Weijmans}, A. and {Bizyaev}, D. and {Brownstein}, J. and {Lane}, R.~R. and {Maiolino}, R. and {Masters}, K. and {Merrifield}, M. and {Nitschelm}, C. and {Pan}, K. and {Roman-Lopes}, A. and {Storchi-Bergmann}, T. and {Schneider}, D.~P.},
        title = "{SDSS-IV MaNGA: Spatially resolved star formation histories in galaxies as a function of galaxy mass and type}",
      journal = {\mnras},
         year = 2017,
        month = apr,
       volume = {466},
       number = {4},
        pages = {4731-4758},
          doi = {10.1093/mnras/stw3371},
archivePrefix = {arXiv},
       eprint = {1612.01546},
 primaryClass = {astro-ph.GA},
       adsurl = {https://ui.adsabs.harvard.edu/abs/2017MNRAS.466.4731G}
}

@ARTICLE{KangX2016,
       author = {{Kang}, Xiaoyu and {Zhang}, Fenghui and {Chang}, Ruixiang and {Wang}, Lang and {Cheng}, Liantao},
        title = "{The star formation history of low-mass disk galaxies: A case study of NGC 300}",
      journal = {\aap},
         year = 2016,
        month = jan,
       volume = {585},
          eid = {A20},
        pages = {A20},
          doi = {10.1051/0004-6361/201527041},
archivePrefix = {arXiv},
       eprint = {1510.04768},
 primaryClass = {astro-ph.GA},
       adsurl = {https://ui.adsabs.harvard.edu/abs/2016A&A...585A..20K}
}

@article{Chung2025,
    author = {Chung, Chul and Park, Seunghyun and Son, Junhyuk and Cho, Hyejeon and Lee, Young-Wook},
    title = {Strong progenitor age bias in supernova cosmology – I. Robust and ubiquitous evidence from a larger sample of host galaxies in a broader redshift range},
    journal = {\mnras},
    volume = {538},
    number = {4},
    pages = {3340-3350},
    year = {2025},
    month = {04},
    issn = {0035-8711},
    doi = {10.1093/mnras/staf497},
    url = {https://doi.org/10.1093/mnras/staf497},
    eprint = {https://academic.oup.com/mnras/article-pdf/538/4/3340/62787583/staf497.pdf},
}

@article{Hoang2026,
      title={Progenitor age-bias-corrected Type Ia supernovae favor a logarithmic luminosity-distance relation}, 
      author={Hoang Ky Nguyen},
      year={2026},
      eprint={2608.00806},
      archivePrefix={arXiv},
      primaryClass={astro-ph.CO},
      url={https://arxiv.org/abs/2608.00806}, 
}

\begin{appendix}
\section{Detailed method of host-mass re-weighting}
\label{appendix}
First, we re-weighted our SN~Ia sample so that its host-mass distribution matches the ZTF SN~Ia DR2 reference sample. The host mass was divided into discrete bins, and in each bin ($k$) we computed the normalised fractions:
\begin{equation}
p_{\rm ZTF}(k)=\frac{N_{\rm ZTF}(k)}{\sum_j N_{\rm ZTF}(j)},
\qquad
p_{\rm our}(k)=\frac{N_{\rm our}(k)}{\sum_j N_{\rm our}(j)} \, ,
\end{equation}
where $N_{\rm ZTF}(k)$ and $N_{\rm our}(k)$ are the numbers of SNe in the ZTF reference sample and in our sample, respectively. Each SN in our sample was then assigned a bin-dependent weight:
\begin{equation}
w_i^{\rm raw}=\frac{p_{\rm ZTF}(k_i)}{p_{\rm our}(k_i)} \, ,
\end{equation}
where $k_i$ denotes the host-mass bin occupied by the $i$-th SN. In this way, host-mass bins that are underrepresented in our sample relative to ZTF are up-weighted, whereas overrepresented bins are down-weighted. To prevent a very small number of objects from dominating the analysis, the raw bin-wise weights were capped at an upper limit of 5 and then re-normalised such that their mean over the full sample equals unity:
\begin{equation}
w_i^{\rm cap}=\min\!\left(w_i^{\rm raw},\,5\right),
\end{equation}
\begin{equation}
w_i=\frac{w_i^{\rm cap}}{\langle w^{\rm cap}\rangle}.
\end{equation}
We then repeated the mass-only, age-only, and joint fits using these sample weights. In practice, this re-weighting enters the fit by modifying the original $\chi^2$ function such that each SN contributes in proportion to its sample weight, $w_i$:
\begin{equation}
\chi^2_{\rm weighted}
=
\sum_{i=1}^{N}
w_i\,
\frac{[\mu_{\text{obs},i}(\alpha, \beta) - \mu_{\text{model}}(z_i)]^2}
{\sigma_{\mu,i}^2 + \sigma_{\text{int}}^2} .
\end{equation}
Here, each SN contributes to the weighted $\chi^2$ through the squared difference between its observed and model distance modulus, scaled by the total variance $\sigma_{\mu,i}^2+\sigma_{\rm int}^2$ and multiplied by the sample weight $w_i$. The best-fit parameters are then obtained by minimising this weighted $\chi^2$ over the full SN sample.

Second, we repeated the same set of fits using the MaNGA volume weight \texttt{ESWEIGHT} from the MaNGA \texttt{drpall} catalogue. In the MaNGA data model, \texttt{ESWEIGHT} is defined for the combined Primary and full Secondary samples and corrects the selection to an effective volume-limited sample. Since our host galaxies were cross-matched against the full \texttt{drpall} catalogue rather than restricted to a narrower MaNGA statistical subsample, we used ESWEIGHT here as an approximate volume-weighting correction for the MaNGA selection function and therefore as a robustness test of our inferred environmental steps.

\section{Method for determining the potential impact of the local LWA on cosmological inference}
\label{appendixB}

\subsection{Cosmological distance model}

The theoretical distance modulus, $\mu_{\text{model}}$, depends on redshift $z$ and the set of cosmological parameters $\theta = \{w, \Omega_\mathrm{m}, H_0\}$. Under the assumption of a flat universe ($\Omega_\mathrm{k} = 0$), its expression is determined by the luminosity distance, $d_L$:
\begin{equation}
\mu_{\text{model}}(z, \theta) = 5 \log_{10} [(1+z) \frac{c}{H_0} \int_0^z \frac{dz'}{E(z'; w, \Omega_\mathrm{m})}] + 25,
\end{equation}
where $E(z)$ is the dimensionless Hubble expansion rate:
\begin{equation}
E(z) = \sqrt{\Omega_\mathrm{m}(1+z)^3 + (1-\Omega_\mathrm{m})(1+z)^{3(1+w)}}.
\end{equation}
In this study, we fixed $H_0 = 70$ km s$^{-1}$ Mpc$^{-1}$ and $\Omega_\mathrm{m} = 0.30$ as fiducial values to focus on observing changes in $w$.

\subsection{Parameterisation of  the local LWA step, likelihood analysis, and bias determination}

To introduce environmental dependence in the fitting, we constructed a modified model. We introduced the step term $\Delta_\mathrm{A}$ and cosmological parameters simultaneously in the fit:
\begin{equation}
\mu_{\text{corr}} = \mu_{\text{obs}} - S(\mathrm{Age}; \Delta_\mathrm{A}),
\end{equation}
where $S(\mathrm{Age}; \Delta_\mathrm{A})$ is defined as follows:
\begin{equation}
S(\mathrm{Age}; \Delta_\mathrm{A}) = \Delta_\mathrm{A} \times [\Theta(Age - Age_{\text{th}}) - 0.5],
\end{equation}
where $\Delta_\mathrm{A}$ is defined as the difference in Hubble residuals between the young environment ($Age < Age_{\mathrm{th}}$) and the old environment ($Age > Age_{\mathrm{th}}$), $\Theta$ is the Heaviside step function, and $Age_{\mathrm{th}}$ is the maximum significance splitting threshold determined in Sect.~\ref{sec:classify}. When the host galaxy age is greater than the threshold, the correction amount is $+\Delta_\mathrm{A}/2$, and conversely $-\Delta_\mathrm{A}/2$. This parameterisation process ensures that $\Delta_\mathrm{A}$ directly corresponds to the absolute luminosity difference between the two subsamples.

We performed the parameter inference using MLE. Under the assumption of Gaussian uncertainties, this is equivalent to minimising the $\chi^2$ statistic:
\begin{equation}
\chi^2(w, \Delta_\mathrm{A}) = \sum_{i=1}^{N} \frac{[\mu_{\text{obs, i}} - S(Age; \Delta) - \mu_{\text{model}}(z_i, w)]^2}{\sigma_{\text{total, i}}^2}.
\end{equation}
To quantify the sensitivity of the inferred $w$ to the environmental
treatment, we performed likelihood profiling under two different assumptions:
1. Baseline case, $\Delta_\mathrm{A} = 0$: ignoring the local LWA step, fitting the case without considering environmental dependence. Its optimal solution is denoted as $w_{\text{base}}$.
2. Local LWA step case, $\Delta_\mathrm{A}$ free: taking the step amplitude $\Delta_\mathrm{A}$ as a free parameter, joint fitting with $w$, minimising $\chi^2$. Its optimal solution is denoted as $w_{\text{corr}}$.

We defined the resulting shift as the difference between the two best-fit values:
\begin{equation}
\Delta w = w_{\text{corr}} - w_{\text{base}}.
\end{equation}

\end{appendix}
\end{document}